\documentclass[12pt]{iopart}

\usepackage{graphicx}
\graphicspath{{Figures/}}
\usepackage{url}
\usepackage{siunitx}
\usepackage{array}
\newcolumntype{L}{>{$}l<{$}} 
\usepackage{xfrac}
\usepackage{textcomp}
\usepackage{bm}

\usepackage{subcaption}
\usepackage{booktabs}
\usepackage{colortbl} 
\usepackage{xcolor}
\usepackage{makecell}
\usepackage{fancyhdr}  
\usepackage{rotating}    
\usepackage{adjustbox}     
\usepackage{float}  

\begin{document}
	
	\title[Numerical analysis of Josephson junction arrays for multi-order quantum voltage steps]{Numerical analysis of Josephson junction arrays for multi-order quantum voltage steps}
	
	\author{Paolo Durandetto}
	\address{Istituto Nazionale di Ricerca Metrologica (INRiM) - Strada delle Cacce, 91 - 10135 Torino (Italy)}
	\ead{p.durandetto@inrim.it}
	\vspace{10pt}
	\begin{indented}
		\item[]April 2025
	\end{indented}
	
	\begin{abstract}
		The dynamics of overdamped Josephson junctions under varying microwave-driving conditions have been studied through numerical simulations using the resistively-shunted junction (RSJ) model, with a focus on primary voltage metrology applications, where a significantly high number of series-connected junctions and stringent uniformity of their electrical parameters are required. The aim is to determine the optimal junction characteristics and external microwave (rf) parameters that maximize the width of quantum voltage levels (Shapiro steps) from order $n=0$ to $n > 1$. Both the rf and dc power requirements, along with the junction parameter spread and power attenuation, are analyzed as key factors that need to be optimized for improved performance of the quantum device. This work aims to advance the development of next-generation Programmable Josephson Voltage Standards (PJVS) with logic architectures that surpass the conventional binary and ternary codifications used in present quantum voltage arrays, while significantly reducing the overall number of junctions as well as the number of sub-arrays and bias lines. Existing technologies exploiting $n = 0$ and $n = \pm 1$ voltage steps are first discussed and analyzed to verify the validity of the simulation model. They are then further investigated to extend their usability with multi-order quantum steps for $n$ up to 3.
		From the simulation results, it follows that present junction technologies may be employed with no modifications for the simultaneous operation of quantum steps up to $n=2$, although optimal power efficiency would require a retrimming of the junction's electrical parameters. On the contrary, extending the highest step order to $n=3$ strictly requires the junction's characteristic parameters to be properly adjusted to maintain sustainable power levels as well as acceptable quantum-locking ranges.

	\end{abstract}

	%
	%
	%
	%
	%


	\section{Introduction}
	The development of quantum voltage standards capable of generating time-dependent voltages has been ongoing since the end of the last century \cite{hamilton1995josephson, benz1996pulse}. Currently, after nearly 30 years of extensive research \cite{rufenacht2018impact}, programmable Josephson voltage standards (PJVS) and Josephson Arbitrary Waveform Synthesizers (JAWS) constructed from low temperature superconductors (LTS) for operation at liquid helium temperatures are available commercially and are still being optimized further.
	Although the primary goals of both PJVS (frequency extension \cite{behr2021ac}) and JAWS (both amplitude and frequency extension \cite{flowers2019development, thomas2024vhf}) present significant challenges, there is still potential for technical simplifications.
	
	All present PJVS implementations use arrays of overdamped Josephson junctions (JJs) segmented in binary, ternary, or mixed sub-arrays, each current-driven either to the zeroth or the first quantum voltage step (positive or negative). The use of second-order quantum steps, either for doubling the PJVS full-scale voltage without affecting resolution or for considerably reducing the number of JJs, sub-arrays and bias lines, was proposed a few years ago and experimentally-tested with a conventional binary-divided array \cite{durandetto2018exploiting, durandetto2019non, durandetto2023recent}. However, the realization of a PJVS device capable of working simultaneously with maximum step order $n_\mathrm{max}>1$ deserves a dedicated study, in order to determine the best operating and physical parameters (critical current, rf-frequency and rf-power) as well as to understand the possible limitations of such unconventional standards, e.g. dissipated power, rf-power attenuation, JJs parameters uniformity, etc. 
	
	In this work, the expected behavior of a rf-current driven overdamped JJ is first studied in the framework of the resistively-shunted junction (RSJ) model \cite{stewart1968current, mccumber1968effect} via computer simulations implemented in \textit{Python}. For those interested, it is worth noting that a free \textit{Python} simulator for easily generating and analyzing the current-voltage ($IV$) characteristics of an ac-biased JJ has recently been made available \cite{rahmonova2024toolkit}.
	However, this toolkit was not employed in the present study, despite sharing some similarities.
	Next, arrays of hundreds or thousands of JJs are examined, considering the variability of JJ electrical parameters, and the power attenuation of the external rf excitation. The objective is to identify the optimal electrical parameters and the dissipated dc and rf power levels, as well as the arrangement of the JJs on the chip, such that the current width of quantum steps up to the second or third order is simultaneously greater than a specified threshold. This requirement stems from the need to ease the accuracy, stability and noise demands on the system components, from the bias electronics to the cryogenic setup, thus ensuring robust voltage quantization in the presence of non-ideal experimental conditions.
	 In Sec.~\ref{sec:single-jj}, a single JJ governed by the RSJ model is simulated for different values of the normalized-units parameters space. Afterward, conversion from normalized to absolute units and real-world effects are taken into account to obtain the expected performances of novel programmable standards, taking as basis LTS PJVS reference both \qty{70}{\giga\hertz} devices fabricated in Europe by Physikalisch-Technische Bundesanstalt (PTB) and Supracon AG \cite{bauer2022josephson, schubert2016dry} and ``low rf-frequency'' PJVSs realized in the United States (US) by the National Institute of Standards and Technology (NIST) \cite{rufenacht2018impact, howe2014nist}.
	This research also aims to support the future design and development of Programmable Josephson Voltage Standards (PJVS) that utilize high-temperature superconductors (HTS) operating at liquid nitrogen temperatures (\qty{77}{\kelvin}). At present, the output voltage capabilities of these systems are limited by technological challenges related to the integration of several HTS JJs into a single array \cite{klushin2020present}. Consequently, exploring higher-order Shapiro steps in conjunction with a reduced number of bias lines presents a promising opportunity.

	\section{Single JJ simulation} \label{sec:single-jj}

	As is customary in the examination of Shapiro steps \cite{benz1996pulse, kautz1995shapiro, maggi1996step, hassel2007millimeter, li2018doubled, sharafiev2018hts, park2021characterization, shelly2020existence}, this simulation relies on the numerical solution of either the RCSJ (resistively-capacitively shunted junction) model or, in cases where the capacitance is negligible, the RSJ nonlinear differential equation for a current-biased JJ that includes both dc and ac (rf) components. It is worth mentioning that this model has recently gained attention for the study of the dual phenomenon of the inverse ac Josephson effect, the Quantum-Phase Slip effect, which lead to quantized current steps known as dual Shapiro steps \cite{shaikhaidarov2022quantized, crescini2023evidence, shaikhaidarov2024feasibility}. 
	Moreover, recent studies have shown that Shapiro steps can also emerge in vortex-based structures, such as superconducting nanobridges and strips, with pinning centers \cite{jelic2015stroboscopic, kozlov2024dynamic}: this Josephson-like behavior is more accurately captured by the time-dependent Ginzburg-Landau equations, which provide a more comprehensive model for vortex dynamics compared to the phase dynamics describing ordinary JJs.
			
	Being the PJVS built on overdamped JJs, the RSJ model has been used to simulate a current-biased JJ shunted by its normal resistance. For the sake of simplicity, junctions size effects are not taken into account in this study, i.e. point-junction model is considered \cite{kautz1995shapiro}. In absolute units, the instantaneous behavior at a given time $t$ of a JJ with critical current $I_\mathrm{c}$ and normal resistance $R_\mathrm{n}$, dc-biased at $I_\mathrm{dc}$ and rf-biased at frequency $f_\mathrm{rf}$ with amplitude $I_\mathrm{rf}$, is governed by
	
	\begin{equation}
		\centering
		I_\mathrm{dc} + I_\mathrm{rf}\,\sin\left( 2\pi\,f_\mathrm{rf}\, t\right)  = I_{\text{c}}\,\sin\left( \varphi(t)\right) + \frac{\Phi_\mathrm{0}}{2\pi\,R_\mathrm{n}}\frac{d\varphi (t)}{dt}
		\label{eq.rsjeq1}
	\end{equation}
	
	\noindent with $\varphi$ the superconducting phase difference across the JJ and $\Phi_\mathrm{0}=h/2e$ the magnetic flux quantum.
	As usual, in the interest of generalization, JJ simulations are performed using normalized units, thereby transforming Eq.~\ref{eq.rsjeq1} into
	\begin{equation}
		\centering
		i_{\text{dc}} + i_{\text{rf}}\sin \left(\Omega_{\text{rf}} \,\tau \right) = \sin \left(\varphi(\tau) \right)  + \frac{d\varphi(\tau)}{d\tau}
		\label{eq.rsjeq2}
	\end{equation}
	
	\noindent being $i_\mathrm{dc}=I_\mathrm{dc}/I_\mathrm{c}$, $i_\mathrm{rf}=I_\mathrm{rf}/I_\mathrm{c}$, $\Omega_\mathrm{rf} = f_\mathrm{rf}/f_\mathrm{c}$ and $\tau=2\pi\,f_\mathrm{c}\, t$, with $f_\mathrm{c} = I_\mathrm{c}\, R_\mathrm{n}/\Phi_\mathrm{0}$ the characteristic frequency of the junction. All variables are dimensionless, except for $\varphi$ and $\tau$, which are expressed in radians.  
	%
	
\begin{figure}[tbhp]
		\centering
		\begin{subfigure}{0.45\textwidth}
			\centering
			\includegraphics[width=\linewidth]{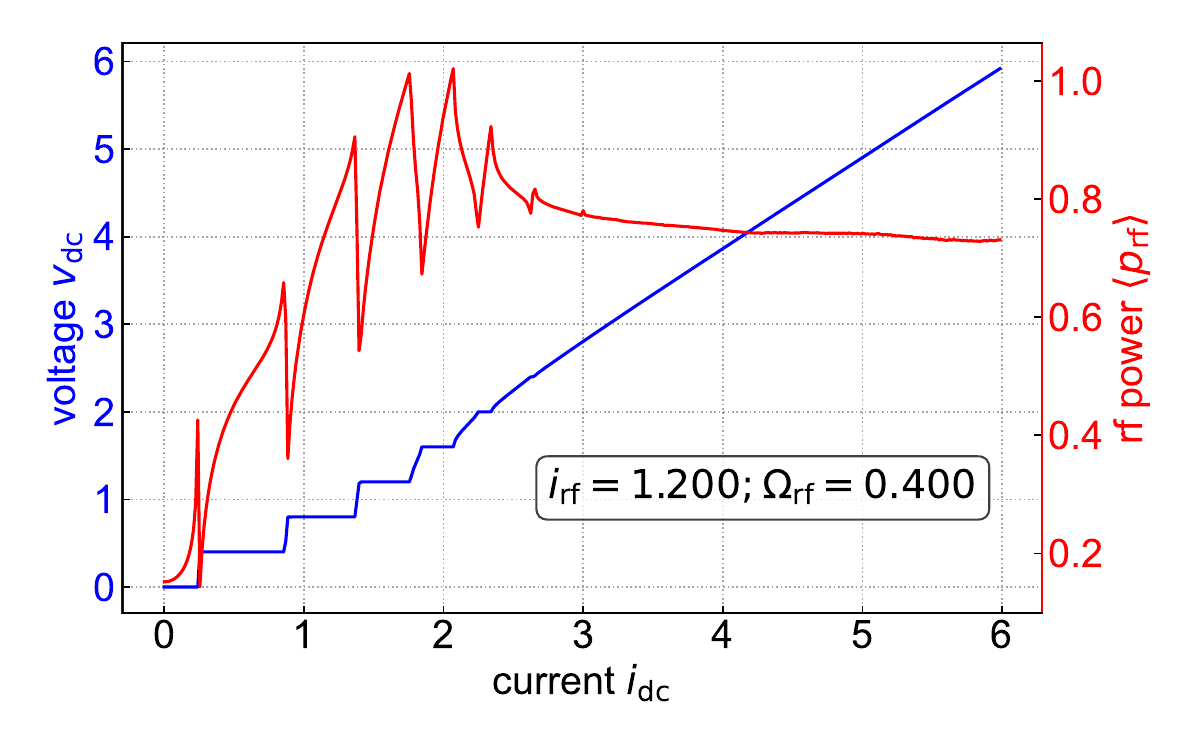}
			\caption{$i_{\mathrm{rf}}=1.2$, $\Omega_{\mathrm{rf}}=0.4$}
			\label{fig:IV_IacP_irf_1.200-Omega_0.400}
		\end{subfigure}
		\hfill
		\begin{subfigure}{0.45\textwidth}
			\centering
			\includegraphics[width=\linewidth]{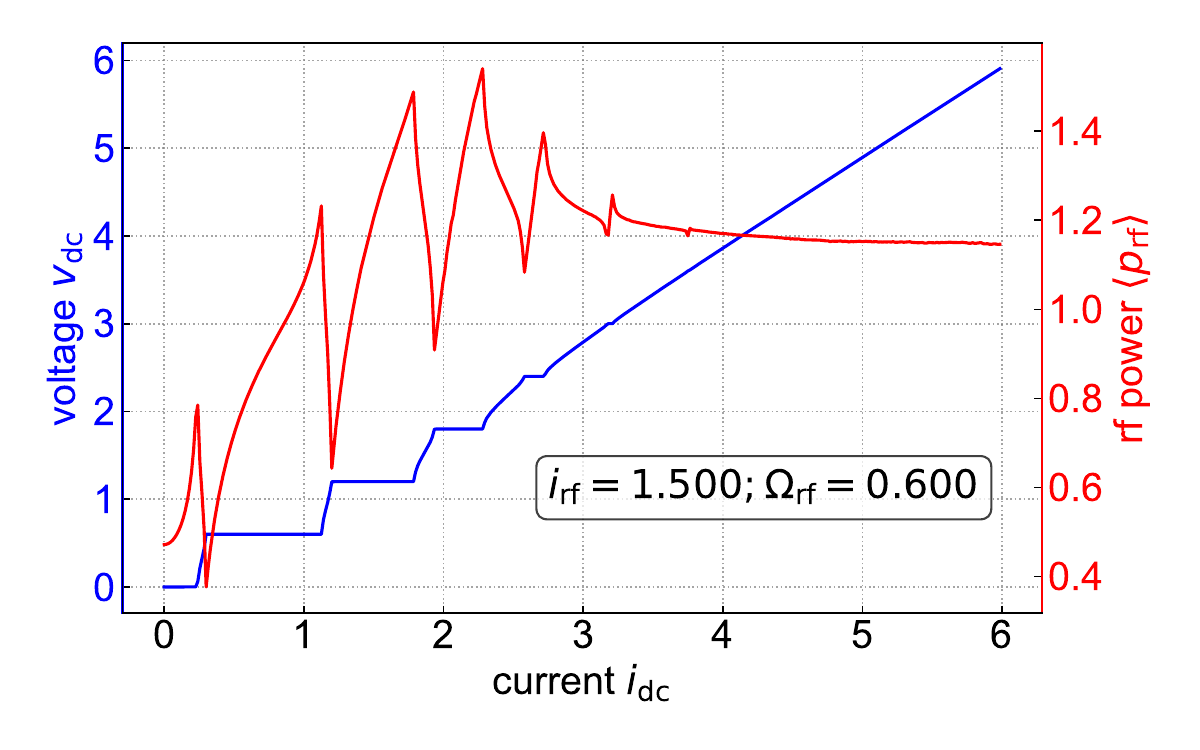}
			\caption{$i_{\mathrm{rf}}=1.5$, $\Omega_{\mathrm{rf}}=0.6$}
			\label{fig:IV_IacP_irf_1.500-Omega_0.600}
		\end{subfigure}
		\hfill
		\begin{subfigure}{0.45\textwidth}
			\centering
			\includegraphics[width=\linewidth]{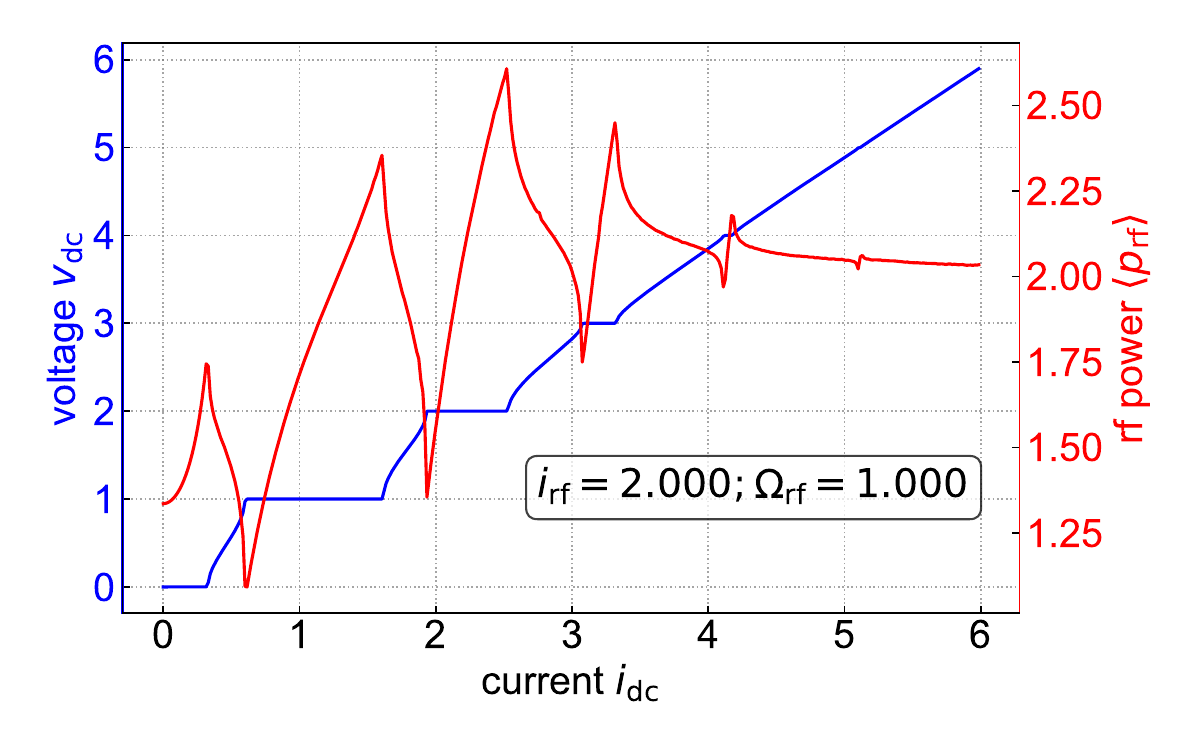}
			\caption{$i_{\mathrm{rf}}=2$, $\Omega_{\mathrm{rf}}=1$}
			\label{fig:IV_IacP_irf_2.000-Omega_1.000}
		\end{subfigure}
		
		\caption{Normalized plots of current-voltage (blue, left y-axis) and  current-average rf power (red, right y-axis) for a single JJ at three ($i_{\mathrm{rf}}$, $\Omega_{\mathrm{rf}}$) pairs.
		}
		\label{fig:IV_IacP_plots}
	\end{figure}

	A \textit{Python} program has been developed to numerically solve Eq.~\ref{eq.rsjeq2} by extracting the instantaneous phase $\varphi(\tau)$ and its time-derivative $d\varphi(\tau) / d\tau$. To this aim, the \textit{odeint} function of the \textit{SciPy} package has been exploited \cite{scipy_odeint}. The JJ $IV$ characteristics have been calculated for several ($i_{\text{rf}}$, $\Omega_{\text{rf}}$) pairs by sweeping the dc current $i_{\text{dc}}$ from 0 to 6 in steps of 0.015 (400 points) and, at each $i_{\text{dc}}$, the normalized dc voltage $v_{\text{dc}} = \langle\frac{d\varphi(\tau)}{d\tau}\rangle$ is calculated by averaging the resulting voltage oscillation over several periods to achieve suitable accuracy. Since the JJ under investigation is non-hysteretic, simulating the decreasing current sweep is not required. Additionally, because of the sinusoidal rf bias, the \textit{IV} curve exhibits symmetry, and only positive dc bias currents are considered, as the negative \textit{IV} portion is the reflection of the positive one.
	The $IV$ characteristics for three ($i_{\text{rf}}$, $\Omega_{\text{rf}}$) pairs are presented in Figure~\ref{fig:IV_IacP_plots} (blue line), where Shapiro steps of order $n = v_\mathrm{dc}/\Omega_\mathrm{rf}$ up to 4 are clearly observable.

	In addition to dc currents and voltages, the instantaneous normalized power $p(\tau)$ has been also evaluated for each ($i_{\mathrm{rf}}$, $\Omega_{\mathrm{rf}}$) pair using the following equation:
	\begin{equation}
		\centering
		p(\tau) = \left[i_{\text{dc}} + i_{\text{rf}}\sin \left(\Omega_{\text{rf}} \,\tau \right)\right]\cdot \frac{d\varphi(\tau)}{d\tau} = \underbrace{p_{\text{dc}}(\tau)}_{i_{\text{dc}}\cdot v_{\text{dc}}} + p_{\text{rf}}(\tau)
		\label{eq.power}
	\end{equation}
	\noindent being $p_\mathrm{dc}(\tau)$ and $p_\mathrm{rf}(\tau)$ its dc and ac power components. The average $\langle p(\tau) \rangle$ thus represents the total power dissipated by the junction, encompassing both dc and rf contributions.
	The average rf-power $\langle p_{\text{rf}} \rangle$ dependence with the bias current $i_\mathrm{dc}$ is illustrated in Figure~\ref{fig:IV_IacP_plots} with the red lines (right y-axis). It can be observed that, in correspondence with quantum voltage steps, the average rf power trend exhibits a monotonically increasing behavior, followed by a decrease when the junction is in the ``off-step'' condition.
	This power value is important for the JJ array discussion in Sec.~\ref{sec:sim_mult_jj}, since it is related to the rf power attenuation of the JJs distributed along the microwave transmission line.

	\subsection{Shapiro steps and rf power}\label{subsec:qsteps_rfpow}
\begin{figure*}[tbhp]
		\centering
		\begin{subfigure}{0.32\textwidth}
			\centering
			\includegraphics[width=\linewidth, trim={0.5cm 0cm 0.5cm 0.5cm}, clip]{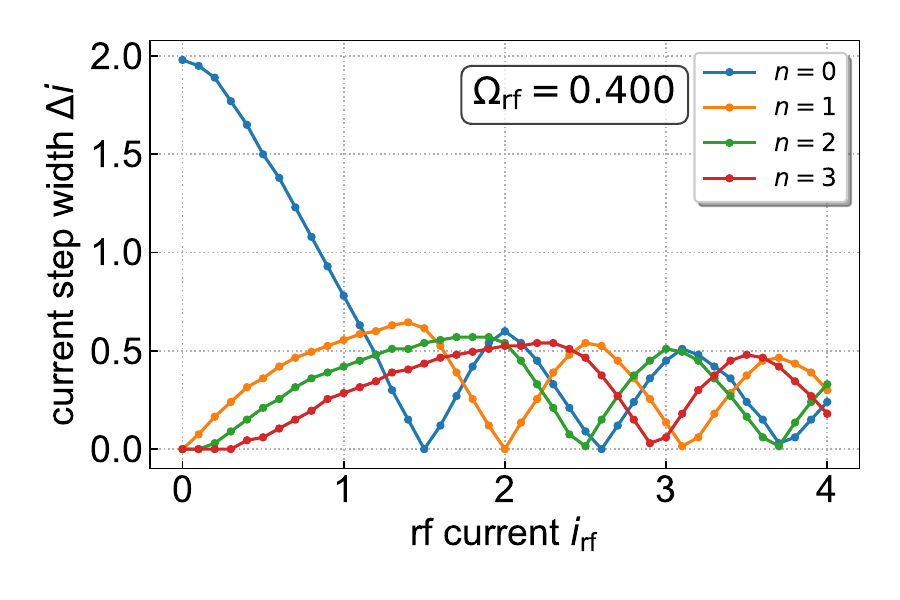}
			\caption{}
			\label{fig:Bessel_Omega_0.400}
		\end{subfigure}
		\begin{subfigure}{0.32\textwidth}
			\centering
			\includegraphics[width=\linewidth, trim={0.3cm 0cm 0.5cm 0.5cm}, clip]{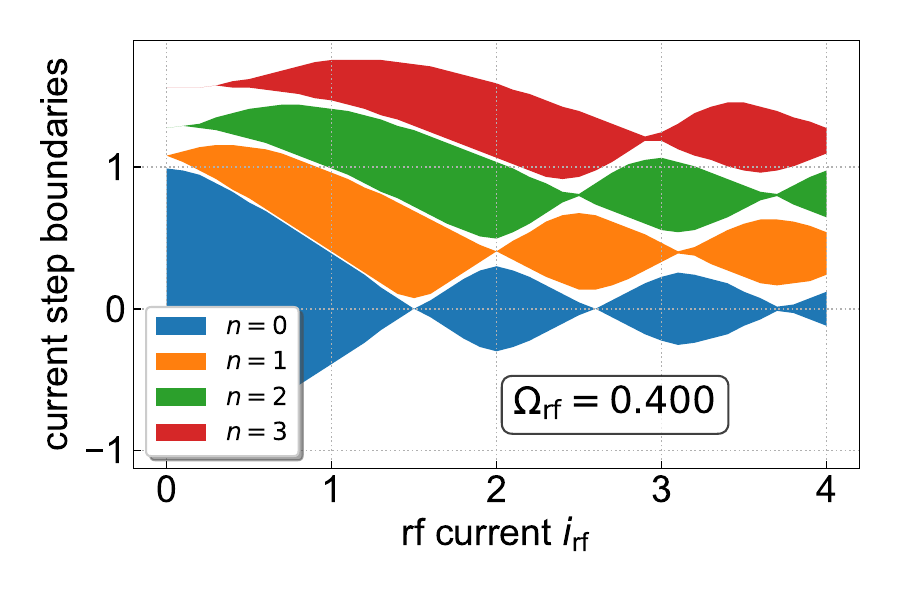}
			\caption{}
			\label{fig:center_width_Omega_0.400}
		\end{subfigure}
		\begin{subfigure}{0.32\textwidth}
			\centering
			\includegraphics[width=\linewidth, trim={0.5cm 0cm 0.5cm 0.5cm}, clip]{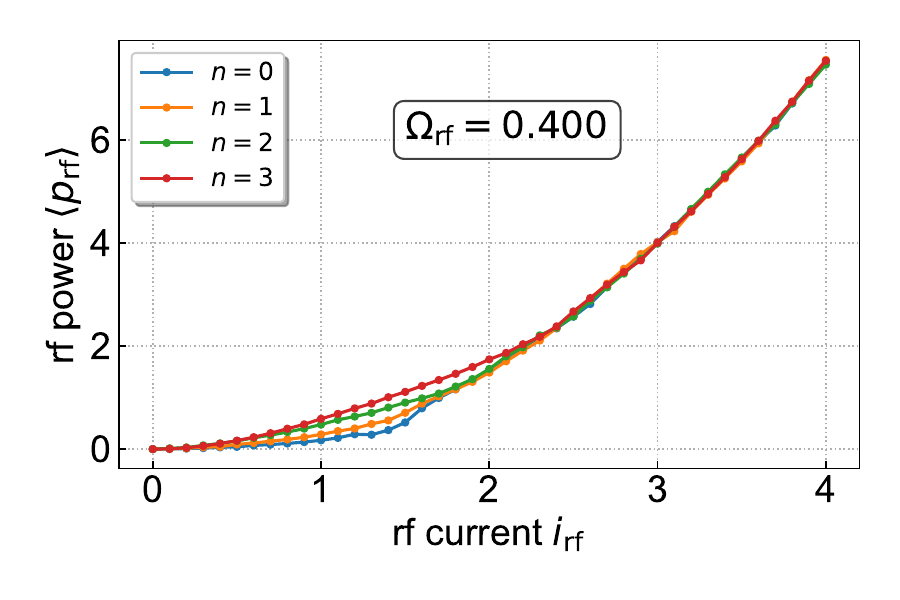}
			\caption{}
			\label{fig:mid_rfpower_Omega_0.400}
		\end{subfigure}
		\begin{subfigure}{0.32\textwidth}
			\centering
			\includegraphics[width=\linewidth, trim={0.5cm 0cm 0.5cm 0.5cm}, clip]{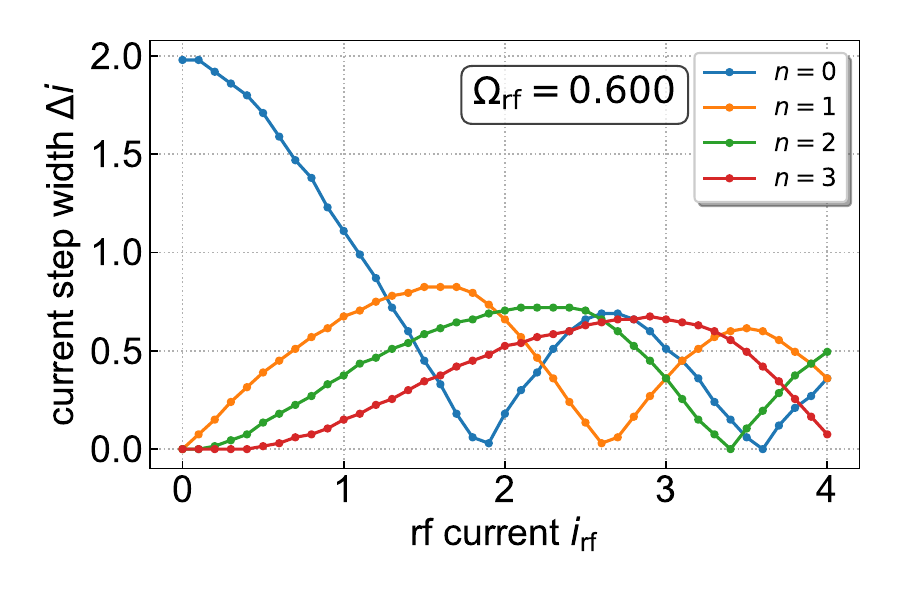}
			\caption{}
			\label{fig:Bessel_Omega_0.600}
		\end{subfigure}
		\begin{subfigure}{0.32\textwidth}
			\centering
			\includegraphics[width=\linewidth, trim={0.5cm 0cm 0.5cm 0.5cm}, clip]{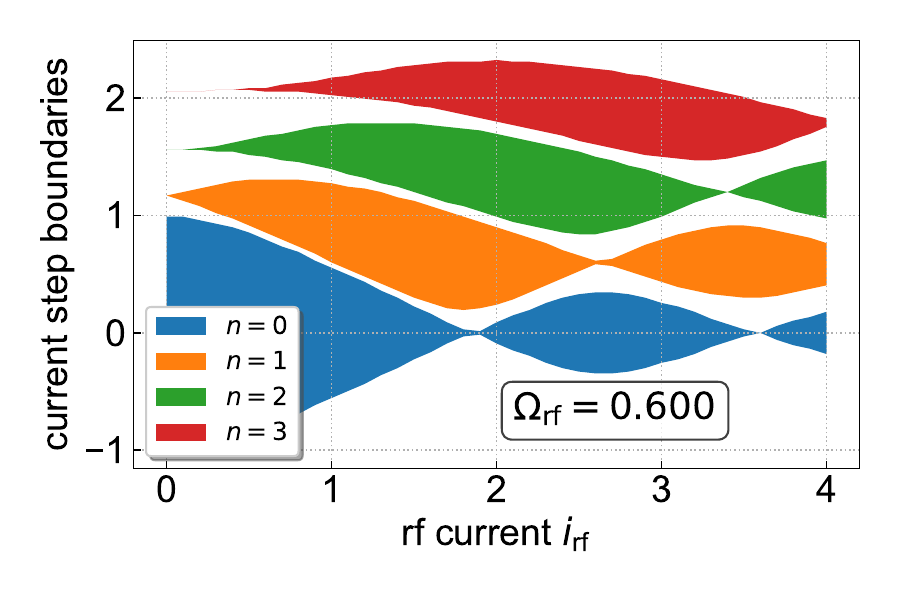}
			\caption{}
			\label{fig:center_width_Omega_0.600}
		\end{subfigure}
		\begin{subfigure}{0.32\textwidth}
			\centering
			\includegraphics[width=\linewidth, trim={0.5cm 0cm 0.5cm 0.5cm}, clip]{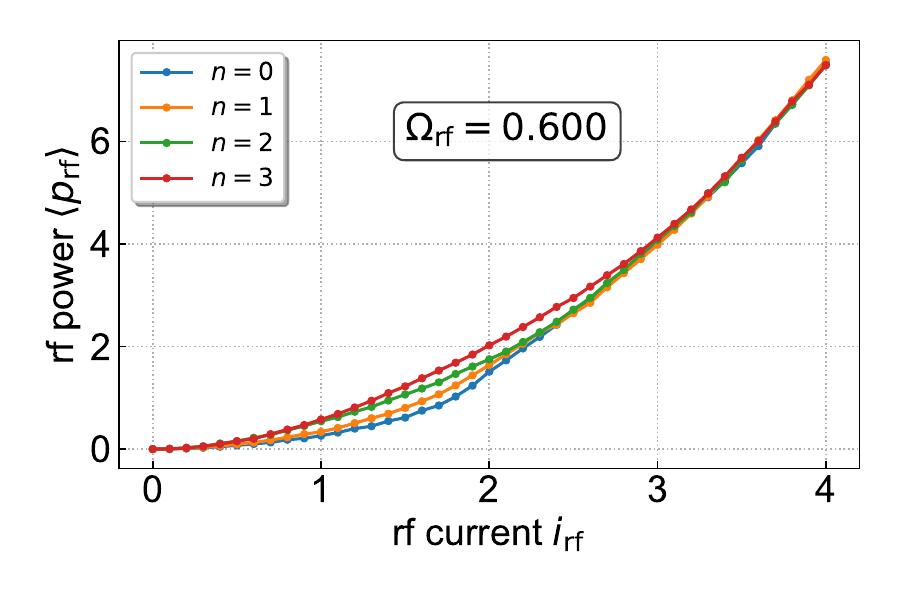}
			\caption{}
			\label{fig:mid_rfpower_Omega_0.600}
		\end{subfigure}    
		\begin{subfigure}{0.32\textwidth}
			\centering
			\includegraphics[width=\linewidth, trim={0.5cm 0cm 0.5cm 0.5cm}, clip]{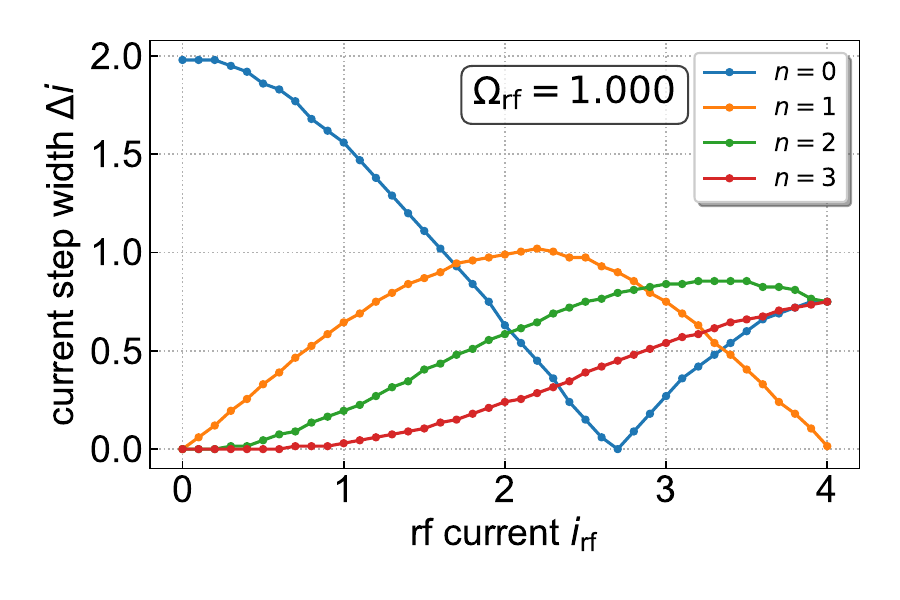}
			\caption{}
			\label{fig:Bessel_Omega_1.000}
		\end{subfigure}
		\begin{subfigure}{0.32\textwidth}
			\centering
			\includegraphics[width=\linewidth, trim={0.5cm 0cm 0.5cm 0.5cm}, clip]{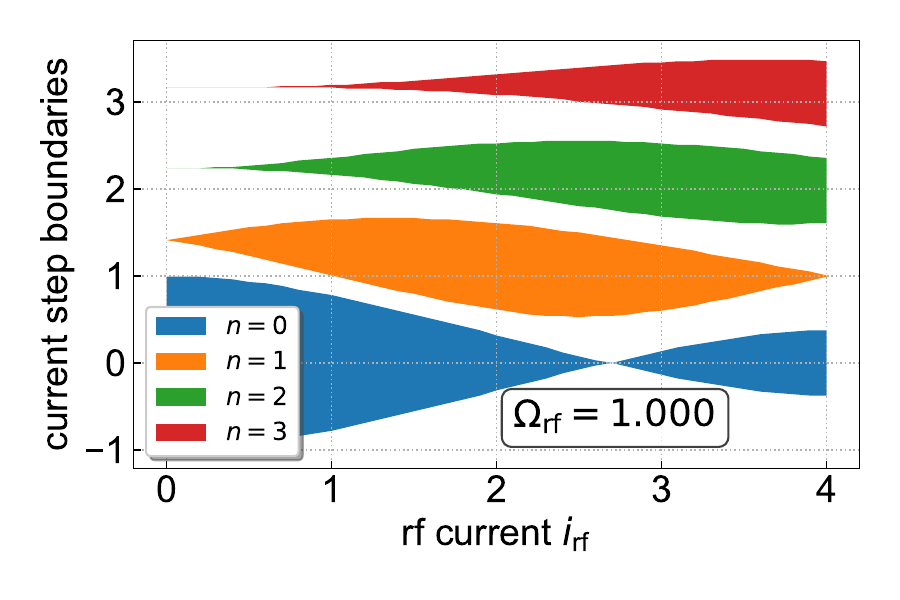}
			\caption{}
			\label{fig:center_width_Omega_1.000}
		\end{subfigure}
		\begin{subfigure}{0.32\textwidth}
			\centering
			\includegraphics[width=\linewidth, trim={0.5cm 0cm 0.5cm 0.5cm}, clip]{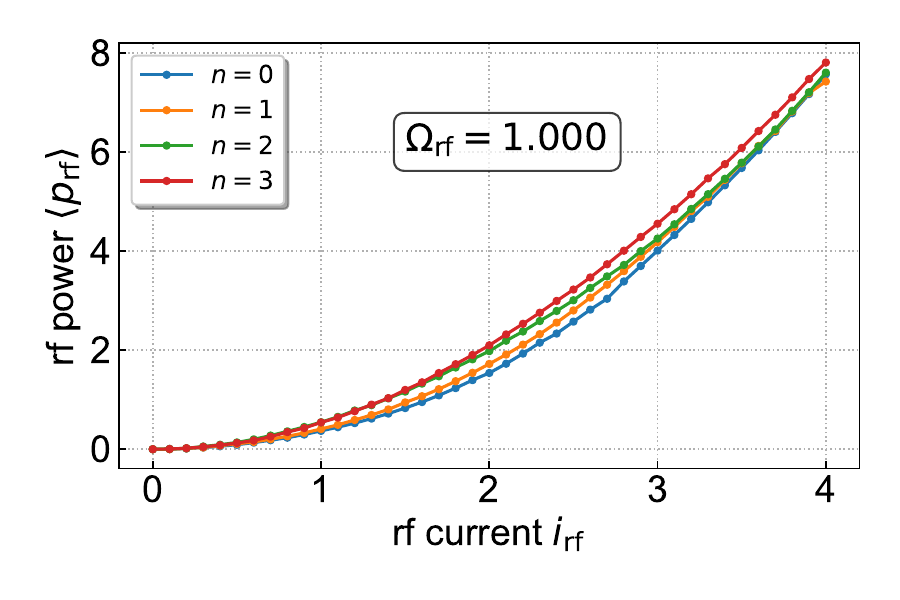}
			\caption{}
			\label{fig:mid_rfpower_Omega_1.000}
		\end{subfigure}
		\caption{Single JJ analysis at different $\Omega_{\mathrm{rf}}$ values. Left column: quantum step width $\Delta i$ for $n$ up to 3 of a single JJ as a function of normalized rf current $i_{\mathrm{rf}}$ for $\Omega_{\mathrm{rf}}=0.4$ (a), $\Omega_{\mathrm{rf}}=0.6$ (d), $\Omega_{\mathrm{rf}}=1$ (g). Center column: normalized current step boundaries for $n$ up to 3 as a function of $i_{\mathrm{rf}}$ for $\Omega_{\mathrm{rf}}=0.4$ (b), $\Omega_{\mathrm{rf}}=0.6$ (e), $\Omega_{\mathrm{rf}}=1$ (h). Left column: Average rf power dissipated within center-biased Shapiro steps for $n$ up to 3 as a function of $i_{\mathrm{rf}}$ for $\Omega_{\mathrm{rf}}=0.4$ (c), $\Omega_{\mathrm{rf}}=0.6$ (f), $\Omega_{\mathrm{rf}}=1$ (i).}
		\label{fig:Bessel_centerstep_rfpower}
	\end{figure*}

	Normalized $IV$ traces have been obtained throughout the overall rf current-frequency parameter space of interest and analyzed to extract the width of Shapiro steps with order up to $n=3$. As shown in Figures~\ref{fig:Bessel_Omega_0.400}, \ref{fig:Bessel_Omega_0.600} and \ref{fig:Bessel_Omega_1.000}, and in agreement with \cite{likharev1986chapter11}, the normalized step width $\Delta i = \Delta I / I_\mathrm{c}$ dependence with $i_\mathrm{rf}$ for three $\Omega_\mathrm{rf}$ values resembles the $n$\textsuperscript{th} order Bessel function of the first kind, which is the analytical solution of a voltage-biased junction \cite{barone1982josephson, tafuri2019fundamentals, Durandetto2019}.
	It can be noticed that, for $n\neq 0$, both maximum step size and corresponding rf current increase with $\Omega_\mathrm{rf}$. Hence, attention has to be paid to achieve suitably-wide quantum voltage steps with sustainable rf power levels. 
	Furthermore, the dc current position of each Shapiro step is crucial when multiple junctions are connected in series to enhance the output voltage. This configuration necessitates that not only is the step size slightly dependent on the rf current, but the bias current must also exhibit restrained deviations in response to the rf excitation. The quantum-locking range for Shapiro steps up to $n=3$ are plotted in Figures~\ref{fig:center_width_Omega_0.400}, \ref{fig:center_width_Omega_0.600}, and \ref{fig:center_width_Omega_1.000}.
	Finally, as illustrated in Figures~\ref{fig:mid_rfpower_Omega_0.400}, \ref{fig:mid_rfpower_Omega_0.600}, and \ref{fig:mid_rfpower_Omega_1.000}, the rf power dissipated by the junction, when biased at the midpoint of a given quantum step, exhibits an exponential-like rise for increasing rf currents, and also shows slight deviations with step order $n$.
	
	\subsection{Rf current and frequency for multi-order Shapiro steps}
	
\begin{figure*}[tbhp]
		\centering
		\begin{subfigure}{0.49\textwidth}
			\centering
			\includegraphics[width=\linewidth]{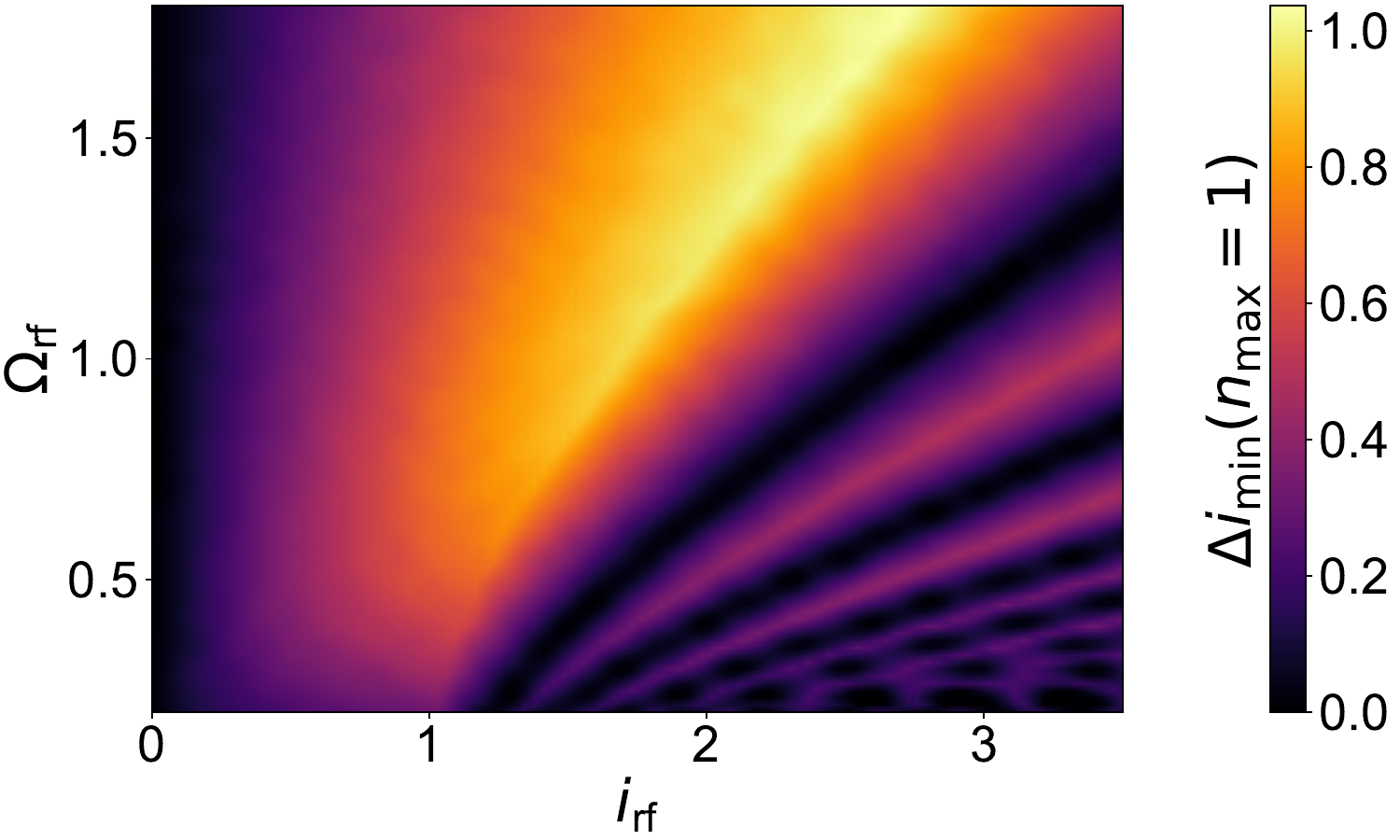}
			\caption{}
			\label{fig:min_step_colormaps_nmax1}
		\end{subfigure}
				\begin{subfigure}{0.49\textwidth}
			\centering
			\includegraphics[width=\linewidth, trim={0cm 0 0cm 0}, clip]{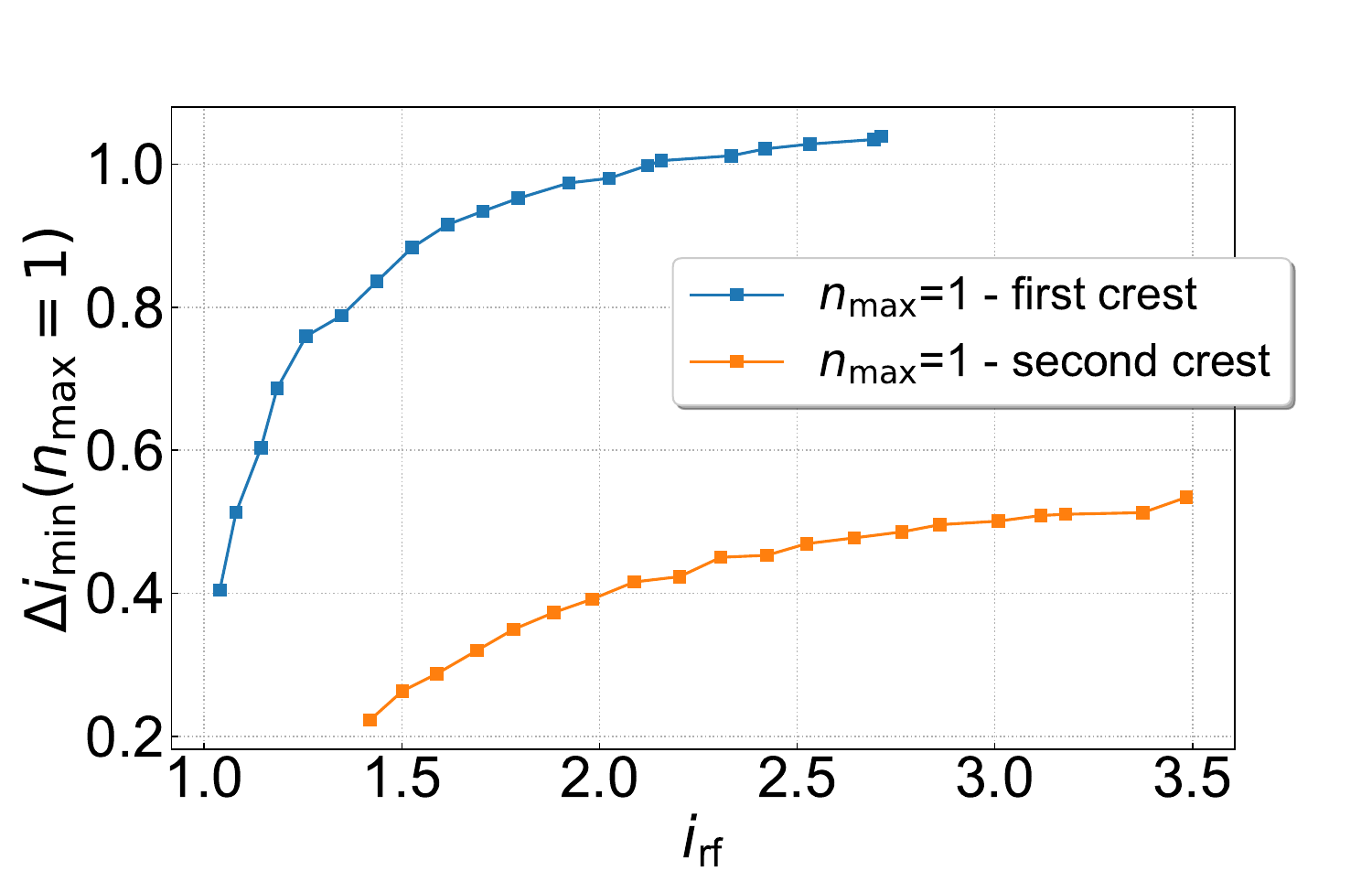}
			\caption{}
			\label{fig:max_step_min_crests_nmax_1}
		\end{subfigure}
		\begin{subfigure}{0.49\textwidth}
			\centering
			\includegraphics[width=\linewidth]{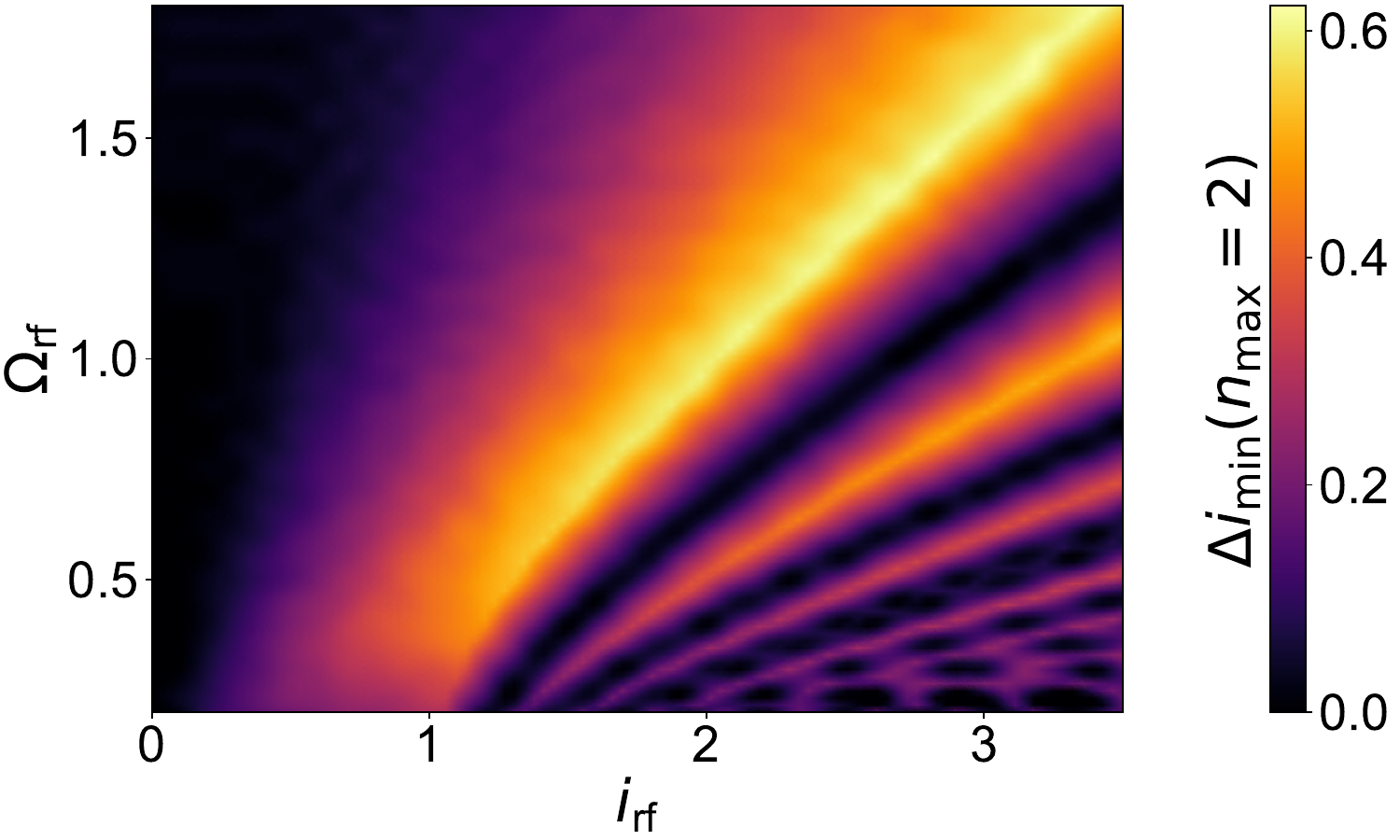}
			\caption{}
			\label{fig:min_step_colormaps_nmax2}
		\end{subfigure}
				\begin{subfigure}{0.49\textwidth}
			\centering
			\includegraphics[width=\linewidth, trim={0cm 0 0cm 0}, clip]{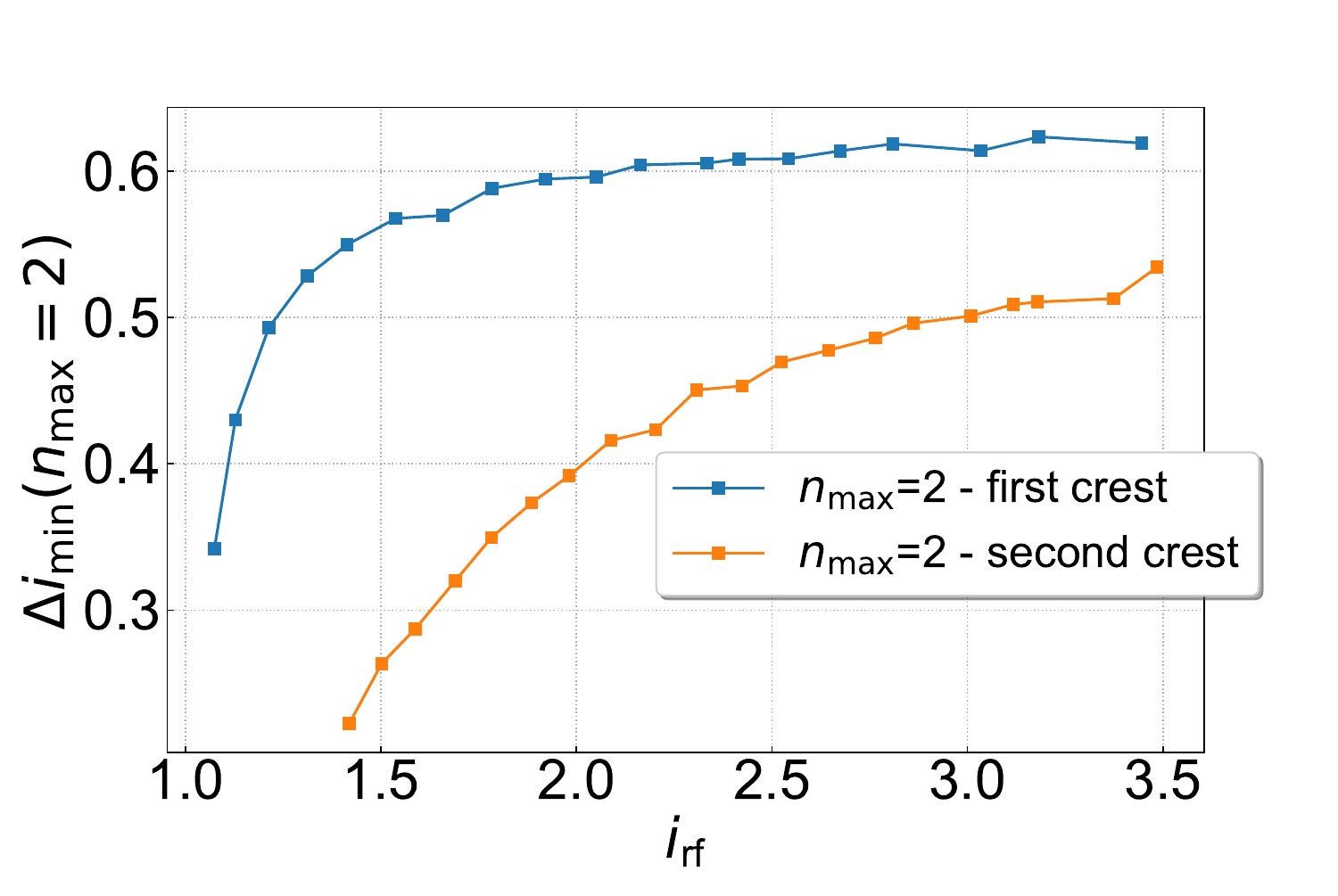}
			\caption{}
			\label{fig:max_step_min_crests_nmax_2}
		\end{subfigure}
		\begin{subfigure}{0.49\textwidth}
			\centering
			\includegraphics[width=\linewidth]{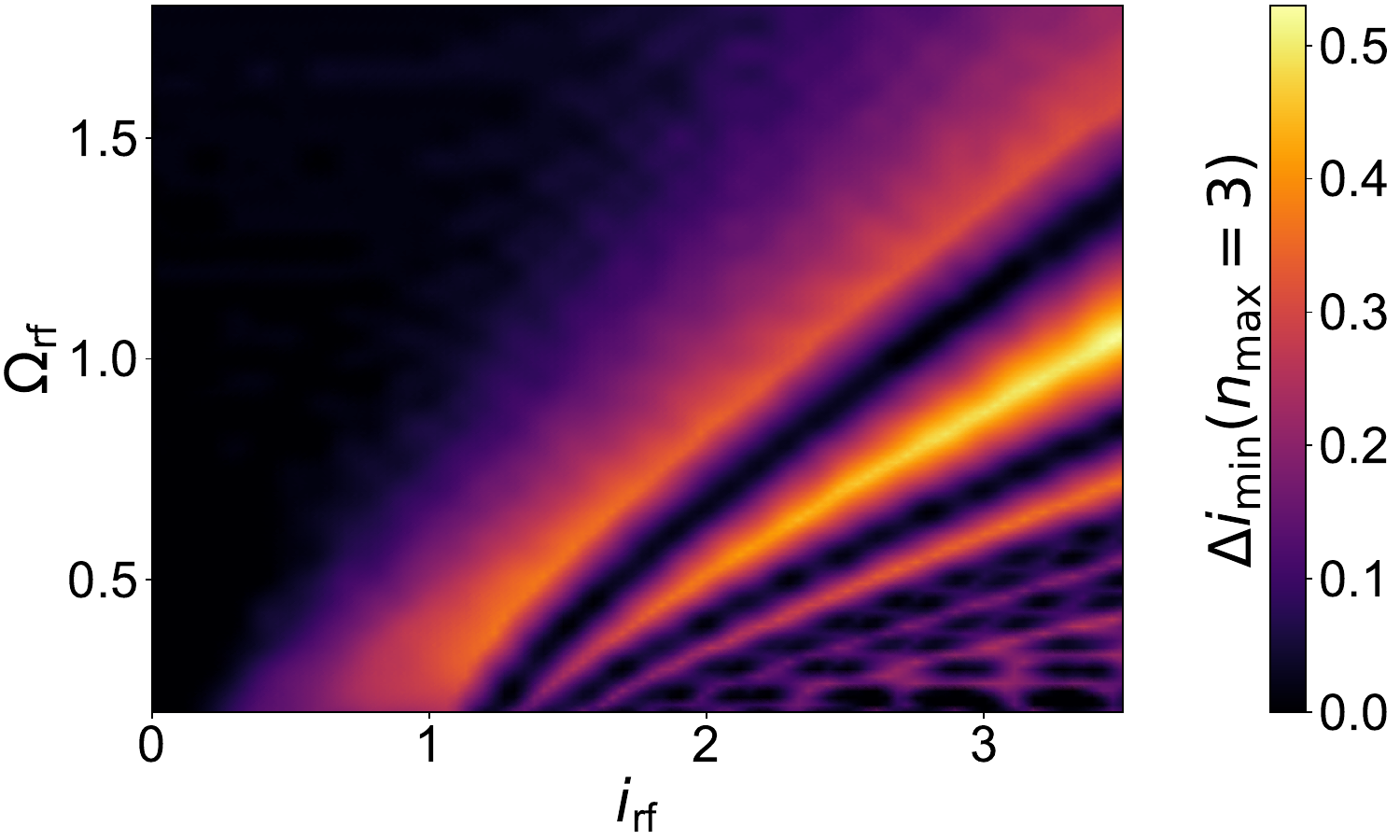}
			\caption{}
			\label{fig:min_step_colormaps_nmax3}
		\end{subfigure}
		\begin{subfigure}{0.49\textwidth}
			\centering
			\centering
			\includegraphics[width=\linewidth, trim={0cm 0 0cm 0}, clip]{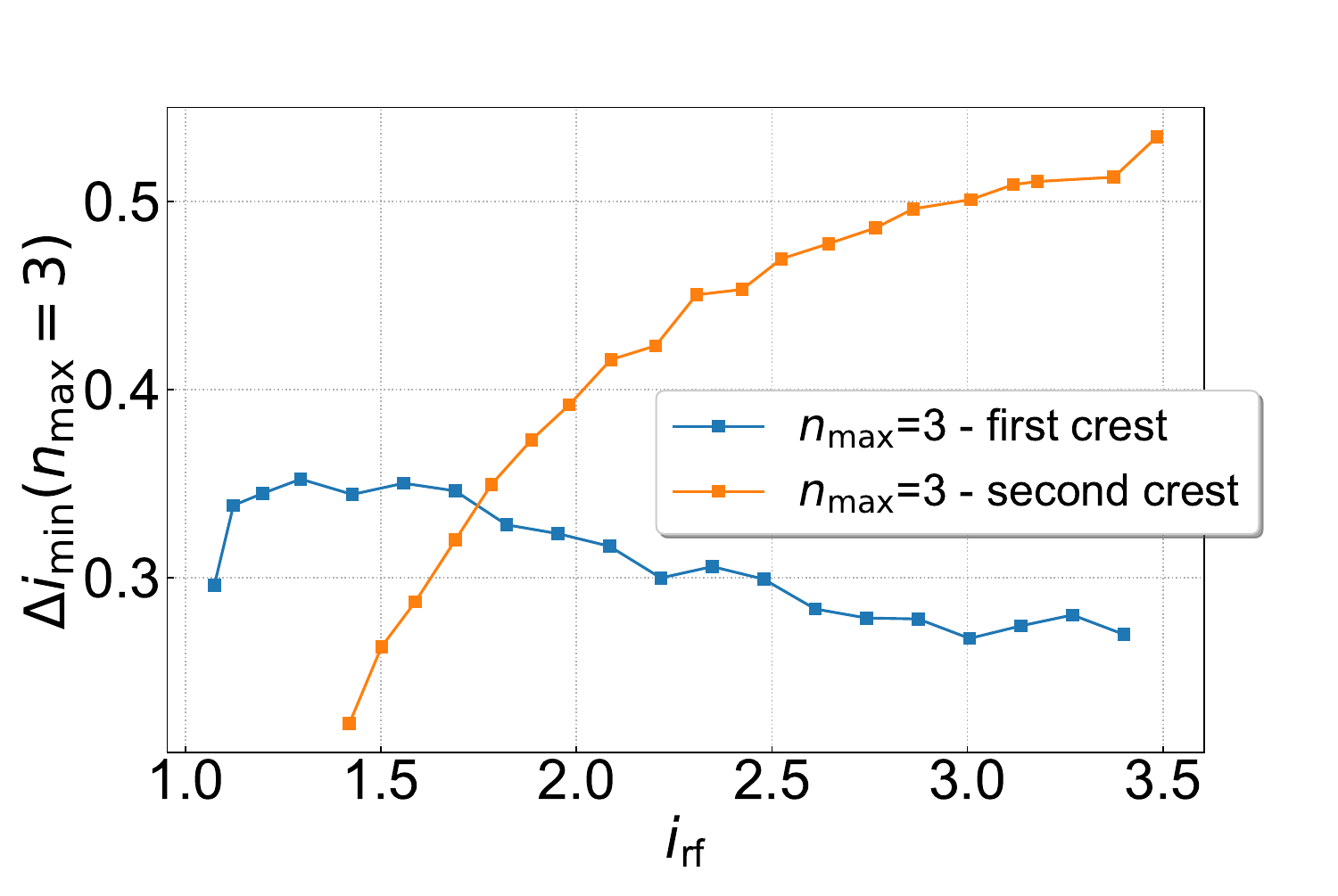}
			\caption{}
			\label{fig:max_step_min_crests_nmax_3}
		\end{subfigure}
		\caption{Single JJ analysis. Left column: Colormaps showing the normalized minimum width $\Delta i_\mathrm{min}$ for step order up to $n_\mathrm{max}$ as a function of rf current $i_{\mathrm{rf}}$ and frequency $\Omega_{\mathrm{rf}}$: a) $n_\mathrm{max}=1$; c) $n_\mathrm{max}=2$; e) $n_\mathrm{max}=3$.
			Right column: normalized step width $\Delta i_\mathrm{min} (n_\mathrm{max})$ in correspondence to the two main ($i_{\mathrm{rf}}$, $\Omega_{\mathrm{rf}}$) local maxima regions (yellow ``crests'') of the related colormap for b) $n_\mathrm{max}=1$; d) $n_\mathrm{max}=2$; f) $n_\mathrm{max}=3$.}
		\label{fig:min_step_cm_oi_wi}
	\end{figure*}

	At this point, information on step width and position, as well as dissipated rf power, has been obtained for each ($i_{\mathrm{rf}}$, $\Omega_{\mathrm{rf}}$) pair examined.
	Since the objective of this work is to optimize both the characteristics of the JJs and the bias parameters for simultaneously maximizing the width of Shapiro steps for orders \( n > 1 \), it is beneficial to reframe the discussion in terms of the minimum width \( \Delta i_{\mathrm{min}}(n_{\mathrm{max}}) \) between steps from 0 to \( n_{\mathrm{max}} \).
	For a fixed normalized frequency $\Omega_{\mathrm{rf}}$, the minimum step width can be easily inferred from the Bessel-like trends (Figure~\ref{fig:Bessel_centerstep_rfpower}(a-b-c)). Instead, to properly compare results at different values of $\Omega_{\mathrm{rf}}$, either an interactive 3-D plot or a 2-D colormap may come to help.
	Interpolated colormaps showing the minimum quantum step width $\Delta i_\mathrm{min} (n_\mathrm{max})$, with $n_\mathrm{max}$ up to 3, are shown in Figures~\ref{fig:min_step_colormaps_nmax1}, \ref {fig:min_step_colormaps_nmax2} and \ref{fig:min_step_colormaps_nmax3}.
	These colormaps show that local maxima of $\Delta i_\mathrm{min}(n_\mathrm{max})$ lie along approximately linearly-related ($i_{\mathrm{rf}}$, $\Omega_{\mathrm{rf}}$) pairs. The 2D plots in Figures~\ref{fig:min_step_colormaps_nmax1}, \ref{fig:min_step_colormaps_nmax2} and \ref{fig:min_step_colormaps_nmax3} explicitly show the normalized steps width along the yellow regions of the colormaps (henceforth referred to as ``crests'') as a function of $i_\mathrm{rf}$. 
	Figures~\ref{fig:min_step_colormaps_nmax1} and \ref{fig:max_step_min_crests_nmax_1} cover the scenario of conventional PJVS exploiting zero and first order Shapiro steps ($n_\mathrm{max}=1$): the maximum normalized step width is $\Delta i_\mathrm{min} \simeq 1$, indicating that the Shapiro step width for both the zero and first steps is roughly equal to the critical current of the junction. This happens at the first crest
	(blue line in Figure~\ref{fig:max_step_min_crests_nmax_1}) for $i_\mathrm{rf}>2$.
	A secondary main crest (orange line) is also present, though with half step width at higher rf current values (above 2.5). 
	A comparable scenario arises for \( n_\mathrm{max}=2 \) (Figures~\ref{fig:min_step_colormaps_nmax2} and \ref{fig:max_step_min_crests_nmax_2}): in this instance, multiple light regions appear, again indicating an almost linear relationship between optimal \( i_{\mathrm{rf}} \) and \( \Omega_{\mathrm{rf}} \) values.
	Nevertheless, the highest step width, which belongs to the first crest, is about $0.6\,I_\mathrm{c}$ for $i_\mathrm{rf}\geq 2.5$. 
	A slightly different scenario shows up instead for $n_\mathrm{max}=3$: as can be seen in Figures~\ref{fig:min_step_colormaps_nmax3} and \ref{fig:max_step_min_crests_nmax_3}, among the two main yellow crests, the second one (orange line) exhibits the largest simultaneous Shapiro step width of about 0.5 for $i_\mathrm{rf}>2.5$. 
	It is worth noting that, in the three \(n_{\mathrm{max}}\) cases, the second crest, represented by the orange line in Figures~\ref{fig:max_step_min_crests_nmax_1}, \ref{fig:max_step_min_crests_nmax_2}, and \ref{fig:max_step_min_crests_nmax_3}, is exactly the same, thus indicating that the minimum step order is either the zeroth or the first.
	
	
	The analysis presented in this section focuses on a single overdamped JJ governed by the RSJ model. In summary, the simultaneous utilization of Shapiro steps up to \( n_\mathrm{max} = 2 \) or \( 3 \) results in approximately a \( \sim\qty{50}{\%} \) reduction in the normalized step width compared to the conventional case of \( n_\mathrm{max} = 1 \), accompanied by more stringent rf current (or power) requirements.
	The potential benefits of investigating higher-order steps become clear when considering large JJ arrays that constitute programmable quantum voltage standards with practical voltages of up to \qty{10}{\volt}, as will be discussed in the following sections.
	
	\section{Present PJVS circuits}
	Currently-available PJVS devices with both dc and ac capability and peak voltage up to $V_\mathrm{J} \simeq \qty{\pm 10}{\volt}$ are made of tens or hundreds of thousands of overdamped JJs. The total number of JJs ($N_\mathrm{JJ}$) to reach the \qty{10}{\volt} target depends indeed on the employed microwave frequency $f_\mathrm{rf}$ according to the well-know frequency-to-voltage relation
	\begin{equation}
		V_{\mathrm{J}} = n_\mathrm{max} \, N_\mathrm{JJ}\,\Phi_\mathrm{0}\,  f_\mathrm{rf}
		\label{eq:Josephon_eq}
	\end{equation}
	\noindent with $n_\mathrm{max}=1$ ($n= -1, 0, 1$).
	The $IV$ characteristics of the full PJVS array should display usable quantum voltage steps, typically with a minimum quantum-locking range of at least \qty{1}{\milli\ampere} \cite{supracon_ac_voltage_standard_array, nist6000, mueller2007improved}. Two main factors prevent the possibility of simply rescaling the voltage-to-current relation of a single JJ by $N_{\mathrm{JJ}}$: i) the rf power attenuation along the transmission line where the junctions are arranged \cite{muller2012nbsi}, preventing them to be biased by the same rf current $I_\mathrm{rf}$; ii) the spread of JJ electrical parameters (critical current $I_\mathrm{c}$ and normal resistance $R_\mathrm{n}$, in turn related to the characteristic frequency $f_\mathrm{c} = I_\mathrm{c}\, R_\mathrm{n} / \Phi_\mathrm{0}$), which is due to technological limitation of array fabrication \cite{kieler2021stacked, lacquaniti2012josephson}.
	In normalized units, both rf power attenuation and electrical parameters' dispersion cause each junction to exhibit distinct values of $i_\mathrm{rf}$ and $\Omega_{\mathrm{rf}}$, leading to different width and position of the induced quantum steps. As a result, these steps may not be preserved when multiple JJs with varying $IV$ curves are combined together.
	
\begin{figure}[tbhp]
		\centering
		\includegraphics[width=0.9\linewidth, clip, trim={0 0cm 0 0cm}]{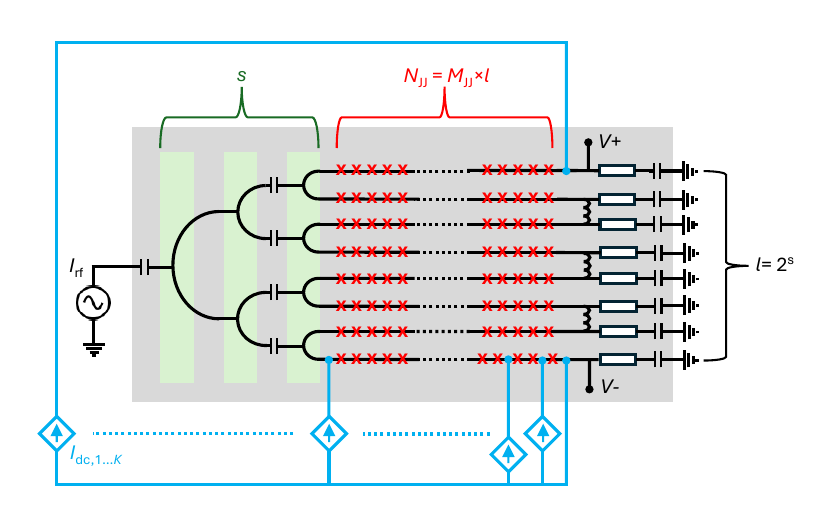}
		\caption{Simplified schematic of PJVS circuit and connections with bias electronics.  The PJVS chip is outlined by the grey rectangular box. A single JJ is represented as a red `x'. Microwave splitting stages $s$ are indicated as green rectangular boxes. The $l$ transmission lines are series-connected via superconducting wirings. The full PJVS voltage is collected at terminals $V_+$ and $V_-$. Bias currents $I_\mathrm{dc}$ are provided by $K$ ideal current sources (blue), with $K$ the number of the PJVS sub-arrays. Rf termination loads are depicted as white boxes at the end of each transmission line.}
		\label{fig:PJVS_scheme}
	\end{figure}

	While parameters spread can be reduced to acceptable levels and is expected to improve over the years thanks to technological progress of fabrication methods, the microwave attenuation along the transmission line is intrinsic to both junction behavior and transmission line effects, thus posing severe limitation to the maximum number of junctions $M_\mathrm{JJ}$ that can be arranged along a single transmission line ($M_\mathrm{JJ}< Z/R_\mathrm{n}\ll N_\mathrm{JJ}$, with $Z$ the characteristic impedance of the line)\cite{dresselhaus2009tapered}.
	The attenuation of the rf power along the transmission line arises from dissipative effects in both the junctions and the line itself.
	Modern PJVS devices employ overdamped, non-hysteretic JJs with negligible capacitance, unlike older JVS systems with underdamped JJs. In the older designs, attenuation was considerably lower because the capacitive shunt across the ideal junction effectively shorted the microwave signal. However, the limitations of these hysteretic arrays, including their slow programmability and the very narrow quantum step width ($<\qty{50}{\micro\ampere}$), drove the development of JVS with non-hysteretic junctions \cite{Durandetto2019}.
	Currently, all PJVS implementations worldwide follow a similar design, schematically represented in Figure~\ref{fig:PJVS_scheme}: the total number of junctions $N_\mathrm{JJ}$ is almost-evenly distributed among $l = 2^s$ parallel transmission lines, each counting $M_\mathrm{JJ}\simeq N_\mathrm{JJ}/l$ junctions. Here, $s$ represents the number of on-chip splitting stages, which enable the input rf power to be equally shared between the $l$ arrays. This series/parallel circuit, similar to that reported in \cite{hamilton2000josephson}, features a network of low and high-pass filters enabling the microwave power to be split
	into $l$ parallel paths while maintaining a dc path in which
	all JJs are series-connected.
	
	\noindent PJVS circuits are commonly classified into two main categories, that will be referred to here as the ``European PJVS'' (EU-PJVS), designed and developed by PTB and Supracon AG \cite{schubert2016dry, mueller20091, muller2014microwave}, and the ``United States PJVS'' (US-PJVS), realized at NIST \cite{dresselhaus201010,burroughs2011nist, rufenacht2014cryocooled}. Both PJVS systems utilize niobium (Nb) as the superconducting material (S), allowing them to operate at approximately \qty{4}{\kelvin} in either liquid helium or dry cryostats. These quantum standards are commercially available and worldwide distributed to National Metrology Institutes (NMIs) and advanced calibration laboratories \cite{schubert2015ac}. The key characteristics of these two PJVS versions are outlined in Table~\ref{tab:EU-US-PJVS}.
	
	\begin{table}[H]
		\centering
		\caption{Typical values of \qty{10}{\volt} EU-PJVS and US-PJVS main characteristic parameters}
	\label{tab:EU-US-PJVS}
	\begin{adjustbox}{width=\textwidth}
		\begin{tabular}{lcc}
			\toprule
			& EU-PJVS \cite{mueller20091} & US-PJVS \cite{dresselhaus201010} \\
			\midrule
			Microwave frequency $f_\mathrm{rf}$& \qty{70}{\giga\hertz} & \qty{18}{\giga\hertz}\\
			Number of junctions $N_\mathrm{JJ}$ & $\sim\num{70000}$ & $\sim\num{270000}$ \\
			Barrier material & Nb\textsubscript{x}Si\textsubscript{1-x} (x $\sim \qty{10}{\%}$) & Nb\textsubscript{x}Si\textsubscript{1-x} (x $\sim \qty{20}{\%}$)  \\
			Maximum step order $n_\mathrm{max}$ & 1  & 1  \\
			Transmission line type & Microstripline & Coplanar waveguide\\
			Transmission line characteristic impedance $Z$ & $\sim\qty{16}{\ohm}$ &$\qty{50}{\ohm}-\qty{23}{\ohm}$ (tapered) \\ 
			Number of rf power splitters $s$ & 7 & 5  \\
			Number of parallel transmission lines $l=2^s$ & 128 & 32\\
			Number of junctions per line $M_\mathrm{JJ} \simeq N_\mathrm{JJ}/l$ & $\sim 550$ & $\sim 8400$ \\
			\bottomrule
		\end{tabular}
	\end{adjustbox}    
\end{table}

A noteworthy feature of both EU and US-PJVS is the use of SNS (Superconductor/Normal metal/Superconductor) JJs of the same kind, with amorphous niobium-silicide (Nb\textsubscript{x}Si\textsubscript{1-x}) as a normal metal (N). In particular, the latter enables the easy tuning of the JJs electrical parameters by modifying its relative content (x) and thickness to enhance performance for the specific rf frequency being utilized \cite{scheller2012sns}, thereby allowing its implementation in both 15-\qty{20}{\giga\hertz} and \qty{70}{\giga\hertz} JVS. Building on its success, researchers at the National Institute of Metrology (NIM) in China have utilized the same Nb\textsubscript{x}Si\textsubscript{1-x} technology to realize \qty{2}{\volt} PJVS chips with enhanced microwave transmission \cite{cao2022realization}.

For the sake of completeness, it is worth mentioning that other PJVS devices have been developed by various groups using different materials, while still sharing some similarities with the EU and US designs. A collaboration between INRiM and PTB led to the development of \qty{1}{\volt} PJVS devices made of \num{8192} JJs of the SNIS technology (I: insulator), which features a non-superconducting barrier made of a thick aluminum layer (N) with a thin oxidation layer (I) \cite{lacquaniti2011}. These devices demonstrate enhanced second-order steps at liquid helium temperatures \cite{trinchera2016synthesis}. 
Another SNS JJ type has been developed in Japan by the National Metrology Institute of Japan / National Institute of Advanced Industrial Science and Technology (NMIJ/AIST), which utilizes niobium nitride (NbN) as the superconductive element, thereby improving the PJVS cryocooler refrigeration at temperatures around \qty{10}{\kelvin} \cite{yamamori200810}. In conclusion, although they have not yet developed into a fully operational Programmable Josephson Voltage Standard (PJVS), tabletop dc JVS systems utilizing high-temperature superconductors (HTS) have been successfully created using bi-crystal JJs made from Yttrium Barium Copper Oxide (YBCO)
\cite{sosso2007metrological, khorshev2019voltage, klushin2020present}. Nowadays, these systems achieve output voltages up to \qty{0.1}{\volt} \cite{klushin2008optimization} and are cooled using either liquid nitrogen or compact cryostats, thereby paving the way for the broader adoption of quantum standards outside NMIs.

Despite the various implementations, this work focuses on the two EU and US-PJVS technologies listed in Table~\ref{tab:EU-US-PJVS}, as they are the most widely used. However, the analysis presented below is limited to considerations related to microwave constraints, specifically the external rf bias frequency and the transmission line used to deliver the signal to the JJ arrays. In particular, it is clear that the selection of the microwave frequency $f_\mathrm{rf}$ is crucial for the performance of the voltage standard device: using a \qty{70}{\giga\hertz} microwave frequency allows reducing the overall number of junctions $N_\mathrm{JJ}$, though with the added complexity of handling sophisticated microwave equipment and low-impedance microstripline attenuation. On the other hand, using a \qty{18}{\giga\hertz} rf signal needs a considerable number of junctions to reach the \qty{10}{\volt} level, partially compensated by low power attenuation along the \qty{50}{\ohm} coplanar waveguide (CPW).

\section{Simulation of multiple JJs} \label{sec:sim_mult_jj}
A simple algorithm to estimate the behavior of several series-connected JJs within a microwave transmission line of fixed characteristic impedance ($Z$), by taking into account both JJs variations and rf attenuation, has been developed and described as follows.

\subsection{Rf current attenuation}
A thorough analysis of the rf dissipation effect requires an accurate model of the JJs within the distributed transmission line, which becomes challenging due to the large number of junctions integrated within the waveguide. In this study, the rf current $I_\mathrm{rf}(k)$ driving the $k$\textsuperscript{th} junction, with $k$ running from 1 to $M_\mathrm{JJ}$, is evaluated one-by-one from the knowledge of the normalized average microwave power $\langle p_\mathrm{rf} \rangle$ introduced in Sec.~\ref{subsec:qsteps_rfpow}. The relation between the absolute rf current of two consecutive JJs is

\begin{equation}
	I_\mathrm{rf} (k+1) = I_\mathrm{rf} (k) \, 10^{\frac{\alpha_\mathrm{JJ}(k)}{20}}
	\label{eq:curr_att}
\end{equation}

\noindent with $\alpha_\mathrm{JJ}(k)$ the JJ rf-attenuation of the $k$-junction in dB. From the incoming power $P_\mathrm{rf} (k)=Z\, I_\mathrm{rf}^2(k)/2$ and the average dissipated power $P_d(k) = \langle p_\mathrm{rf} (k) \rangle \, R_\mathrm{n}\, I_\mathrm{c}^2$, the attenuation $\alpha_\mathrm{JJ}(k)$ can be written as
\begin{eqnarray}
	\alpha_\mathrm{JJ}(k) &= 20\, \log_{10}\left(\frac{I_\mathrm{rf} (k+1)}{ I_\mathrm{rf} (k)}\right)= \nonumber\\
	&= 10\, \log_{10}\left(\frac{P_\mathrm{rf} (k+1)}{ P_\mathrm{rf} (k)}\right)= \nonumber\\
	&= 10\, \log_{10}\left(\frac{P_\mathrm{rf} (k) - P_\mathrm{d}(k)}{ P_\mathrm{rf} (k)}\right) = \nonumber\\
	&= 10\, \log_{10}\left(1 - \frac{P_\mathrm{d}(k)}{ P_\mathrm{rf} (k)}\right) = \nonumber\\
	&= 10\, \log_{10}\left(1 - \frac{2\,P_\mathrm{d}(k)}{ Z\, I_\mathrm{rf}^2 (k)}\right) = \nonumber\\
	&= 10\, \log_{10}\left(1 - \frac{2\,\langle p_\mathrm{rf} (k) \rangle \, R_\mathrm{n}(k)\, I_\mathrm{c}^2(k)}{ Z\, i_\mathrm{rf}^2 (k)\,I_\mathrm{c}^2(k)}\right)= \nonumber\\
	&=10\, \log_{10}\left(1 - \frac{2\,\langle p_\mathrm{rf} (k) \rangle}{ z(k)\, i_\mathrm{rf}^2 (k)}\right)\nonumber\\
	\label{eq:att}
\end{eqnarray}
\noindent with $z(k)=Z/R_\mathrm{n}(k)$ the normalized transmission line impedance.
As shown in Figure~\ref{fig:IV_IacP_plots} and Figures~\ref{fig:mid_rfpower_Omega_0.400}-\ref{fig:mid_rfpower_Omega_0.600}-\ref{fig:mid_rfpower_Omega_1.000}, besides the known $i_\mathrm{rf}$ value, the average dissipated power depends on step order $n\leq n_\mathrm{max}$ and bias current $I_\mathrm{dc}$, which are not known \textit{a priori}. Therefore, for this analysis, each JJ is assumed to be biased at the center of its $n_\mathrm{max}$ Shapiro step. 
In this way, the attenuation of each junction is evaluated from the knowledge of its absolute electrical parameters ($I_\mathrm{c}$ and $R_\mathrm{n}$) and of the impedance of the transmission line $Z$.

\subsection{Spread of junctions electrical parameters}
Ideally, all $N_\mathrm{JJ}$ junctions in the superconducting circuit are identical, sharing the same $I_\mathrm{c}$ and $R_\mathrm{n}$ values. However, imperfections during the fabrication process result in a small non-uniformity in the electrical parameters, typically within a few percent.
As a result, the position and width of the Shapiro steps of JJs driven by the same microwave excitation, with current $I_\mathrm{rf}$ and frequency $f_\mathrm{rf}$, may lie in different regions of the ($i_\mathrm{rf}$, $\Omega_\mathrm{rf}$) map. Consequently, they may exhibit distinct $IV$ characteristics, which, when combined, could lead to reduced Shapiro steps or, in the worst case, no steps at all.

In this study, nominal JJ electrical parameters $I_\mathrm{c}$ and $R_\mathrm{n}$ along a single transmission line are assumed to follow a uniform random distribution within the ranges $[I_\mathrm{c} - \delta I_\mathrm{c}, I_\mathrm{c} + \delta I_\mathrm{c}]$ and $[R_\mathrm{n} - \delta R_\mathrm{n}, R_\mathrm{n} + \delta R_\mathrm{n}]$, with $0<\delta I_\mathrm{c}/I_\mathrm{c}\ll1$ and $0<\delta R_\mathrm{n}/R_\mathrm{n}\ll1$. Consequently, the characteristic frequency $f_\mathrm{c}$ is randomly distributed between $[(I_\mathrm{c} - \delta I_\mathrm{c})(R_\mathrm{n} - \delta R_\mathrm{n})/\Phi_\mathrm{0}, (I_\mathrm{c} + \delta I_\mathrm{c})(R_\mathrm{n} + \delta R_\mathrm{n})/\Phi_\mathrm{0}]$. It is worth noting that parameters spread also affects the attenuation $\alpha(k)$ in Eq.~\ref{eq:att}, being $z(k)$, $i_\mathrm{rf}(k)$ and $\langle p_\mathrm{rf} (k) \rangle$ normalized with respect to $I_\mathrm{c}(k)$ and $R_\mathrm{n}(k)$.

\subsection{Shapiro steps evaluation of multiple junctions}\label{subsec:procedure}
Outlined below is the iterated step-by-step procedure for determining steps width and position for all the $M_\mathrm{JJ}$ junctions within the transmission line, rf-driven at $f_\mathrm{rf}$, with $k$ running from 1 to $M_\mathrm{JJ}$:\\
\begin{enumerate}
	\item The electrical parameters of the $k$-th junction $I_\mathrm{c}(k)$ and $R_\mathrm{n}(k)$ are randomly extracted from the ranges $[I_\mathrm{c} - \delta I_\mathrm{c}, I_\mathrm{c} + \delta I_\mathrm{c}]$ and $[R_\mathrm{n} - \delta R_\mathrm{n}, R_\mathrm{n} + \delta R_\mathrm{n}]$, respectively. The incoming rf current $I_\mathrm{rf}(k)$ is known: for $k=1$, it corresponds to the externally  supplied current, while for $k>1$, it is the output rf current evaluated in step 3;\\
	
	\item Normalized units $i_\mathrm{rf}(k) =I_\mathrm{rf}(k)/I_\mathrm{c}(k) $, $\Omega_\mathrm{rf}(k)= \Phi_\mathrm{0}\,f_\mathrm{rf}/(I_\mathrm{c}(k)\,R_\mathrm{n}(k)$) and $z(k)=Z/R_\mathrm{n}(k)$ are computed and exploited to get normalized Shapiro steps width $\Delta i(k, n)$ and position $i_\mathrm{dc}(k, n)$ as well as normalized dissipated power $\langle p_\mathrm{rf} \rangle (k, n_\mathrm{max})$ from single-JJ simulated data (Sec.~\ref{sec:single-jj});\\
	
	\item Absolute values for step width $\Delta I(k, n)$ and center bias current $I_\mathrm{dc}(k, n)$ for $n$ up to $n_\mathrm{max}$ are then evaluated. For each $n\leq n_\mathrm{max}$, the dc bias current ranges overlapping with all the previous ($k-1$) JJs are determined. The attenuation $\alpha(k, n_\mathrm{max})$ is evaluated by means of Eq.~\ref{eq:att} and exploited to determine the rf current level $I_\mathrm{rf}(k+1)$ entering the following junction (Eq.~\ref{eq:curr_att}).\\This procedure is repeated for the next junction until $k=M_\mathrm{JJ}$.\\
\end{enumerate}

\section{Simulation of present PJVS devices}

As described above, the multi-junctions analysis requires switching from normalized to absolute units, since the knowledge of each JJ attenuation in \qty{}{\decibel} and transmission line impedance in \qty{}{\ohm} is strictly necessary.
The multi-junction analysis is first conducted in the conventional case of $n_\mathrm{max}=1$ of present \qty{10}{\volt} EU-PJVS and US-PJVS introduced above (Table~\ref{tab:EU-US-PJVS}). The fixed parameters are the microwave frequency $f_\mathrm{rf}$, the number of junctions $N_\mathrm{JJ}$ to reach the \qty{10}{\volt} target voltage, and the transmission line characteristic impedance $Z$. All the other parameters are spanned across suitable ranges with the aim of finding the optimal condition where steps from 0 to $n_\mathrm{max}$ are simultaneously maximized and larger than a specified threshold. Power dissipation is also considered, as it is crucial to minimize for the reliable operation of the superconducting device in a cryogenic environment.

\subsection{\qty{10}{\volt} EU-PJVS for $n_{\mathrm{max}}=1$} \label{sec:eu-pjvs-nmax1}
As summarized in Table~\ref{tab:EU-US-PJVS}, PJVS arrays for rf frequencies around \qty{70}{\giga\hertz} require about $N_\mathrm{JJ}\sim\num{70000}$ overdamped JJs to achieve the \qty{10}{\volt} target. Owing to rf power attenuation, the number of junctions along a single stripline is about $M_\mathrm{JJ}\sim 550$. Consequently, $l=128$ parallel rf-lines are needed, with $s=7$ on-chip splitting stages.
Following the procedure described in Sec.~\ref{subsec:procedure}, an array of $M_\mathrm{JJ}= 550$ junctions is simulated for different ($i_{\mathrm{rf}}$, $\Omega_{\mathrm{rf}}$) pairs and for typical critical current values. A \qty{5}{\%} dispersion of both $I_\mathrm{c}$ and $R_\mathrm{n}$ is assumed ($\delta I_\mathrm{c}/I_\mathrm{c} = \delta R_\mathrm{n}/R_\mathrm{n}=0.05$). Microwave frequency $f_\mathrm{rf}=\qty{70}{\giga\hertz}$ and line impedance $Z=\qty{16}{\ohm}$ are fixed.

\begin{figure}[hbtp]
	\centering
	\begin{subfigure}{0.48\textwidth}
		\centering
		\includegraphics[width=\linewidth]{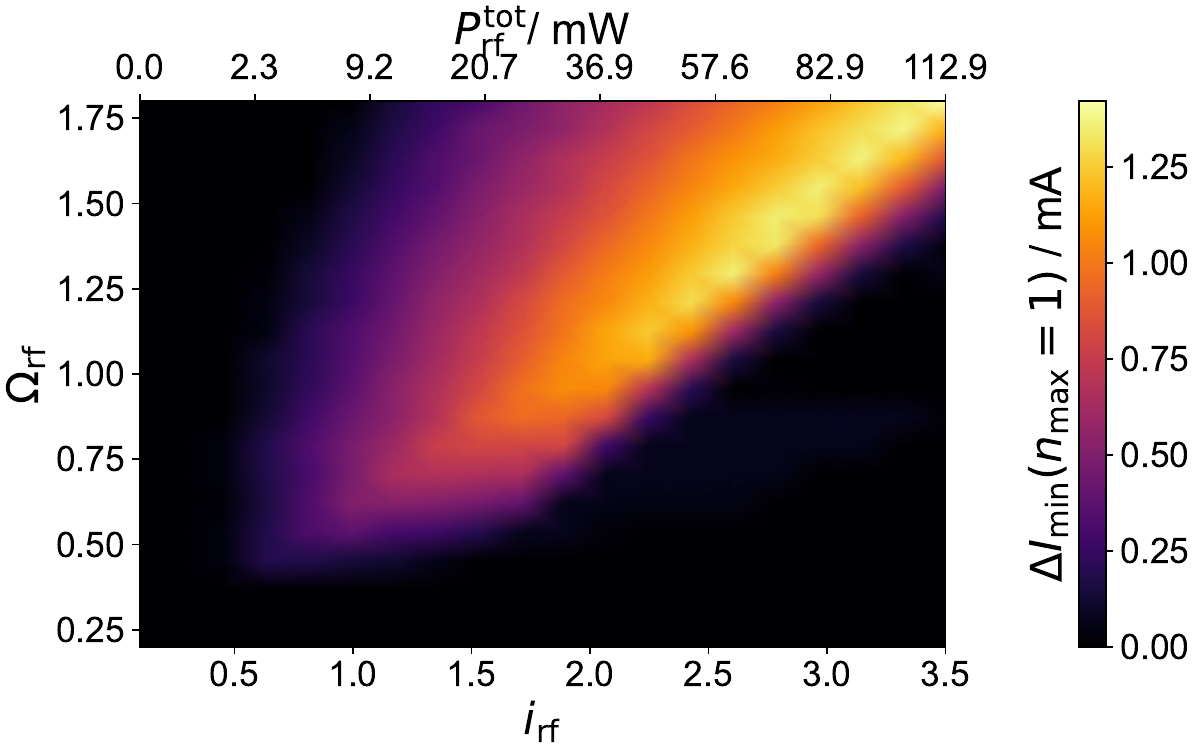}
		\caption{$I_\mathrm{c}=\qty{3}{\milli\ampere}$, $M_\mathrm{JJ}=550$}
		\label{fig:EU-PJVS-cmap_Ptot-fc-DeltaI-nmax_1-sigma_0.05-Ic_3mA-Nstrips_128}
	\end{subfigure}
	\begin{subfigure}{0.48\textwidth}
		\centering
		\includegraphics[width=\linewidth]{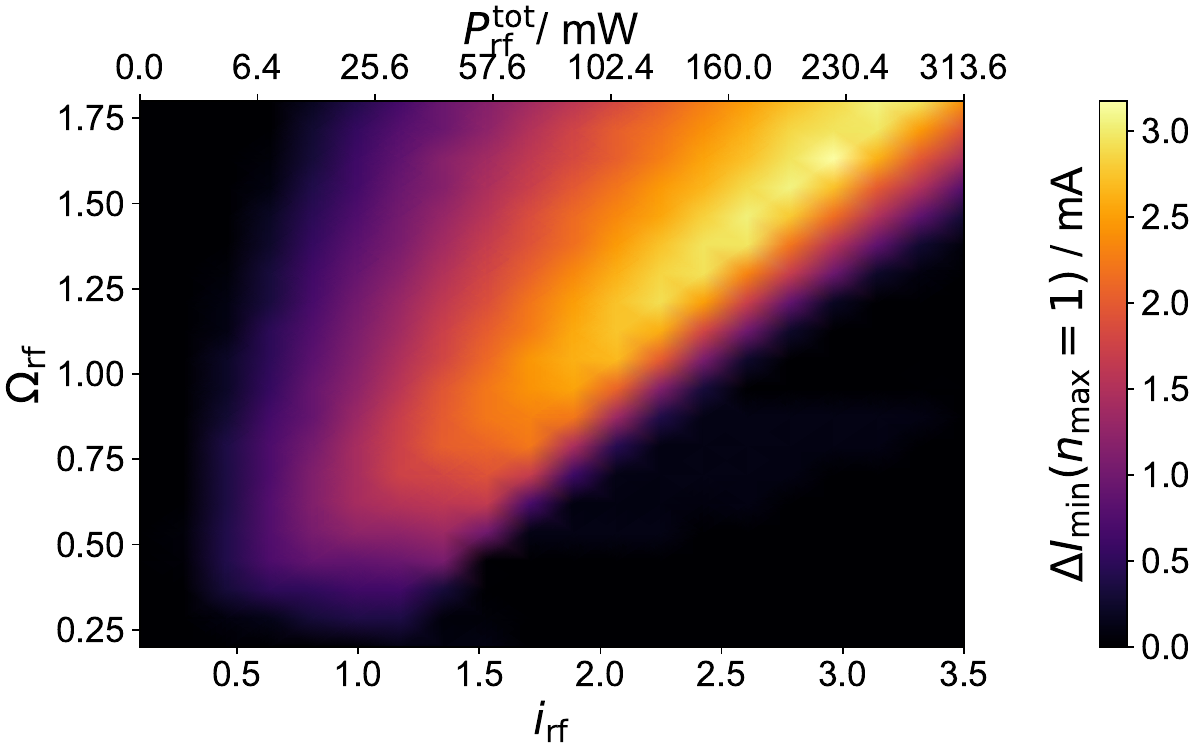}
		\caption{$I_\mathrm{c}=\qty{5}{\milli\ampere}$, $M_\mathrm{JJ}=550$}
		\label{fig:EU-PJVS-cmap_Ptot-fc-DeltaI-nmax_1-sigma_0.05-Ic_5mA-Nstrips_128}
	\end{subfigure}
	\caption{Simulation of conventional \qty{10}{\volt} EU-PJVS ($n_\mathrm{max}=1$): $M_\mathrm{JJ}=550$ JJs arranged along a $Z=\qty{16}{\ohm}$ transmission line and rf-driven at $f_\mathrm{rf}=\qty{70}{\giga\hertz}$, for nominal critical current $I_\mathrm{c}=\qty{3}{\milli\ampere}$ (left) and $I_\mathrm{c}=\qty{5}{\milli\ampere}$ (right). Electrical parameters spread is \qty{5}{\%}. Colormaps (a) and (b) show the absolute minimum step width $\Delta I_\mathrm{min}(n_\mathrm{max}=1)$ in \qty{}{\milli\ampere} as a function of normalized rf current $i_\mathrm{rf}$ (bottom x-axis), total rf power $P_\mathrm{rf}^\mathrm{tot}$ in \qty{}{\milli\watt} (top x-axis) and normalized frequency $\Omega_\mathrm{rf}$.
	}
	\label{fig:EU-PJVS-nmax1}
\end{figure}

The results of the simulation are shown in Figure~\ref{fig:EU-PJVS-nmax1}.
The total microwave power driving the full PJVS device is evaluated as the absolute power reaching the first JJ ($k=1$), multiplied per the number of transmission lines $l$:
\begin{equation}
	P_\mathrm{rf}^\mathrm{tot} = l\,Z\,\frac{I_\mathrm{rf}^2(k=1)}{2} = l\,Z\,\frac{[i_\mathrm{rf}(k=1)\,I_\mathrm{c}(k=1)]^2}{2}
	\label{eq:ptot}
\end{equation}

Power splitter attenuation and differences between the parallel transmission lines are not taken into account in this study.
The total attenuation of the $M_\mathrm{JJ}$ junctions is:

\begin{equation}
	\alpha_\mathrm{tot} = 20\,\log_{10}\left[\frac{i_\mathrm{rf}(k=M_\mathrm{JJ})}{i_\mathrm{rf}(k=1)}\right]
	\label{eq:tot_att}
\end{equation}

\noindent In accordance with \cite{kautz1995shapiro, borovitskii1985increasing}, for both $I_\mathrm{c}$ values, the optimal characteristic frequency $f_\mathrm{c}$ is approximately equal to $f_\mathrm{rf}$ ($\Omega_\mathrm{rf} \sim 1$), where the step width $\Delta I (n_\mathrm{max} = 1)$ is maximized with minimal power consumption. The highest step width is approximately \qty{1}{\milli\ampere} for $I_\mathrm{c}=\qty{3}{\milli\ampere}$ and $P_\mathrm{rf}^\mathrm{tot}\sim\qty{40}{\milli\watt}$ (Figure~\ref{fig:EU-PJVS-cmap_Ptot-fc-DeltaI-nmax_1-sigma_0.05-Ic_3mA-Nstrips_128}), in agreement with PTB/NIST findings in \cite{mueller20091}. The total power attenuation value $\alpha_\mathrm{tot}$ is about \qty{-8.4}{\decibel}, resulting from the distribution of the single JJ attenuation values $\alpha_\mathrm{JJ}$ (Figure~\ref{fig:attenuation_histo_EUPJVS_3mA_550JJs}, appendix section), with average of about \qty{-0.016}{\decibel}/JJ. At the same time, the normalized rf current $i_\mathrm{rf}$ progressively decreases by half from the first to the last JJ in the array, as shown in Figure~\ref{fig:Irf_vs_k_EUPJVS_3mA_550JJs}. Quantum-locking ranges exceeding \qty{2}{\milli\ampere} can be obtained by raising the nominal critical current, although this requires a microwave power amount approaching \qty{100}{\milli\watt} (Figure~\ref{fig:EU-PJVS-cmap_Ptot-fc-DeltaI-nmax_1-sigma_0.05-Ic_5mA-Nstrips_128}).


A similar investigation has been carried out for the US-PJVS, where thousands of junctions are arranged along a single CPW and driven at a frequency of \( f_{\mathrm{rf}} = \qty{18}{\giga\hertz} \). This topic is further elaborated in the Appendix section (\ref{appsec:us-pjvs-nmax1}).

\begin{figure*}[btp]
	\centering
	\begin{subfigure}{0.49\textwidth}
		\centering
		\includegraphics[width=\linewidth]{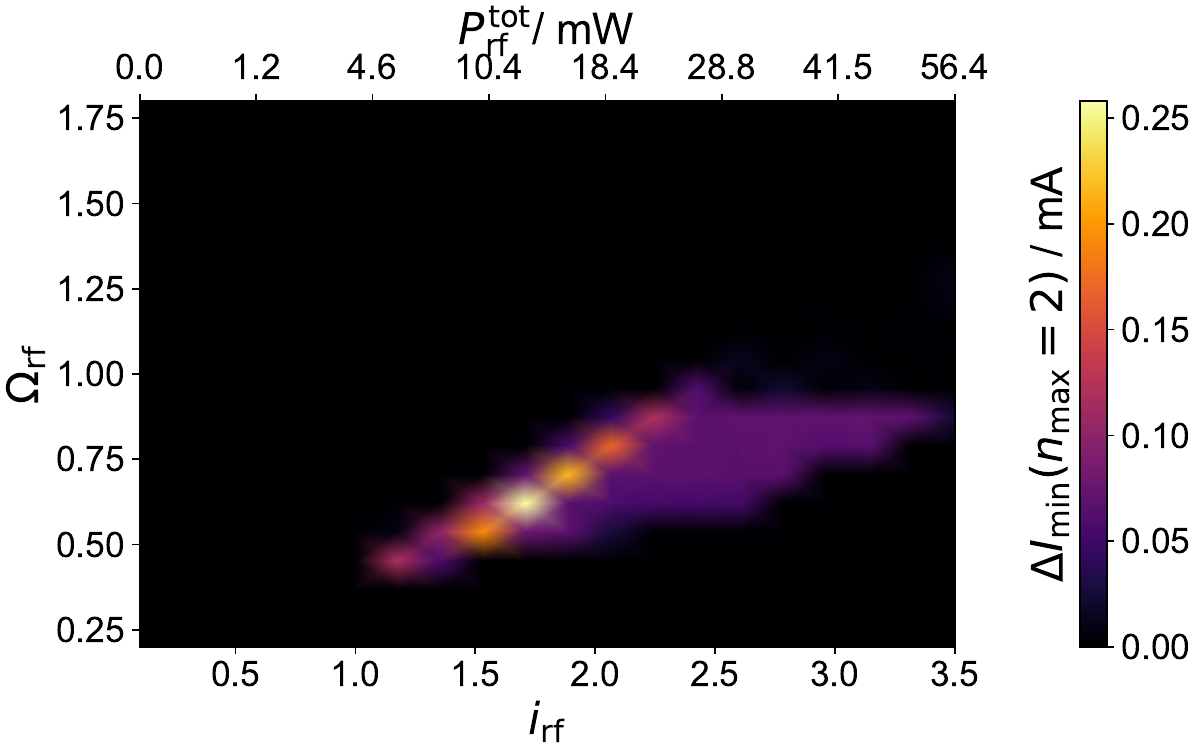}
		\caption{$I_\mathrm{c}=\qty{3}{\milli\ampere}$, $M_\mathrm{JJ}=550$}
		\label{fig:EU-PJVS-cmap_Ptot-fc-DeltaI-nmax_2-sigma_0.05-Ic_3mA-Nstrips_64}
	\end{subfigure}
		\begin{subfigure}{0.49\textwidth}
		\centering
		\includegraphics[width=\linewidth]{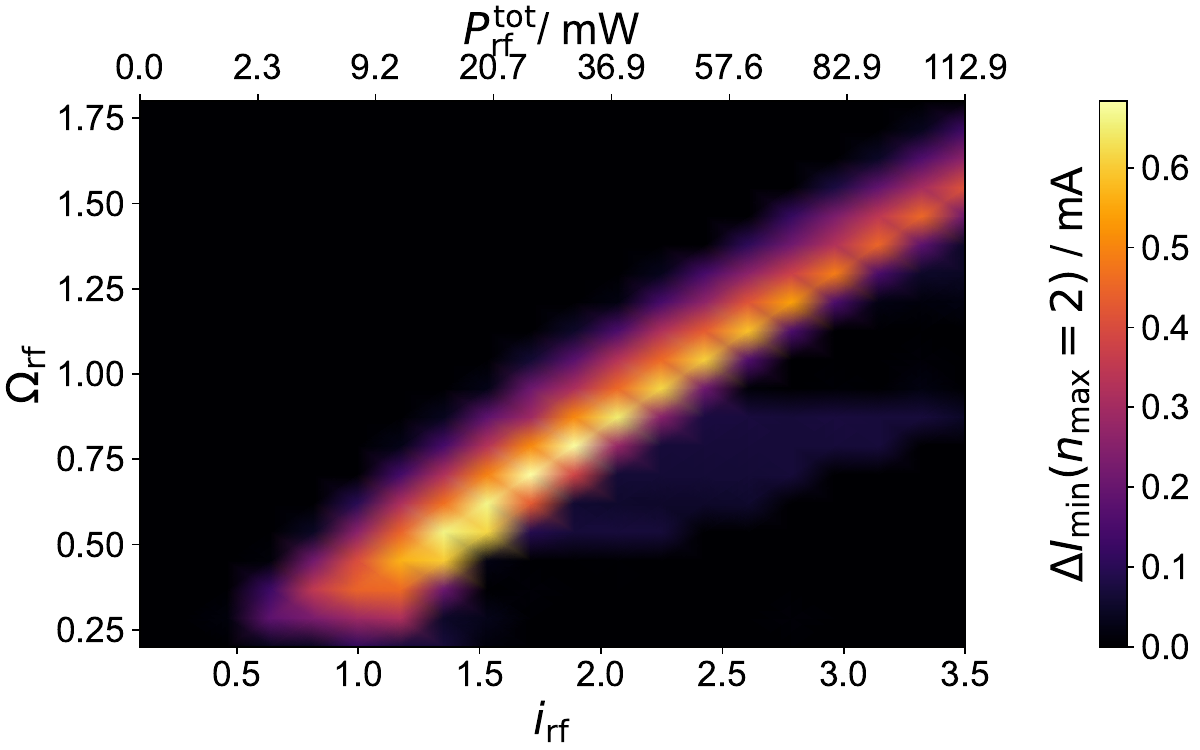}
		\caption{$I_\mathrm{c}=\qty{3}{\milli\ampere}$, $M_\mathrm{JJ}=275$}
		\label{fig:EU-PJVS-cmap_Ptot-fc-DeltaI-nmax_2-sigma_0.05-Ic_3mA-Nstrips_128}
	\end{subfigure}
	\begin{subfigure}{0.49\textwidth}
		\centering
		\includegraphics[width=\linewidth]{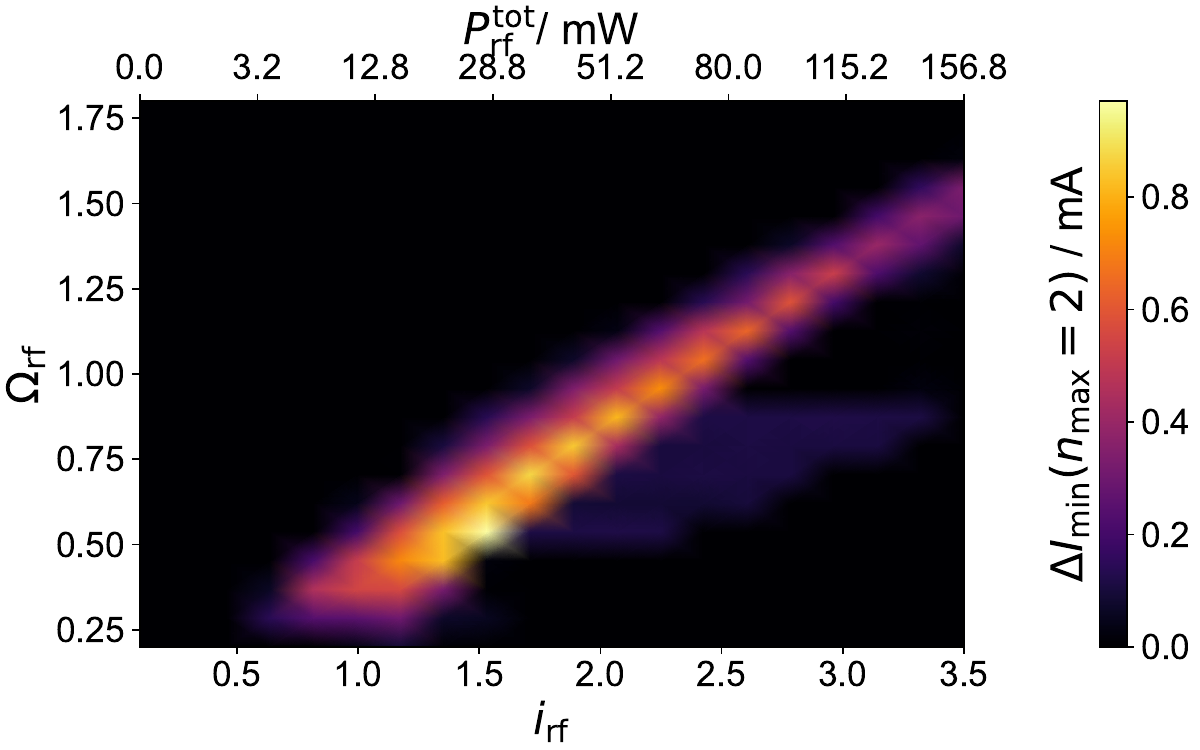}
		\caption{$I_\mathrm{c}=\qty{5}{\milli\ampere}$, $M_\mathrm{JJ}=550$}
		\label{fig:EU-PJVS-cmap_Ptot-fc-DeltaI-nmax_2-sigma_0.05-Ic_5mA-Nstrips_64}
	\end{subfigure}
		\begin{subfigure}{0.49\textwidth}
		\centering
		\includegraphics[width=\linewidth]{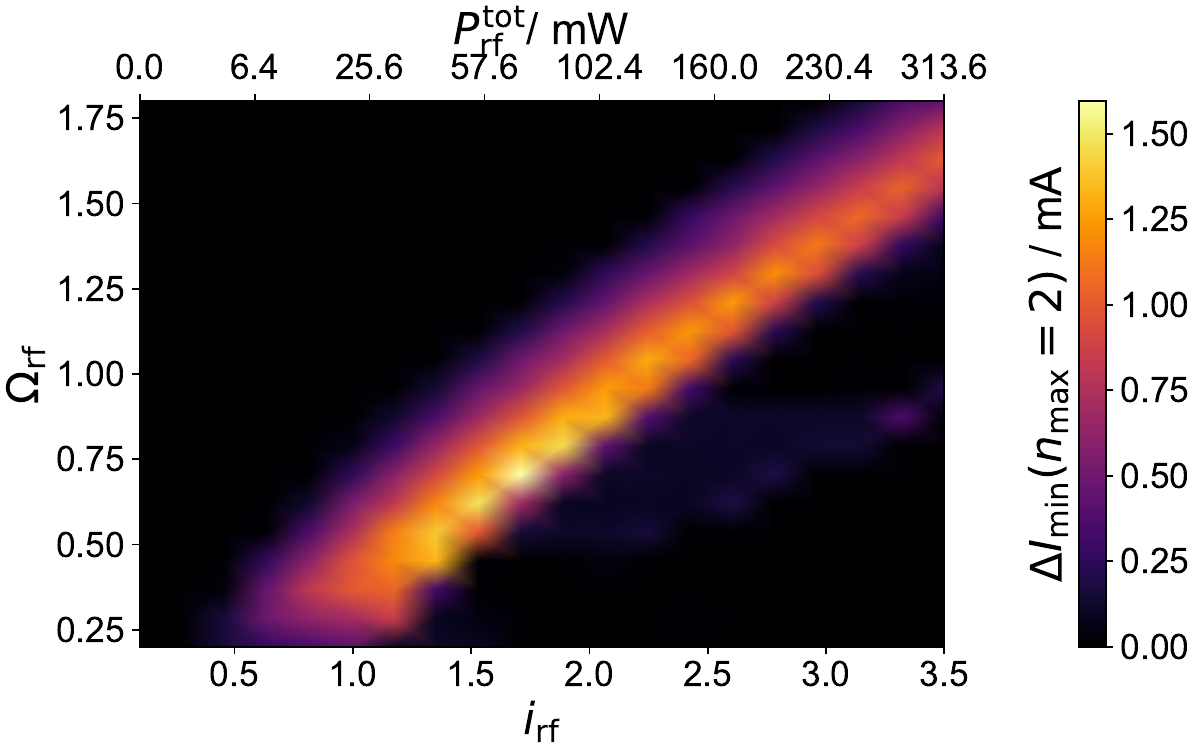}
		\caption{$I_\mathrm{c}=\qty{5}{\milli\ampere}$, $M_\mathrm{JJ}=275$}
		\label{fig:EU-PJVS-cmap_Ptot-fc-DeltaI-nmax_2-sigma_0.05-Ic_5mA-Nstrips_128}
	\end{subfigure}
	\begin{subfigure}{0.49\textwidth}
		\centering
		\includegraphics[width=\linewidth]{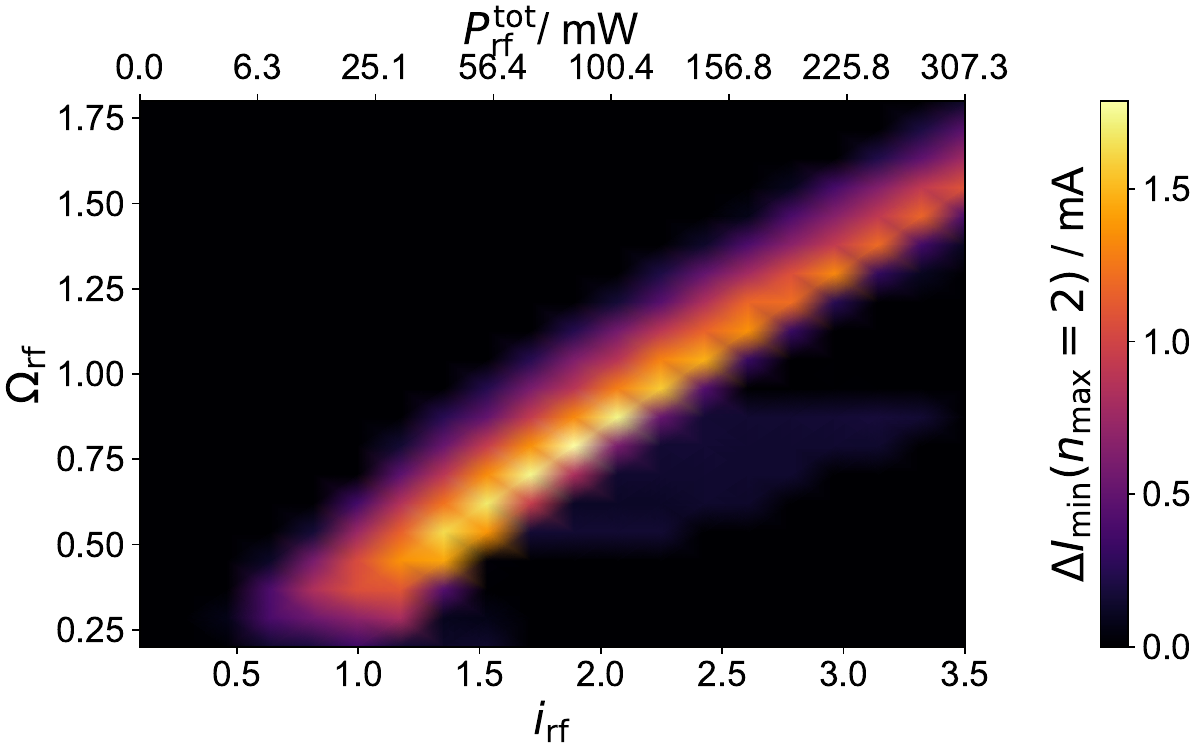}
		\caption{$I_\mathrm{c}=\qty{7}{\milli\ampere}$, $M_\mathrm{JJ}=550$}
		\label{fig:EU-PJVS-cmap_Ptot-fc-DeltaI-nmax_2-sigma_0.05-Ic_7mA-Nstrips_64}
	\end{subfigure}
	\begin{subfigure}{0.49\textwidth}
		\centering
		\includegraphics[width=\linewidth]{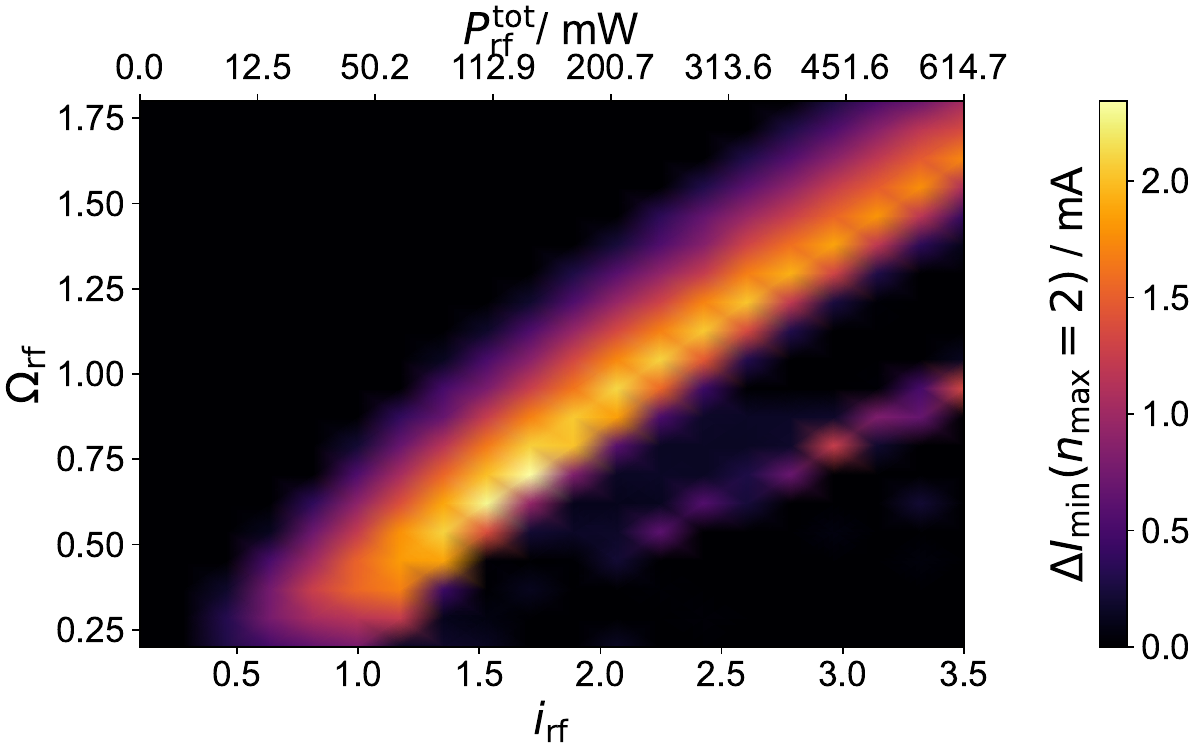}
		\caption{$I_\mathrm{c}=\qty{7}{\milli\ampere}$, $M_\mathrm{JJ}=275$}
		\label{fig:EU-PJVS-cmap_Ptot-fc-DeltaI-nmax_2-sigma_0.05-Ic_7mA-Nstrips_128}
	\end{subfigure}    
	\caption{Analysis of EU-PJVS type $M_\mathrm{JJ}$ Josephson junctions for $n_\mathrm{max}=2$, driven at $f_\mathrm{rf}=\qty{70}{\giga\hertz}$ and integrated in a $Z=\qty{16}{\ohm}$ transmission line, for three $I_\mathrm{c}$ values and two $M_\mathrm{JJ}$ values. Top row (a, b): $I_\mathrm{c}=\qty{3}{\milli\ampere}$. Center row (c, d): $I_\mathrm{c}=\qty{5}{\milli\ampere}$. Bottom row (e, f): $I_\mathrm{c}=\qty{7}{\milli\ampere}$. Left column (a, c, e): $M_\mathrm{JJ}=550$ ($l=64$, $s=6$). Right column (b, d, f): $M_\mathrm{JJ}=275$ ($l=128$, $s=7$).}
	\label{fig:EU-PJVS_3-5-7mA_nmax2}
\end{figure*}


\section{Novel PJVS with simultaneous multi-order steps}
Following the validation of the simulation consistency in the \(n_{\mathrm{max}}=1\) scenario with real PJVS implementations, the analysis of simultaneous operations for \(n_{\mathrm{max}}>1\) has been conducted using both EU and US parameters. The aim is to reduce the total number of junctions \(N_{\mathrm{JJ}}\) and transmission lines \(l\) on the chip while preserving voltage full-scale and resolution, as well as limiting total power dissipation. The \(n_{\mathrm{max}}=2\) scenario is treated as follows, whereas the \(n_{\mathrm{max}}=3\) case is presented in the Appendix (Sec.~\ref{subsec:pjvs_nmax3}).

\subsection{\qty{10}{\volt} EU-PJVS with $n_{\mathrm{max}}=2$} \label{subsec:eu-pjvs-nmax2}
In comparison to the conventional case (Sec.~\ref{sec:eu-pjvs-nmax1}), the further availability of second-order Shapiro steps results in a reduction of the total number of junctions by half. Additionally, if $M_\mathrm{JJ}$ is equivalent to the $n_{\text{max}}=1$ case, the number of transmission lines $l$ may also be halved.
The simulation has been performed across the whole ($i_\mathrm{rf}$, $\Omega_\mathrm{rf}$) parameters space for critical current values $I_\mathrm{c}=\qty{3}{\milli\ampere}$, $I_\mathrm{c}=\qty{5}{\milli\ampere}$ and $I_\mathrm{c}=\qty{7}{\milli\ampere}$, and for $M_\mathrm{JJ}=550$ ($l=64$, $s=6$) and $M_\mathrm{JJ}=275$ ($l=128$, $s=7$). The results are summarized in  Figure~\ref{fig:EU-PJVS_3-5-7mA_nmax2}.

In accordance with the single-junction analysis (Sec.~\ref{sec:single-jj}), the optimal pair of characteristic frequency and rf power that achieves a suitably wide quantum step with minimal power requirements is found at $i_\mathrm{rf}\sim 1.5$ and $\Omega_\mathrm{rf} \sim 0.6$ ($f_\mathrm{c} \sim \qty{117}{\giga\hertz}$).
By keeping the same number of junctions per transmission line of the $n_\mathrm{max}=1$ case ($M_\mathrm{JJ}\sim 550$), small steps below \qty{300}{\micro\ampere} are obtained for $I_\mathrm{c}=\qty{3}{\milli\ampere}$ (Figure~\ref{fig:EU-PJVS-cmap_Ptot-fc-DeltaI-nmax_2-sigma_0.05-Ic_3mA-Nstrips_64}). The \qty{1}{\milli\ampere} target is attained for $I_\mathrm{c}$ between \qty{5}{\milli\ampere} and \qty{7}{\milli\ampere} and with suitably low power amounts (\qtyrange{30}{50}{\milli\watt}). On the other hand, halving $M_\mathrm{JJ}$ enables a lower critical current for the \qty{1}{\milli\ampere} threshold ($I_\mathrm{c}\simeq \qty{5}{\milli\ampere}$). Again, the total power $P_\mathrm{rf}^\mathrm{tot}$ is only about \qty{50}{\milli\watt} (Figure~\ref{fig:EU-PJVS-cmap_Ptot-fc-DeltaI-nmax_2-sigma_0.05-Ic_5mA-Nstrips_128}).

\begin{figure}[tbp]
	\centering
	\begin{subfigure}{0.48\textwidth}
		\centering
		\includegraphics[width=\linewidth]{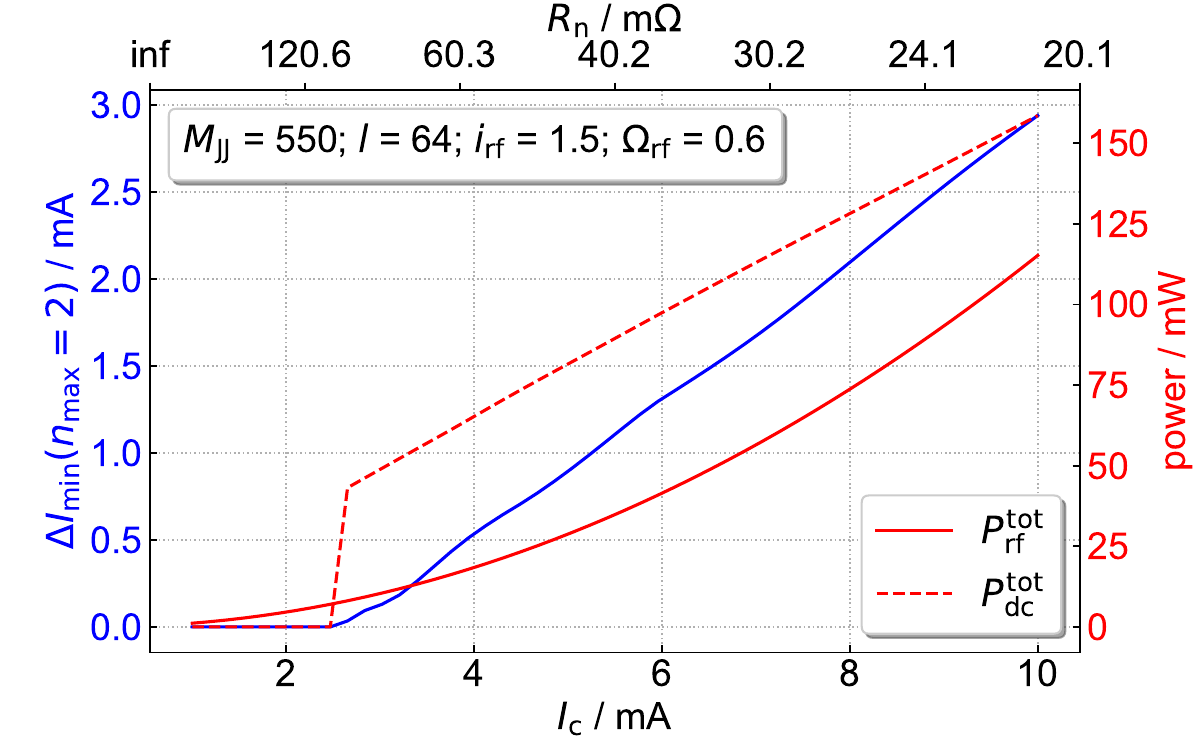}
		\caption{}
		\label{fig:EU-PJVS-nmax2-Ic_Ptot_for_1mA-Nstrips_64-irf_1.5-Orf_0.6}
	\end{subfigure}    
	\begin{subfigure}{0.48\textwidth}
		\centering
		\includegraphics[width=\linewidth]{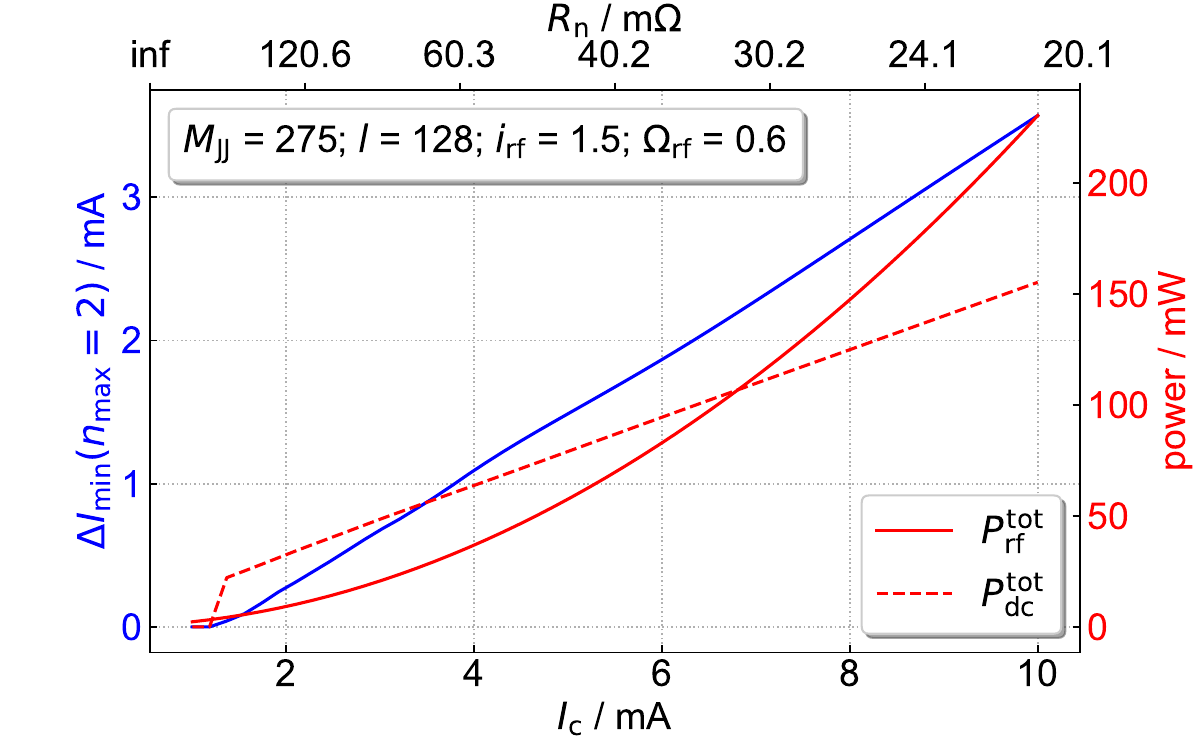}
		\caption{}
		\label{fig:EU-PJVS-nmax2-Ic_Ptot_for_1mA-Nstrips_128-irf_1.5-Orf_0.6}
	\end{subfigure}
	\caption{\qty{10}{\volt} EU-PJVS: minimum step width $\Delta I_\mathrm{min}(n_\mathrm{max}=2)$ (blue curve, left y-axis), total rf power $P_\mathrm{rf}^\mathrm{tot}$ and total dc power $P_\mathrm{dc}^\mathrm{tot}$ (red curves, right y-axis) as a function of nominal critical current $I_\mathrm{c}$ (bottom x-axis) and normal resistance $R_\mathrm{n}$ (top x-axis) for $n_\mathrm{max}=2$ with $i_\mathrm{rf}=1.5$ and $\Omega_\mathrm{rf}=0.6$. a) $M_\mathrm{JJ}=550$ and $l=64$; b) $M_\mathrm{JJ}=275$ and $l=128$.}
	\label{fig: Ic_DeltaI_Ptot_Rn_nmax2_optimal}
\end{figure}

In Figures~\ref{fig:EU-PJVS-nmax2-Ic_Ptot_for_1mA-Nstrips_64-irf_1.5-Orf_0.6} and \ref{fig:EU-PJVS-nmax2-Ic_Ptot_for_1mA-Nstrips_128-irf_1.5-Orf_0.6}, the minimum quantum step width $\Delta I_{\mathrm{min}}$ for $n_\mathrm{max}=2$ is illustrated as a function of $I_{\mathrm{c}}$ under the optimal conditions of $i_{\mathrm{rf}}=1.5$ and $\Omega_{\mathrm{rf}}=0.6$. Also, corresponding values of total rf power $P_\mathrm{rf}^\mathrm{tot}$, total dc power $P_\mathrm{dc}^\mathrm{tot}$, and normal resistance $R_\mathrm{n}= \Phi_\mathrm{0}\,f_\mathrm{rf}\, /\, (\Omega_\mathrm{rf}\,I_\mathrm{c})$ are shown.
\noindent It is apparent that the use one fewer power splitter stage $s$ comes with a higher critical current compared to the $s=7$ case with halved JJs per transmission line ($M_\mathrm{JJ}=275$). No significant differences in terms of rf power appear instead between the two possible scenario ($\sim \qty{40}{\milli\watt}$), though a somewhat higher dc power dissipation occurs for more JJs and fewer striplines. Also, compared to the $n_\mathrm{max}=1$ case, the rf power necessary to get at least \qty{1}{\milli\ampere} step does not change significantly.
Should a different step width threshold be required, the expected values of $I_\mathrm{c}$, $R_\mathrm{n}$, $P_\mathrm{rf}^\mathrm{tot}$ and $P_\mathrm{dc}^\mathrm{tot}$ can be derived from this model.

\begin{figure}[htp]
	\centering
	\begin{subfigure}{0.48\textwidth}
		\centering
		\includegraphics[width=\linewidth]{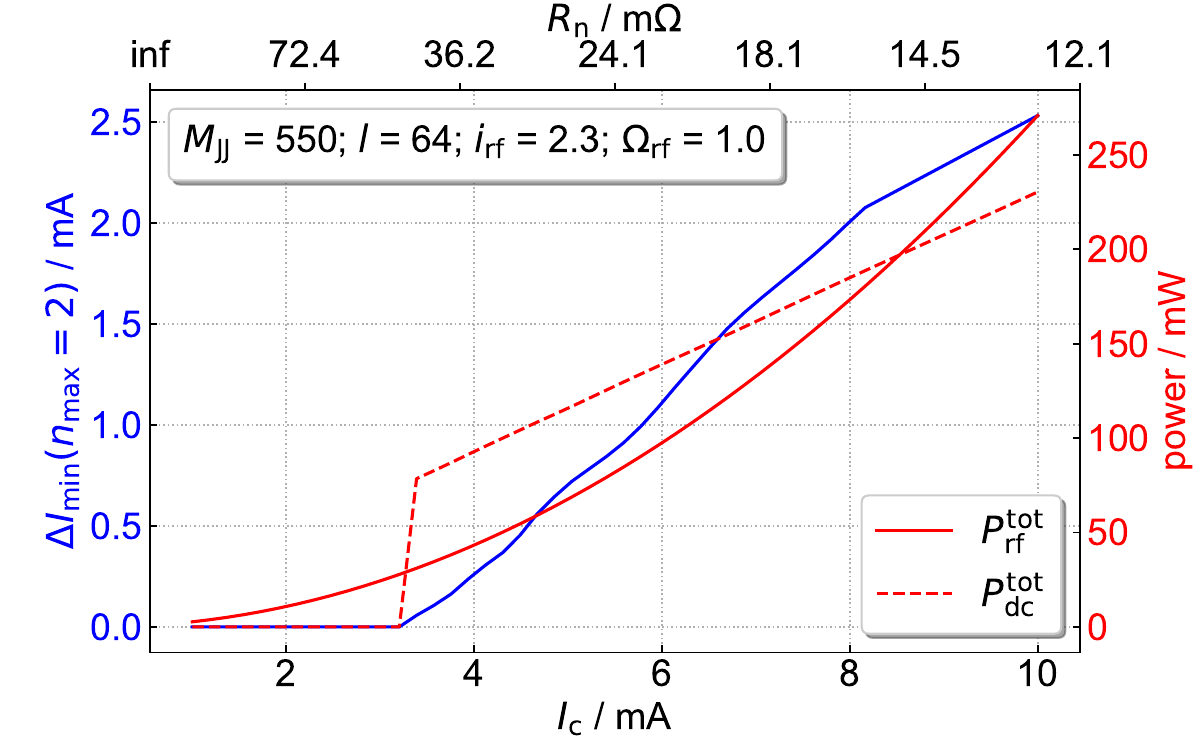}
		\caption{}
		\label{fig:EU-PJVS-nmax2-Ic_Ptot_for_1mA-Nstrips_64-irf_2.3-Orf_1.0}
	\end{subfigure}    
	\begin{subfigure}{0.48\textwidth}
		\centering
		\includegraphics[width=\linewidth]{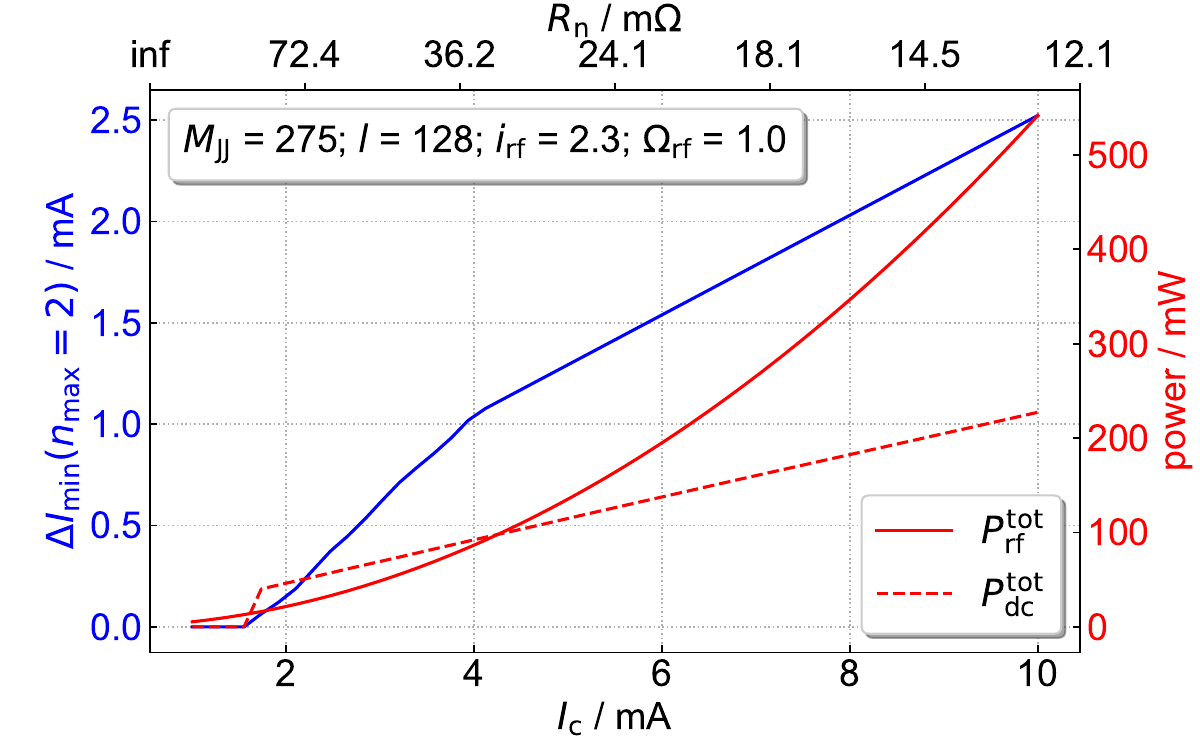}
		\caption{}
		\label{fig:EU-PJVS-nmax2-Ic_Ptot_for_1mA-Nstrips_128-irf_2.3-Orf_1.0}
	\end{subfigure}
	\caption{Analysis of the \qty{10}{\volt} EU-PJVS: minimum step width $\Delta I_\mathrm{min}(n_\mathrm{max}=2)$ (blue curve, left y-axis), total microwave power $P_\mathrm{rf}^\mathrm{tot}$ and total dc power $P_\mathrm{dc}^\mathrm{tot}$ (red curves, right y-axis) as a function of nominal critical current $I_\mathrm{c}$ (bottom x-axis) and normal resistance $R_\mathrm{n}$ (top x-axis) for $n_\mathrm{max}=2$ with $i_\mathrm{rf}=2.3$ and $\Omega_\mathrm{rf}=1$. a) $M_\mathrm{JJ}=550$ and $l=64$; b) $M_\mathrm{JJ}=275$ and $l=128$.}
	\label{fig: Ic_DeltaI_Ptot_Rn_nmax2_non_optimal}
\end{figure}

One could ask whether the electrical parameters of present SNS junctions, optimized for \( n_{\mathrm{max}} = 1 \) with \( \Omega_{\mathrm{rf}} \simeq 1 \), can be leveraged to extend the applicability of the PJVS to higher step orders \( n \) simply by increasing the rf power driving them. In reference to Figure~\ref{fig:EU-PJVS_3-5-7mA_nmax2}, it is evident that JJs operating at \( \Omega_{\mathrm{rf}}=1 \) produce the widest steps when \( i_{\mathrm{rf}} \simeq 2.3 \). The same analysis performed for the optimal \((i_{\mathrm{rf}}, \Omega_{\mathrm{rf}})\) pair has been reiterated for \( \Omega_{\mathrm{rf}}=1 \) and \( i_{\mathrm{rf}}= 2.3 \). The results shown in Figures~\ref{fig:EU-PJVS-nmax2-Ic_Ptot_for_1mA-Nstrips_64-irf_2.3-Orf_1.0} and \ref{fig:EU-PJVS-nmax2-Ic_Ptot_for_1mA-Nstrips_128-irf_2.3-Orf_1.0} demonstrate that rf power levels between \qtyrange{80}{100}{\milli\watt} are required to reach the \qty{1}{\milli\ampere} threshold. This is more than double the power required in the optimal ($i_{\mathrm{rf}}$, $\Omega_{\mathrm{rf}}$) configuration, while the minimum critical current remains nearly the same.


\subsection{\qty{10}{\volt} US-PJVS with $n_{\mathrm{max}}=2$}
\begin{figure}[htp]
	\centering
	\begin{subfigure}{0.48\textwidth}
		\centering
		\includegraphics[width=\linewidth]{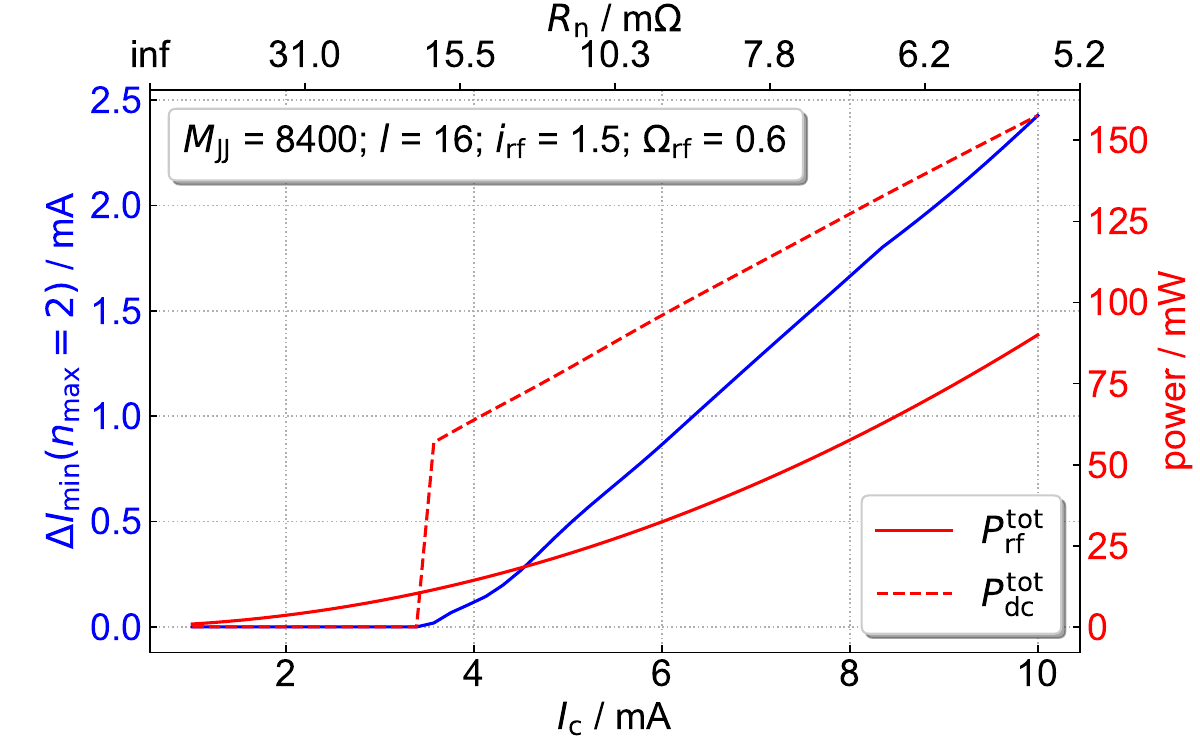}
		\caption{}
		\label{fig:US-PJVS-nmax2-Ic_Ptot_for_1mA-Nstrips_16-irf_1.5-Orf_0.6}
	\end{subfigure}    
	\begin{subfigure}{0.48\textwidth}
		\centering
		\includegraphics[width=\linewidth]{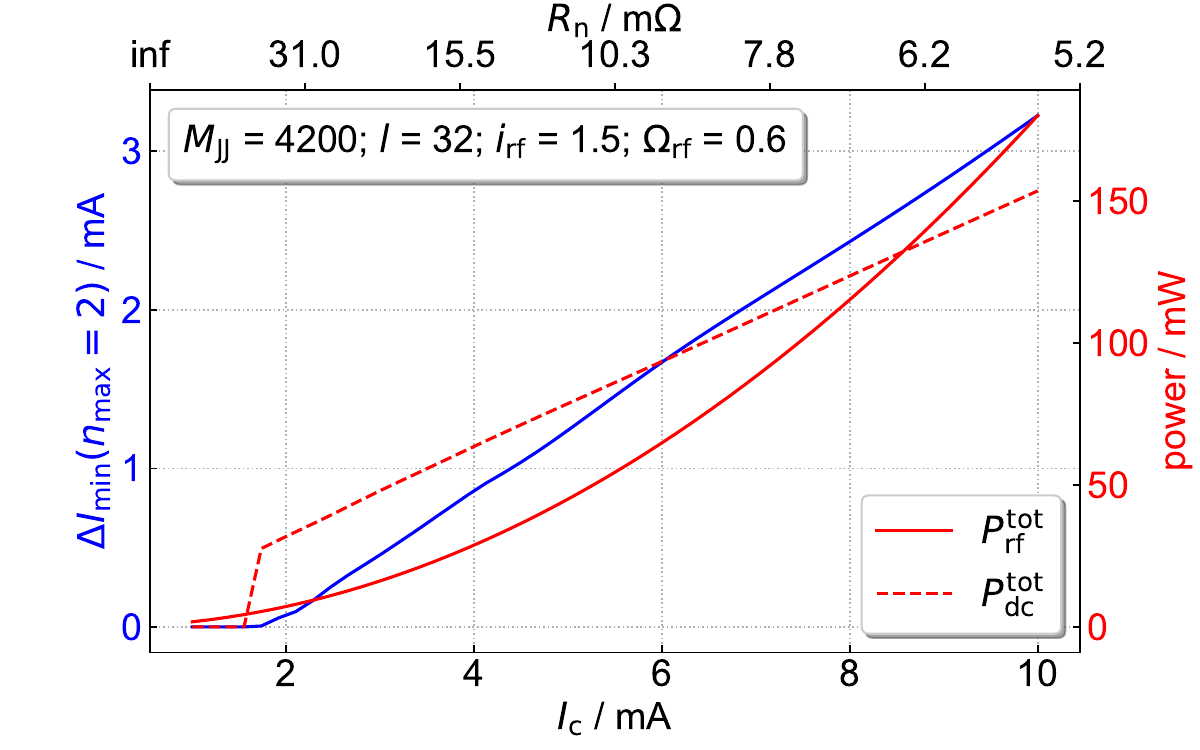}
		\caption{}
		\label{fig:US-PJVS-nmax2-Ic_Ptot_for_1mA-Nstrips_32-irf_1.5-Orf_0.6}
	\end{subfigure}
	\caption{Analysis of the \qty{10}{\volt} US-PJVS: minimum step width $\Delta I_\mathrm{min}(n_\mathrm{max}=2)$ (blue curve, left y-axis), total microwave power $P_\mathrm{rf}^\mathrm{tot}$ and total dc power $P_\mathrm{dc}^\mathrm{tot}$ (red curves, right y-axis) as a function of nominal critical current $I_\mathrm{c}$ (bottom x-axis) and normal resistance $R_\mathrm{n}$ (top x-axis) for $n_\mathrm{max}=2$ with $i_\mathrm{rf}=1.5$ and $\Omega_\mathrm{rf}=0.6$. a) $M_\mathrm{JJ}=8400$ and $l=16$; b) $M_\mathrm{JJ}=4200$ and $l=32$.}
	\label{fig:US-PJVS-Ic_DeltaI_Ptot_Rn_nmax2_optimal}
\end{figure}

\begin{figure}[bthp]
	\centering
	\begin{subfigure}{0.48\textwidth}
		\centering
		\includegraphics[width=\linewidth]{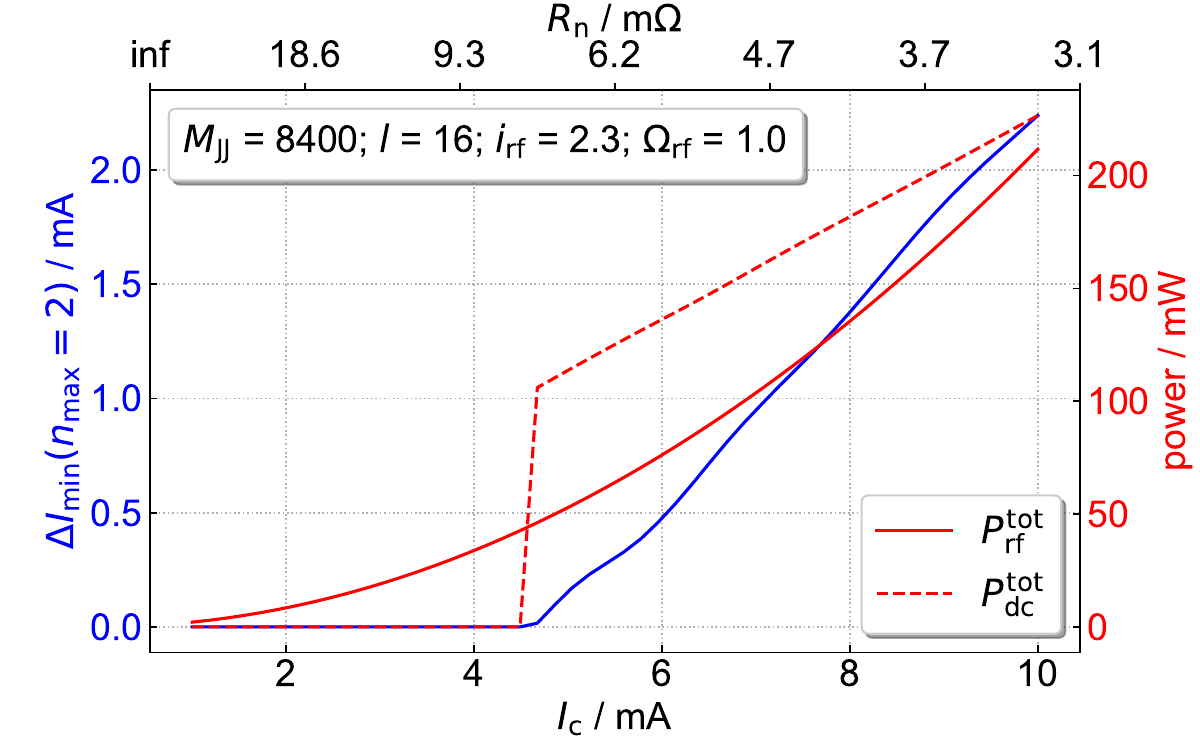}
		\caption{}
		\label{fig:US-PJVS-nmax2-Ic_Ptot_for_1mA-Nstrips_16-irf_2.3-Orf_1.0}
	\end{subfigure}
	\begin{subfigure}{0.48\textwidth}
		\centering
		\includegraphics[width=\linewidth]{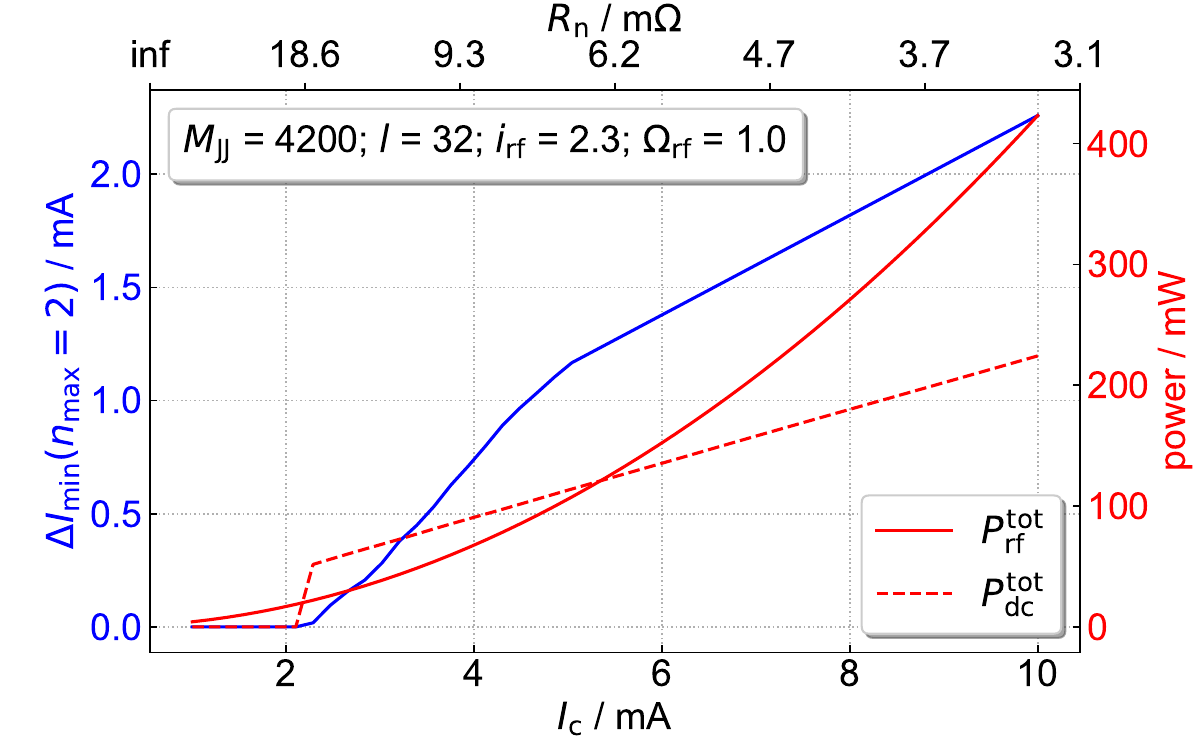}
		\caption{}
		\label{fig:US-PJVS-nmax2-Ic_Ptot_for_1mA-Nstrips_32-irf_2.3-Orf_1.0}
	\end{subfigure}
	\caption{Analysis of the \qty{10}{\volt} US-PJVS: minimum step width $\Delta I_\mathrm{min}(n_\mathrm{max}=2)$ (blue curve, left y-axis), total microwave power $P_\mathrm{rf}^\mathrm{tot}$ and total dc power $P_\mathrm{dc}^\mathrm{tot}$ (red curves, right y-axis) as a function of nominal critical current $I_\mathrm{c}$ (bottom x-axis) and normal resistance $R_\mathrm{n}$ (top x-axis) for $n_\mathrm{max}=2$ with $i_\mathrm{rf}=2.3$ and $\Omega_\mathrm{rf}=1$. a) $M_\mathrm{JJ}=8400$ and $l=16$; b) $M_\mathrm{JJ}=4200$ and $l=32$}
	\label{fig:US-PJVS-Ic_DeltaI_Ptot_Rn_nmax2_non_optimal}
\end{figure}

Josephson arrays of the US-type with quantum steps up to $n=2$ are simulated at the optimal $\Omega_\mathrm{rf}$ value of 0.6.
According to Table~\ref{tab:EU-US-PJVS}, the advantage of using the $n=2$ steps allows the number of splitting stages to be reduced from 5 to 4 by keeping the same $M_\mathrm{JJ}\sim\num{8400}$ along each line. On the other hand, $M_\mathrm{JJ}$ can be halved if $s=5$. 
Results in Figure~\ref{fig:US-PJVS-Ic_DeltaI_Ptot_Rn_nmax2_optimal} show that, for both $l=16$ and $l=32$, the required rf power is only around \qty{40}{\milli\watt} with $I_\mathrm{c}\simeq \qty{6}{\milli\ampere}$ and $I_\mathrm{c}\simeq \qty{4.5}{\milli\ampere}$, respectively. 
Instead, with unoptimized $\Omega_\mathrm{rf}=1$ and $i_\mathrm{rf}=2.3$ (Figure~\ref{fig:US-PJVS-Ic_DeltaI_Ptot_Rn_nmax2_non_optimal}) the required rf power for \qty{1}{\milli\ampere} step width is again more than doubled ($\sim\qty{100}{\milli\watt}$), with $I_\mathrm{c} \simeq \qty{7}{\milli\ampere}$ for $l=16$ and $I_\mathrm{c} \simeq \qty{4.5}{\milli\ampere}$ for $l=32$. As with EU-PJVS (Sec.~\ref{subsec:eu-pjvs-nmax2}), the total dc power $P_\mathrm{dc}^\mathrm{tot}$ is higher for reduced spitting stages and transmission lines.

The main parameters for both PJVS versions with \( n_{\text{max}}=2 \) and for minimum quantum step width of \qty{1}{\milli\ampere} are summarized in Table~\ref{tab:comparison_PJVS_nmax2}. The table indicates that achieving this target is feasible while keeping the total power levels (dc + rf) below \qty{150}{\milli\watt}, provided that both JJ electrical parameters and rf power are adjusted to ensure \( \Omega_{\text{rf}} = 0.6 \) and \( i_{\text{rf}} \simeq 1.5 \). This will require modifications to the overdamped JJs currently in use. However, present junction technology can still be employed, either by significantly increasing total power requirements (exceeding \qty{170}{\milli\watt}), or by operating at lower cryogenic temperatures to meet the condition $\Omega_\mathrm{rf} \simeq 0.6$, albeit with more challenging experimental conditions.

\begin{table}[H]
	\centering
	\caption{Comparison of simulated EU-PJVS and US-PJVS for $n_\mathrm{max}=2$ and minimum $\Delta I_\mathrm{min}=\qty{1}{\milli\ampere}$ for different normalized rf frequencies $\Omega_\mathrm{rf}$, rf currents $i_\mathrm{rf}$, number of transmission lines $l$ (as well as junctions per line $M_\mathrm{JJ}$ and number of splitting stages $s$).}    
	\label{tab:comparison_PJVS_nmax2}    
	\begin{adjustbox}{width=\textwidth}
		
		\begin{tabular}{r|c c|c c|c c|c c}
			\toprule
			& \multicolumn{4}{c|}{\textbf{EU-PJVS}} & \multicolumn{4}{c}{\textbf{US-PJVS}} \\
			\cmidrule(lr){2-5} \cmidrule(lr){6-9}
			& \multicolumn{2}{c|}{$M_\mathrm{JJ}=550$, $l=64$, $s=6$} & \multicolumn{2}{c|}{$M_\mathrm{JJ}=275$, $l=128$, $s=7$} 
			& \multicolumn{2}{c|}{$M_\mathrm{JJ}=8400$, $l=16$, $s=4$} & \multicolumn{2}{c}{$M_\mathrm{JJ}=4200$, $l=32$, $s=5$} \\
			\midrule
			& \makecell{$\Omega_\mathrm{rf}=0.6$\\$i_\mathrm{rf} = 1.5$} & \makecell{$\Omega_\mathrm{rf}=1$\\$i_\mathrm{rf} = 2.3$} & 
			\makecell{$\Omega_\mathrm{rf}=0.6$\\$i_\mathrm{rf} = 1.5$} & \makecell{$\Omega_\mathrm{rf}=1$\\$i_\mathrm{rf} = 2.3$} &
			\makecell{$\Omega_\mathrm{rf}=0.6$\\$i_\mathrm{rf} = 1.5$} & \makecell{$\Omega_\mathrm{rf}=1$\\$i_\mathrm{rf} = 2.3$} & 
			\makecell{$\Omega_\mathrm{rf}=0.6$\\$i_\mathrm{rf} = 1.5$} & \makecell{$\Omega_\mathrm{rf}=1$\\$i_\mathrm{rf} = 2.3$} \\
			\midrule
			$I_\mathrm{c}$ / \qty{}{\milli\ampere}  & 5.2 & 5.8 & 3.8 & 3.9 & 6.3 & 7.1 & 4.4 & 4.6\\
			$R_\mathrm{n}$ / \qty{}{\milli\ohm}     & 46.4 & 25& 63.5 & 37.1 & 9.8& 5.2& 14.1 & 8.1\\
			$P_\mathrm{rf}^\mathrm{tot}$ / \qty{}{\milli\watt}& 32& 90& 33& 82 & 36 & 107 & 35& 88\\
			$P_\mathrm{dc}^\mathrm{tot}$ / \qty{}{\milli\watt}                    & 85& 134& 60 & 90 & 101& 162 & 70 & 104 \\
			$P_\mathrm{rf}^\mathrm{tot} + P_\mathrm{dc}^\mathrm{tot}$ / \qty{}{\milli\watt}& 117& 224& 93 & 172 & 137& 269 & 105& 192\\			
			\bottomrule
		\end{tabular}
	\end{adjustbox}
\end{table}

A comparable study has been performed for \(n_{\text{max}} = 3\), which is detailed in the appendix (Sec.~\ref{subsec:pjvs_nmax3}): in this scenario, seven Shapiro steps (both positive and negative) are required to be simultaneously wider than \qty{1}{\milli\ampere}.
According to \cite{durandetto2019non}, in addition to sparing JJs and enabling the use of fewer parallel transmission lines, the further availability of wide second-order steps facilitates a reduction in the total number of sub-arrays and bias lines by a factor of two compared to conventional PJVS (from 17 to 8 in EU design). However, the inclusion of third-order steps does not yield a significant improvement in these respects, as the count of sub-arrays would only decrease by one relative to the \(n_{\text{max}}=2\) scenario.

\section{Conclusion}
The dynamics of multiple series-connected overdamped Josephson junctions have been investigated through numerical simulations based on the RSJ model implemented in \textit{Python}. The goal was to establish the optimal relationship between rf power and JJs electrical parameters enabling the generation of large quantum voltage steps up to $n_{\text{max}} = 2$ or $n_{\text{max}} = 3$, which can be harnessed in next-generation Programmable Josephson Voltage Standards (PJVS). The analysis considered realistic physical properties akin to present LTS-based PJVS technologies and confirmed agreement for the conventional $n_{\text{max}} = 1$ case. Both rf power attenuation and non-uniformity in the electrical parameters were examined to identify the maximum number of series-connected JJs that can generate usable multi-order quantum voltage steps, while minimizing the total power required to operate the array.

The findings suggest that leveraging second-order quantum steps is feasible for both \qty{18}{\giga\hertz} and \qty{70}{\giga\hertz} PJVS  architectures. This can be achieved either by adjusting the electrical parameters of the JJs to optimize total power consumption or by maintaining present JJ technologies, albeit at the cost of higher power requirements. On the other hand, the usability of third-order Shapiro steps in PJVS systems with \qty{1}{\milli\ampere} quantum-locking ranges would necessitate a rethinking of JJ electrical parameters to ensure sustainable power levels. While transitioning from PJVS binary/ternary structures ($n = -1, 0, +1$) to a quinary configuration ($n = -2, -1, 0, +1, +2$) demonstrates significant advantages, further extending the segmentation to a septary arrangement ($n = -3, -2, -1, 0, +1, +2, +3$) offers only marginal improvements.
Regarding HTS JVS, it could be beneficial to explore more complex combinations of Shapiro step orders to increase the overall quantum-locking range. For instance, one might investigate the option of omitting a specific step order, as suggested in \cite{klushin1999toward}, to enhance the current-width of the remaining step orders without sacrificing the potential to increase the overall voltage.


\section*{Acknowledgments}
The author thanks Emanuele Enrico and Andrea Sosso (INRiM) for valuable discussions.

The project (23FUN05 AQuanTEC) has received funding from the European Partnership on Metrology,
co-financed from the European Union’s Horizon Europe Research and Innovation Programme and by
the Participating States.

The project (23RPT01 WAC) has received funding from the European Partnership on Metrology,
co-financed from the European Union’s Horizon Europe Research and Innovation Programme and by
the Participating States.




\appendix
\section{\qty{10}{\volt} EU-PJVS with $n_\mathrm{max}=1$}
\begin{figure}[htbp]
	\centering
	\begin{subfigure}{0.48\textwidth}
		\centering
		\includegraphics[width=\linewidth]{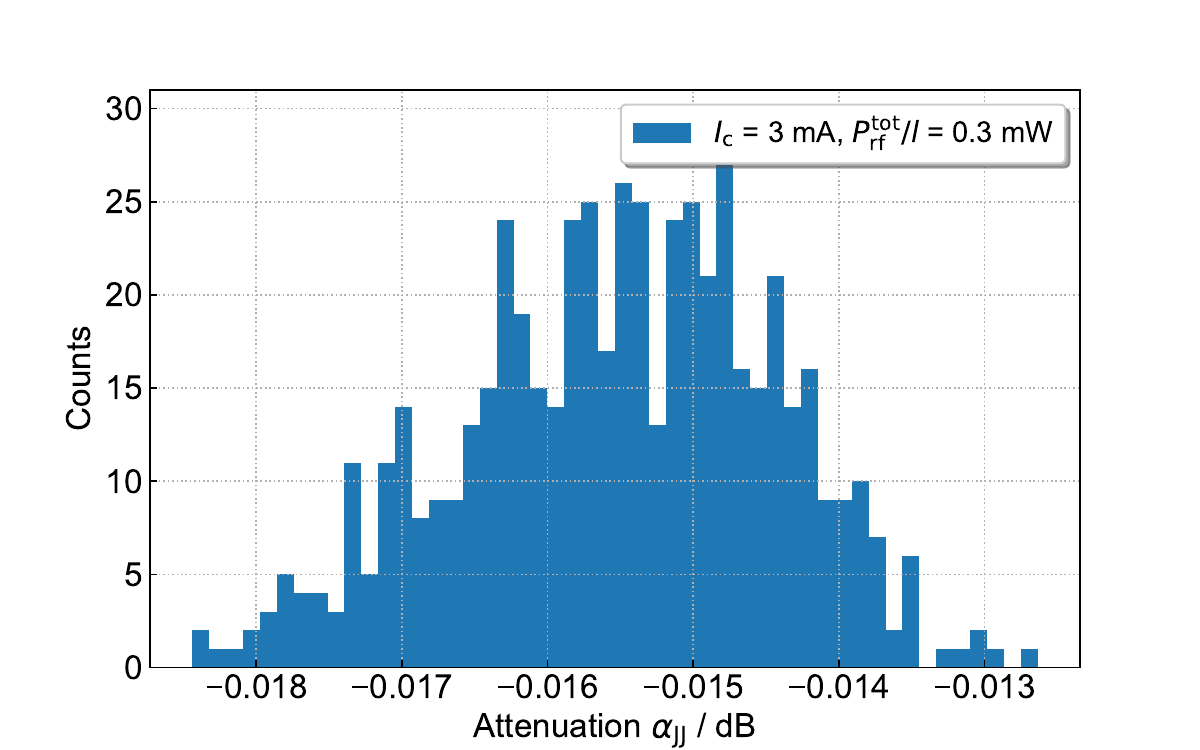}
		\caption{}
		\label{fig:attenuation_histo_EUPJVS_3mA_550JJs}
	\end{subfigure}
	\begin{subfigure}{0.48\textwidth}
		\centering
		\includegraphics[width=\linewidth]{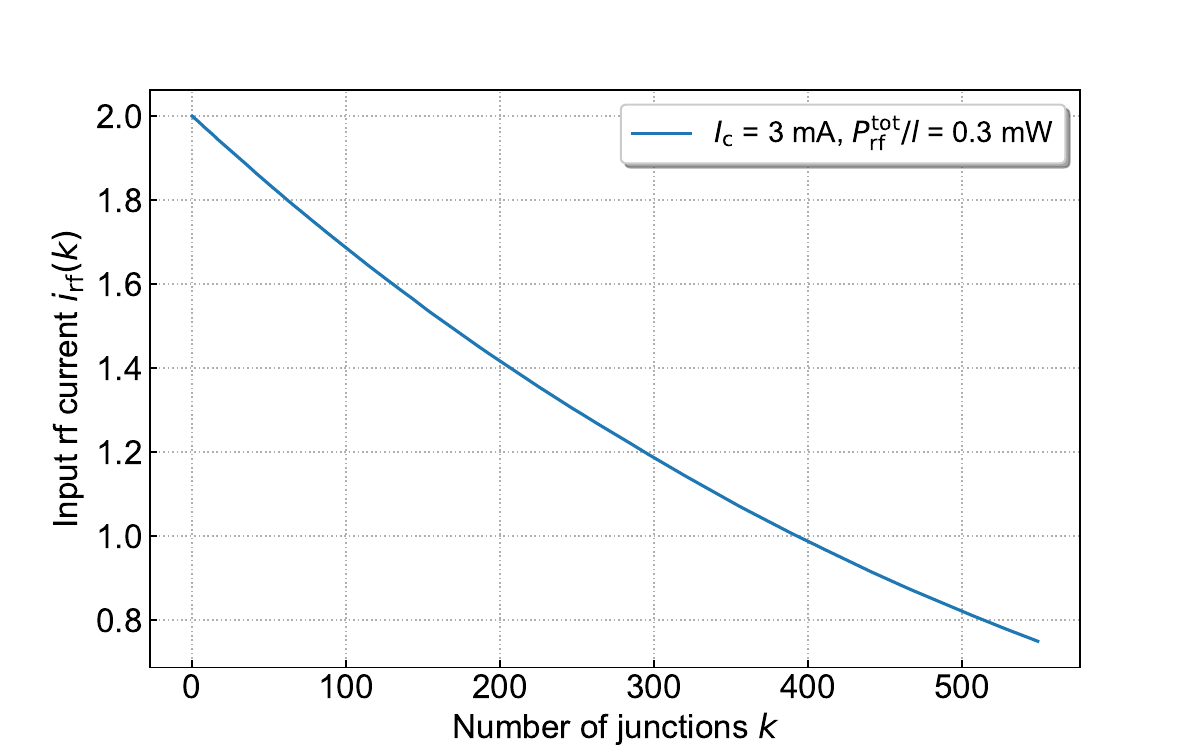}
		\caption{}
		\label{fig:Irf_vs_k_EUPJVS_3mA_550JJs}
	\end{subfigure}
	\caption{Simulation of conventional \qty{10}{\volt} EU-PJVS ($n_\mathrm{max}=1$): $M_\mathrm{JJ}=550$ JJs arranged along a $Z=\qty{16}{\ohm}$ transmission line and rf-driven at $f_\mathrm{rf}=\qty{70}{\giga\hertz}$, nominal critical current $I_\mathrm{c}=\qty{3}{\milli\ampere}$, nominal characteristic frequency $f_\mathrm{c}=\qty{70}{\giga\hertz}$ ($\Omega_\mathrm{rf}=1$), rf-powered at $P_\mathrm{rf}^\mathrm{tot}/l = \qty{0.3}{\milli\watt}$ ($i_\mathrm{rf}=2$, $l=128$), with electrical parameters spread of \qty{5}{\%} and for $n_\mathrm{max}=1$. a) Distribution of rf power attenuation $\alpha_\mathrm{JJ}$ of each JJ in the array. b) Normalized rf current $i_\mathrm{rf}(k)$ progression along the JJ array.
	}
	\label{fig:att_EUPJVS_3mA_550JJs}
\end{figure}

The histogram in Figure~\ref{fig:attenuation_histo_EUPJVS_3mA_550JJs} shows the JJ attenuation $\alpha_\mathrm{JJ}$ for $M_\mathrm{JJ}=550$ junctions ($l=128$, $s=7$) with nominal critical current $I_\mathrm{c}=\qty{3}{\milli\ampere}$ of the EU-PJVS architecture presented in Sec.~\ref{sec:eu-pjvs-nmax1}, with $i_\mathrm{rf}=2$ for $k=1$, corresponding to single line rf power $P_\mathrm{rf}^\mathrm{tot}/l\simeq \qty{0.3}{\milli\watt}$, and nominal $\Omega_\mathrm{rf}=1$. Figure~\ref{fig:Irf_vs_k_EUPJVS_3mA_550JJs} plots the corresponding normalized rf current $i_\mathrm{rf}$ evolution along the $M_\mathrm{JJ}$ junctions within the \qty{16}{\ohm} rf transmission line.

\section{\qty{10}{\volt} US-PJVS with $n_\mathrm{max}=1$} \label{appsec:us-pjvs-nmax1}
\begin{figure}[btp]
	\centering
	\begin{subfigure}{0.48\textwidth}
		\centering
		\includegraphics[width=\linewidth]{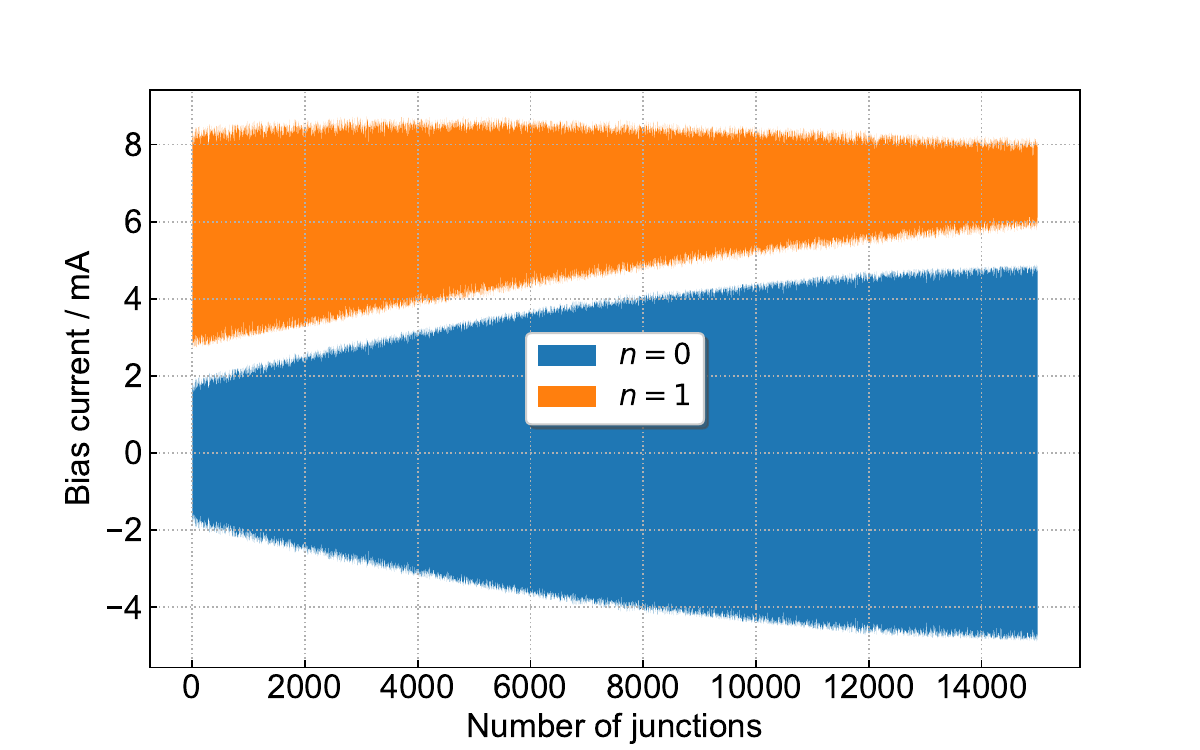}
		\caption{}
		\label{fig:US-PJVS-steps_vs_jj-nmax1_Ic5mA}
	\end{subfigure}
	\begin{subfigure}{0.48\textwidth}
		\centering
		\includegraphics[width=\linewidth]{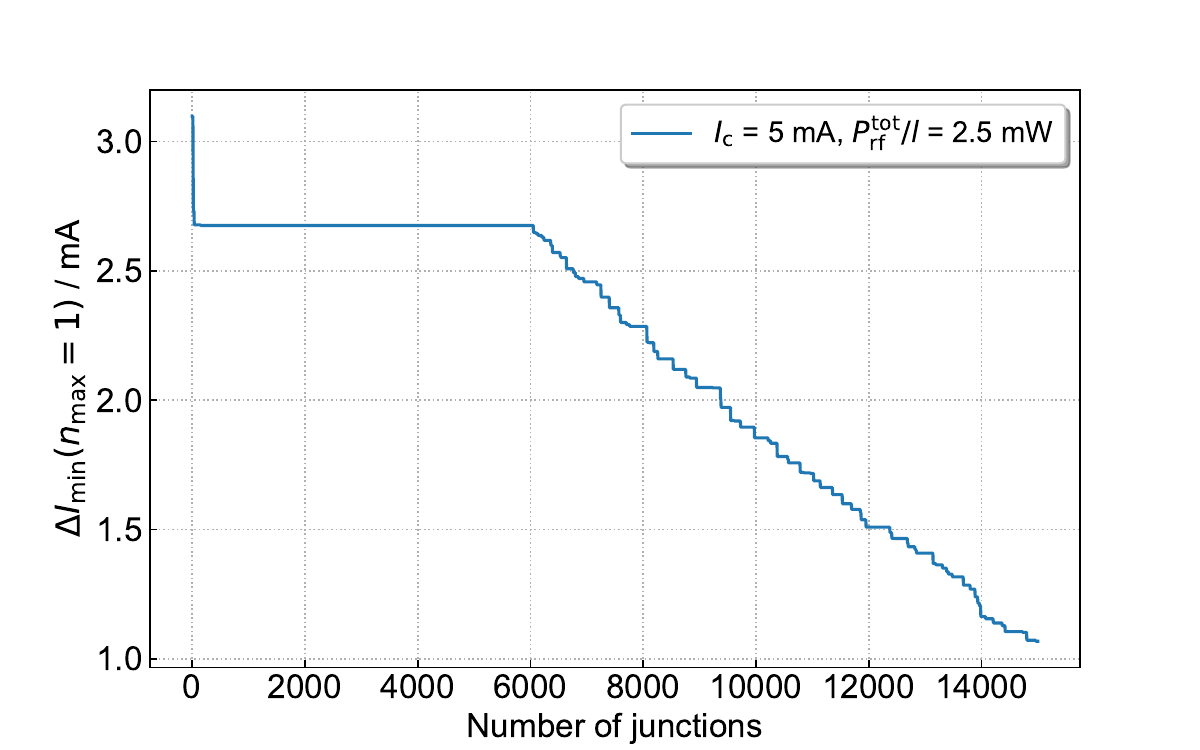}
		\caption{}
		\label{fig:US-PJVS-min_step_width_vs_jj-nmax1_Ic5mA}
	\end{subfigure}
	\caption{Analysis of $M_\mathrm{JJ}=\num{15000}$ series-connected junctions of the US-PJVS type with nominal critical current $I_\mathrm{c}=\qty{5}{\milli\ampere}$, nominal characteristic frequency $f_\mathrm{c}=\qty{18}{\giga\hertz}$ ($\Omega_\mathrm{rf}=1$), rf-powered at $P_\mathrm{rf}^\mathrm{tot}/l = \qty{2.5}{\milli\watt}$ ($i_\mathrm{rf}=2$, $l=32$), with electrical parameters spread of \qty{5}{\%} and for $n_\mathrm{max}=1$. a) Evolution of quantum voltage steps for $n=0$ (blue) and $n=1$ (orange) along the array. b) Evolution of the minimum quantum voltage step width $\Delta I_\mathrm{min}(n_\mathrm{max}=1)$.}
	\label{fig:US-PJVS_5mA_nmax1}
\end{figure}

NIST-PJVS devices features tapered transmission lines \cite{dresselhaus2009tapered}, where the characteristic impedance $Z$ gradually decreases from \qty{50}{\ohm} to \qty{23}{\ohm} to mitigate the rf current reduction.
In this analysis, a constant line impedance of \qty{50}{\ohm} and a \qty{5}{\%} spread of electrical parameters are considered.
Voltage steps evolution of a $M_\mathrm{JJ}=\num{15000}$ junction-array with nominal $I_\mathrm{c} = \qty{5}{\milli\ampere}$, $f_\mathrm{c} = \qty{18}{\giga\hertz}$ and with nominal $\Omega_\mathrm{rf}=1$ and $i_\mathrm{rf}(k=1)=2$ (corresponding to single line rf power $P_\mathrm{rf}^\mathrm{tot}/l\simeq \qty{2.5}{\milli\watt}$) are shown in Figure~\ref{fig:US-PJVS_5mA_nmax1}. Owing to the higher $Z/R_\mathrm{n}$ ratio compared to the EU-PJVS, the average JJ attenuation is considerably reduced to about \qty{-0.0008}{\decibel}/JJ.
Plots in Figure~\ref{fig:US-PJVS_5mA_nmax1} show the $n=0$ and $n=+1$ quantum steps progression of $M_\mathrm{JJ}=\num{15000}$ junctions of the US-PJVS type with nominal $I_\mathrm{c}=\qty{5}{\milli\ampere}$ and the corresponding minimum step width $\Delta I_\mathrm{min}(n_\mathrm{max}=1)$.

\noindent Junction after junction, the minimum step width decreases because of both attenuation and JJ parameter spread. In this case, the step width reduction after the $k\simeq 6000$ junction is due to the decrease of $n=1$ step (Figure~\ref{fig:US-PJVS-steps_vs_jj-nmax1_Ic5mA}). As from Table~\ref{tab:EU-US-PJVS}, the overall $N_\mathrm{JJ}$ junctions are evenly divided in $l=32$ CPW lines, with $M_\mathrm{JJ}\simeq \num{8400}$ JJs each: it can be observed from Figure~\ref{fig:US-PJVS-min_step_width_vs_jj-nmax1_Ic5mA} that, at  $k\simeq\num{8400}$, the quantum step width is approximately \qty{2}{\milli\ampere}, in good agreement with \cite{howe2014nist}.
Provided that the quantum-locking range threshold is \qty{1}{\milli\ampere}, the number of junctions per line can be raised to approximately $M_\mathrm{JJ}\sim \num{15000}$. Consequently, having $l=18$ parallel CPW lines would suffice, although it would not be enough to decrease the number of splitting stages $s$ by one ($\log_\mathrm{2}18\simeq 4.17$).
Therefore, higher microwave power, with a possible increase of $I_\mathrm{c}$, would be needed to reduce the number of splitting stages $s$ by one and to halve the parallel lines $l$.

\section{\qty{10}{\volt} EU-PJVS with $n_{\mathrm{max}}=3$} \label{subsec:pjvs_nmax3}

\begin{figure*}[tbhp]
	\centering
	\begin{subfigure}{0.49\textwidth}
		\centering
		\includegraphics[width=\linewidth]{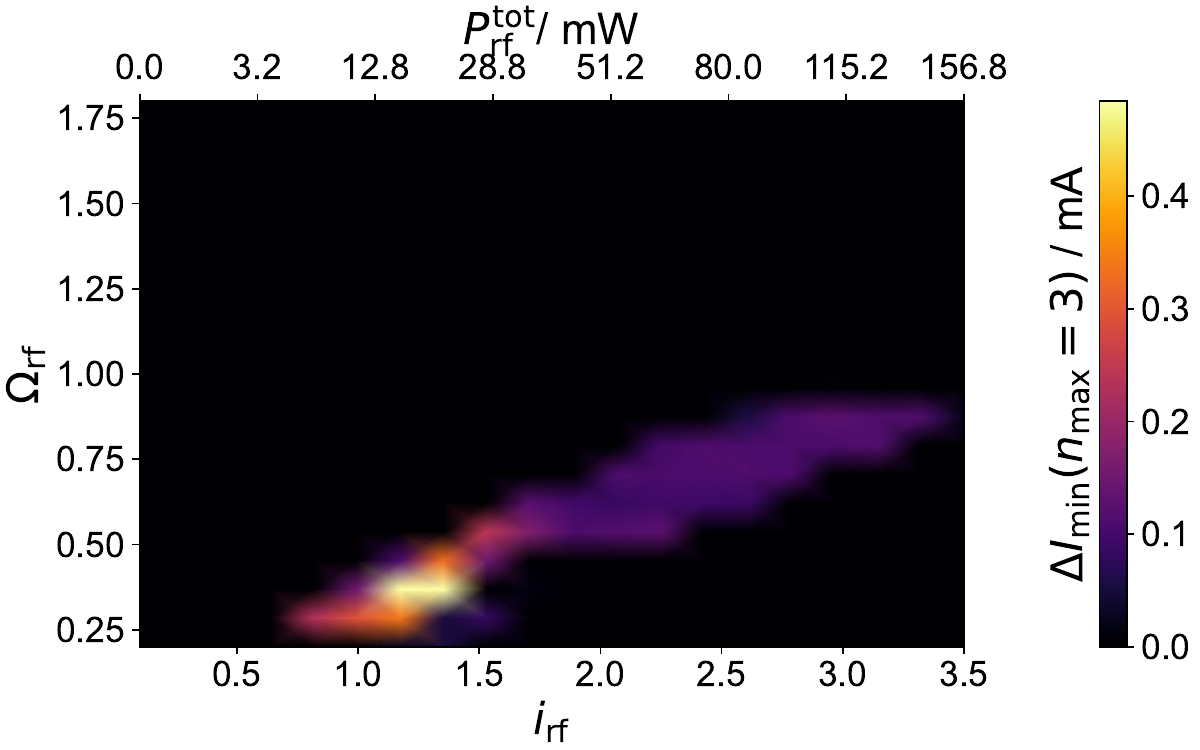}
		\caption{$I_\mathrm{c}=\qty{5}{\milli\ampere}$, $M_\mathrm{JJ}=365$}
		\label{fig:EU-PJVS-cmap_Ptot-fc-DeltaI-nmax_3-sigma_0.05-Ic_5mA-Nstrips_64}
	\end{subfigure}
		\begin{subfigure}{0.49\textwidth}
		\centering
		\includegraphics[width=\linewidth]{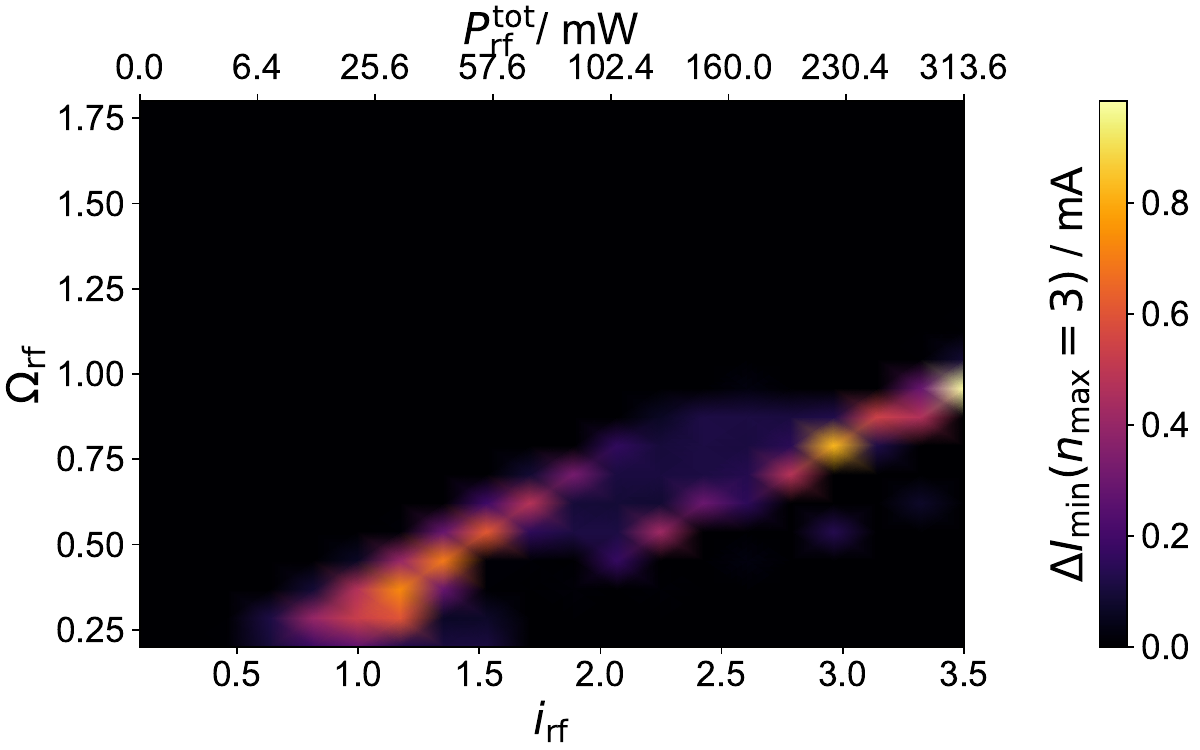}
		\caption{$I_\mathrm{c}=\qty{5}{\milli\ampere}$, $M_\mathrm{JJ}=185$}
		\label{fig:EU-PJVS-cmap_Ptot-fc-DeltaI-nmax_3-sigma_0.05-Ic_5mA-Nstrips_128}
	\end{subfigure}
	\begin{subfigure}{0.49\textwidth}
		\centering
		\includegraphics[width=\linewidth]{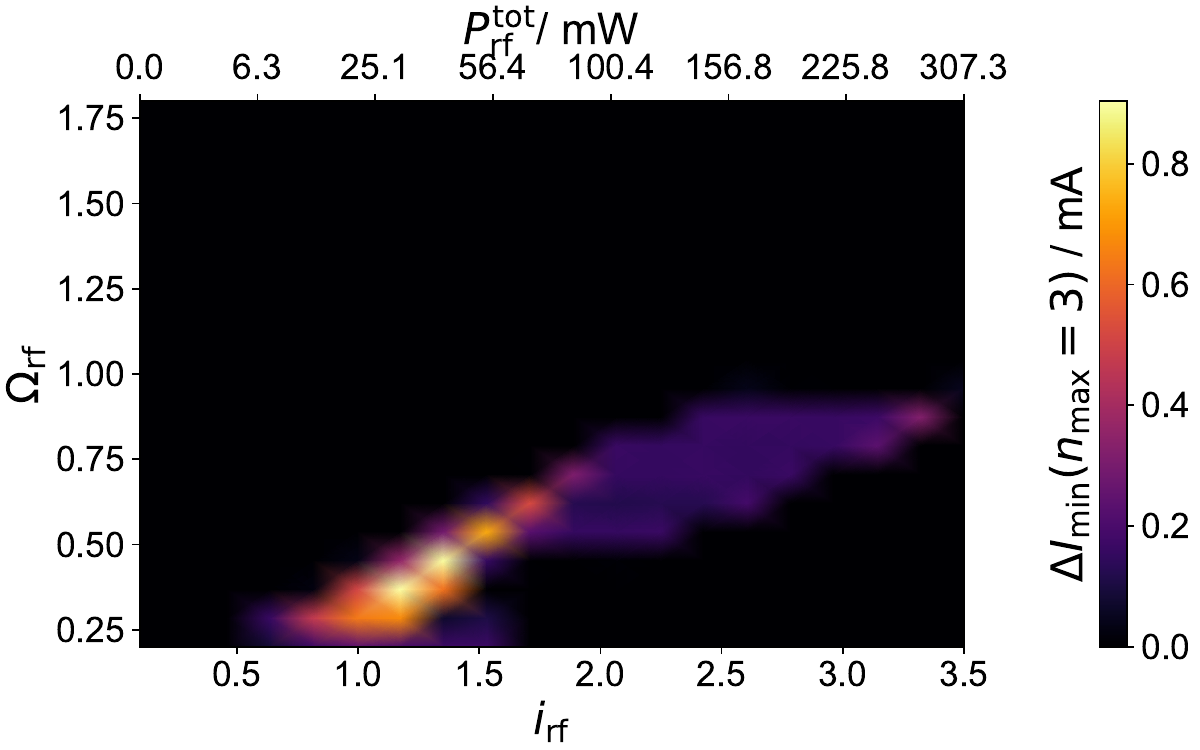}
		\caption{$I_\mathrm{c}=\qty{7}{\milli\ampere}$, $M_\mathrm{JJ}=365$}
		\label{fig:EU-PJVS-cmap_Ptot-fc-DeltaI-nmax_3-sigma_0.05-Ic_7mA-Nstrips_64}
	\end{subfigure}
		\begin{subfigure}{0.49\textwidth}
		\centering
		\includegraphics[width=\linewidth]{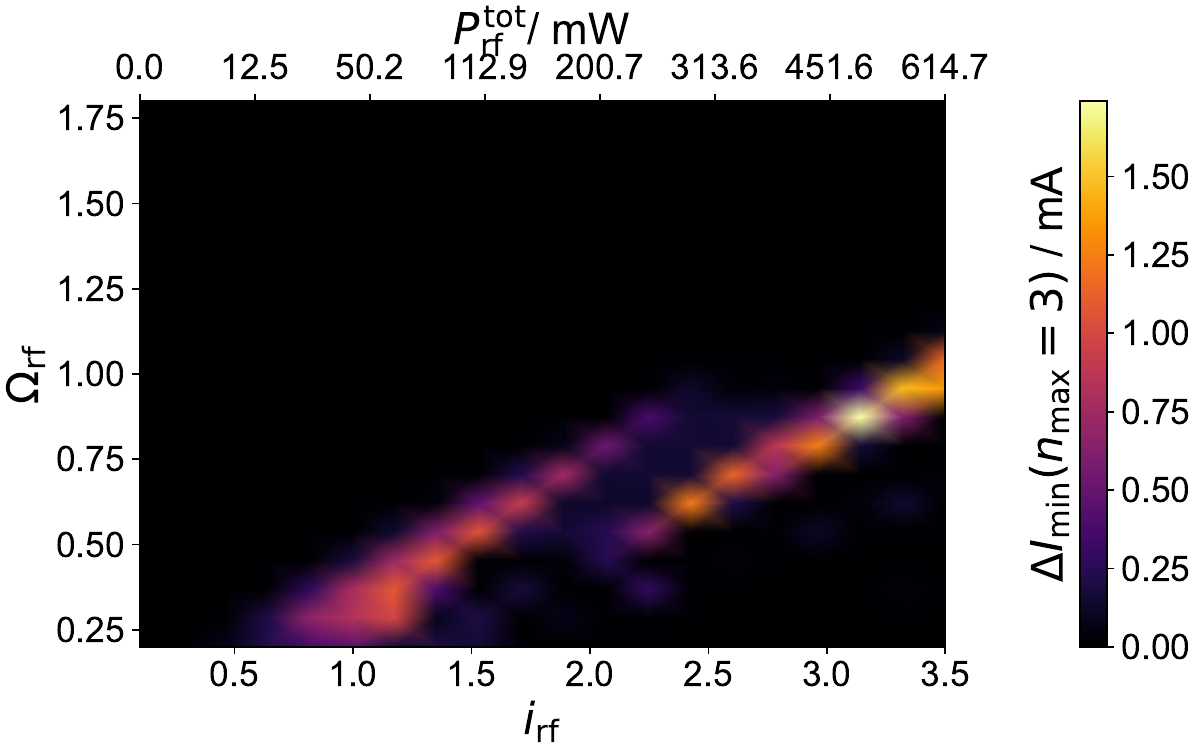}
		\caption{$I_\mathrm{c}=\qty{7}{\milli\ampere}$, $M_\mathrm{JJ}=185$}
		\label{fig:EU-PJVS-cmap_Ptot-fc-DeltaI-nmax_3-sigma_0.05-Ic_7mA-Nstrips_128}
	\end{subfigure}
	\begin{subfigure}{0.49\textwidth}
		\centering
		\includegraphics[width=\linewidth]{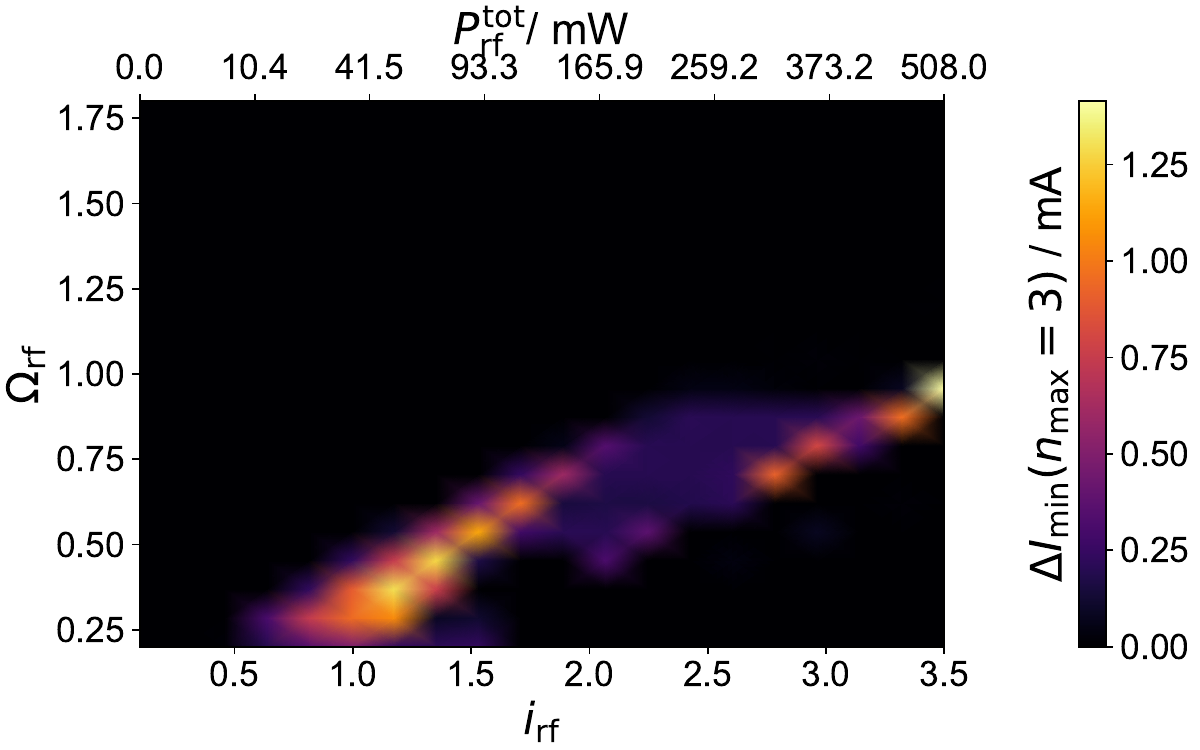}
		\caption{$I_\mathrm{c}=\qty{9}{\milli\ampere}$, $M_\mathrm{JJ}=365$}
		\label{fig:EU-PJVS-cmap_Ptot-fc-DeltaI-nmax_3-sigma_0.05-Ic_9mA-Nstrips_64}
	\end{subfigure}
	\begin{subfigure}{0.49\textwidth}
		\centering
		\includegraphics[width=\linewidth]{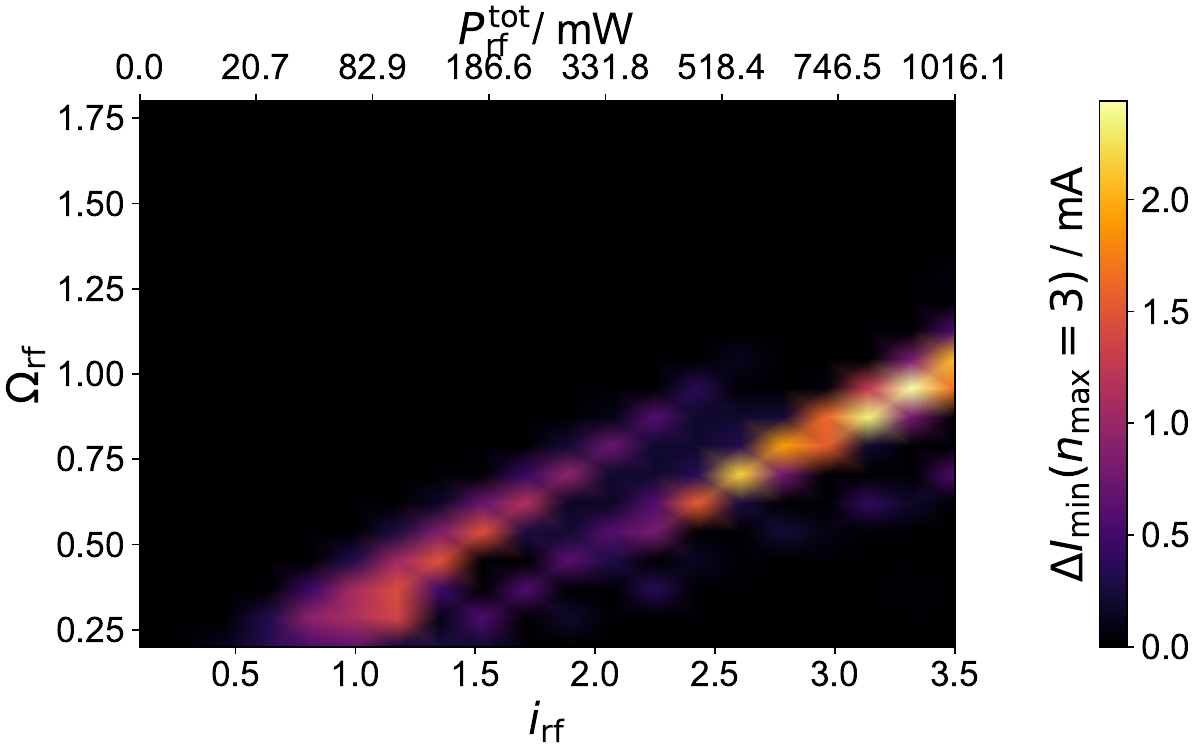}
		\caption{$I_\mathrm{c}=\qty{9}{\milli\ampere}$, $M_\mathrm{JJ}=185$}
		\label{fig:EU-PJVS-cmap_Ptot-fc-DeltaI-nmax_3-sigma_0.05-Ic_9mA-Nstrips_128}
	\end{subfigure}    
	\caption{Analysis of EU-PJVS type $M_\mathrm{JJ}$ Josephson junctions for $n_\mathrm{max}=3$, driven at $f_\mathrm{rf}=\qty{70}{\giga\hertz}$ and integrated in a $Z=\qty{16}{\ohm}$ transmission line, for three $I_\mathrm{c}$ values and two $M_\mathrm{JJ}$ values. Top row (a, b): $I_\mathrm{c}=\qty{5}{\milli\ampere}$. Center row (c, d): $I_\mathrm{c}=\qty{7}{\milli\ampere}$. Bottom row (e, f): $I_\mathrm{c}=\qty{9}{\milli\ampere}$. Left column (a, c, e): $M_\mathrm{JJ}=365$ ($l=64$, $s=6$). Right column (b, d, f): $M_\mathrm{JJ}=185$ ($l=128$, $s=7$)}
	\label{fig:EU-PJVS_5-7-9mA_nmax3}
	
\end{figure*}

The scenario with \(n_\mathrm{max}=3\) for the EU-PJVS is examined here. The results are summarized in Figure~\ref{fig:EU-PJVS_5-7-9mA_nmax3} for \(I_\mathrm{c}=\qty{5}{\milli\ampere}\), \(I_\mathrm{c}=\qty{7}{\milli\ampere}\), and \(I_\mathrm{c}=\qty{9}{\milli\ampere}\). As anticipated, increasing \(n_\mathrm{max}\) necessitates a higher level of critical current compared to the previous scenario, as four-order steps (\(n=0\), 1, 2, 3) must be maximized simultaneously. Here, \(M_\mathrm{JJ}\) is approximately 185 with \(l=128\) parallel lines (\(s=7\)), whereas it is around 365 with \(l=64\) and \(s=6\).

\noindent It can be observed that, for all the tested combinations, the best operating point is at $\Omega_\mathrm{rf}\simeq 0.4$ and $i_\mathrm{rf}\simeq 1.2$. From the single-junction case depicted in Figure~\ref{fig:min_step_colormaps_nmax3}, this point lies on the first crest (blue curve in Figure~\ref{fig:max_step_min_crests_nmax_3}), though operating points on the second crest also appear for larger $I_\mathrm{c}$ values, though with demanding rf power levels above \qty{300}{\milli\watt}.

\begin{figure}[htp]
	\centering
	\begin{subfigure}{0.48\textwidth}
		\centering
		\includegraphics[width=\linewidth]{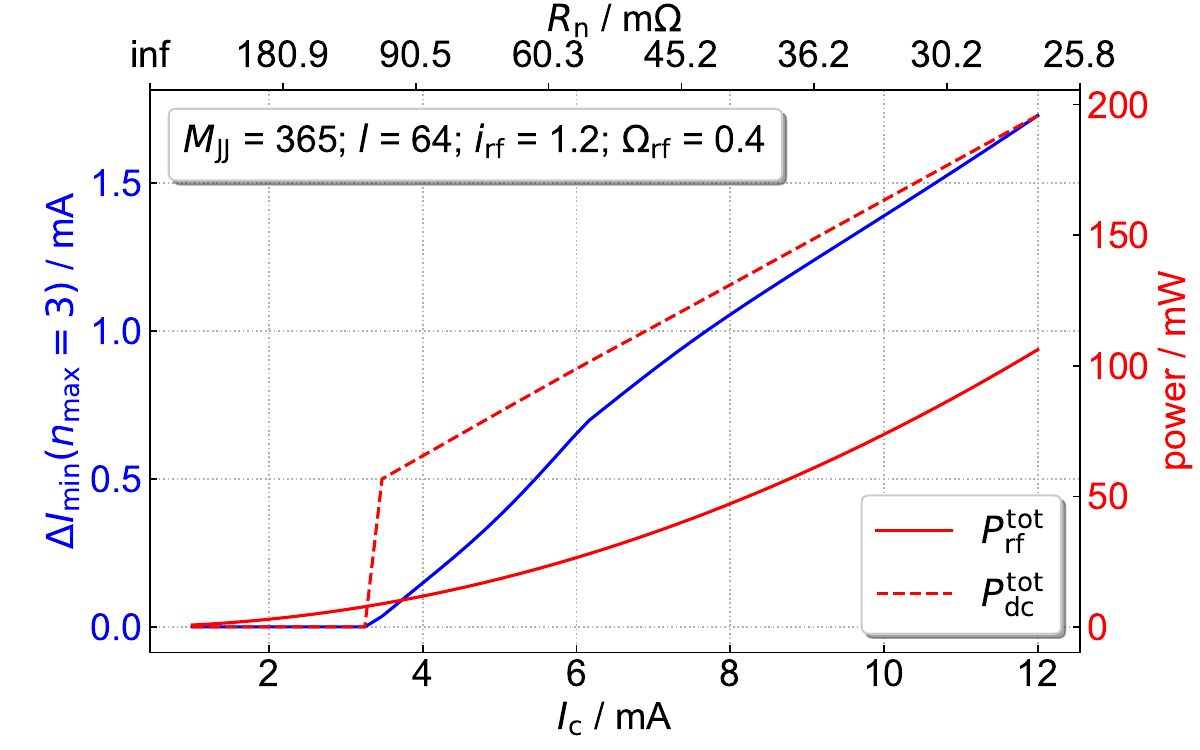}
		\caption{}
		\label{fig:EU-PJVS-nmax3-Ic_Ptot_for_1mA-Nstrips_64-irf_1.2-Orf_0.4}
	\end{subfigure}    
	\begin{subfigure}{0.48\textwidth}
		\centering
		\includegraphics[width=\linewidth]{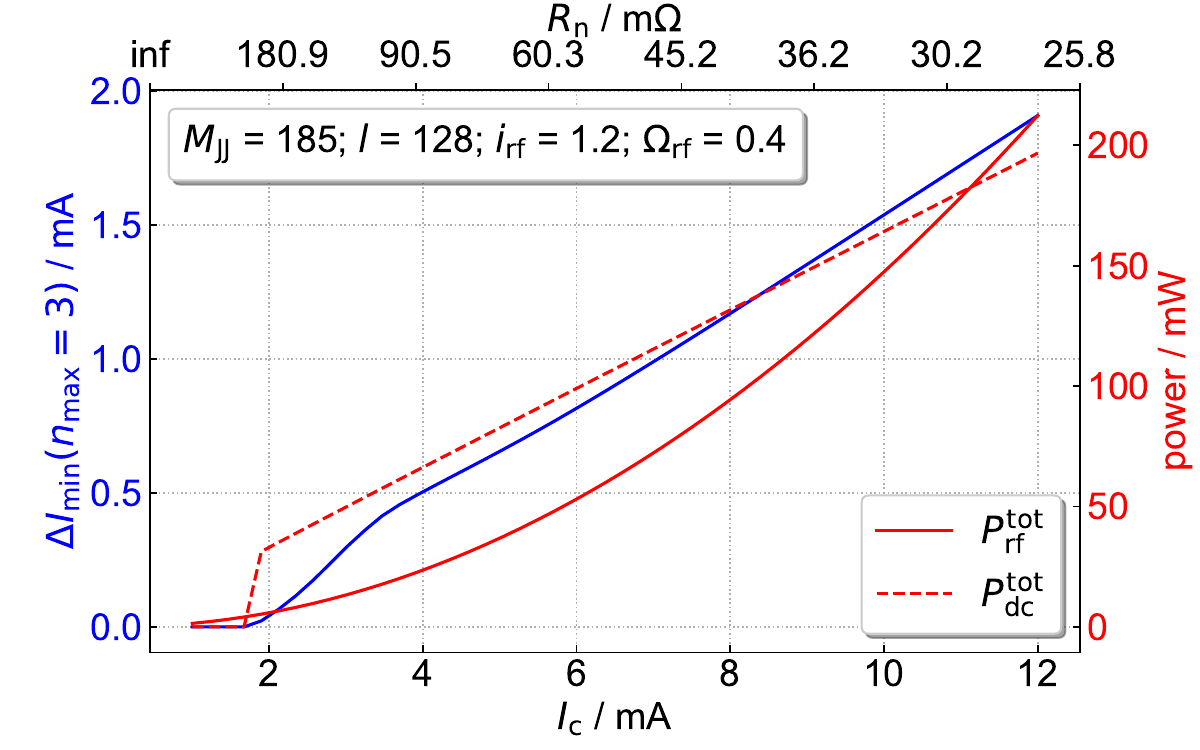}
		\caption{}
		\label{fig:EU-PJVS-nmax3-Ic_Ptot_for_1mA-Nstrips_128-irf_1.2-Orf_0.4}
	\end{subfigure}
	\caption{Analysis of the \qty{10}{\volt} EU-PJVS: minimum step width $\Delta I_\mathrm{min}(n_\mathrm{max}=3)$ (blue curve, left y-axis), total microwave power $P_\mathrm{rf}^\mathrm{tot}$ and total dc power $P_\mathrm{dc}^\mathrm{tot}$ (red curves, right y-axis) as a function of nominal critical current $I_\mathrm{c}$ (bottom x-axis) and normal resistance $R_\mathrm{n}$ (top x-axis) for $n_\mathrm{max}=3$ with $i_\mathrm{rf}=1.2$ and $\Omega_\mathrm{rf}=0.4$. a) $M_\mathrm{JJ}=365$ and $l=64$; b) $M_\mathrm{JJ}=185$ and $l=128$.}
	\label{fig: Ic_DeltaI_Ptot_Rn_nmax3_optimal}
\end{figure}

With \( l = 64 \) (Figure~\ref{fig:EU-PJVS-nmax3-Ic_Ptot_for_1mA-Nstrips_64-irf_1.2-Orf_0.4}), the minimum critical current \( I_\mathrm{c} \) for the \qty{1}{\milli\ampere} target is approximately \qty{8}{\milli\ampere}, accompanied by an acceptable total rf power \( P_\mathrm{rf}^\mathrm{tot} \sim \qty{50}{\milli\watt} \). A lower \( I_\mathrm{c} \) value of \qty{7}{\milli\ampere} is feasible with \( l = 128 \) parallel lines, where \( P_\mathrm{rf}^\mathrm{tot} \sim \qty{75}{\milli\watt} \) (Figure~\ref{fig:EU-PJVS-nmax3-Ic_Ptot_for_1mA-Nstrips_128-irf_1.2-Orf_0.4}).

\begin{figure}[tbh]
	\centering
	\begin{subfigure}{0.48\textwidth}
		\centering
		\includegraphics[width=\linewidth]{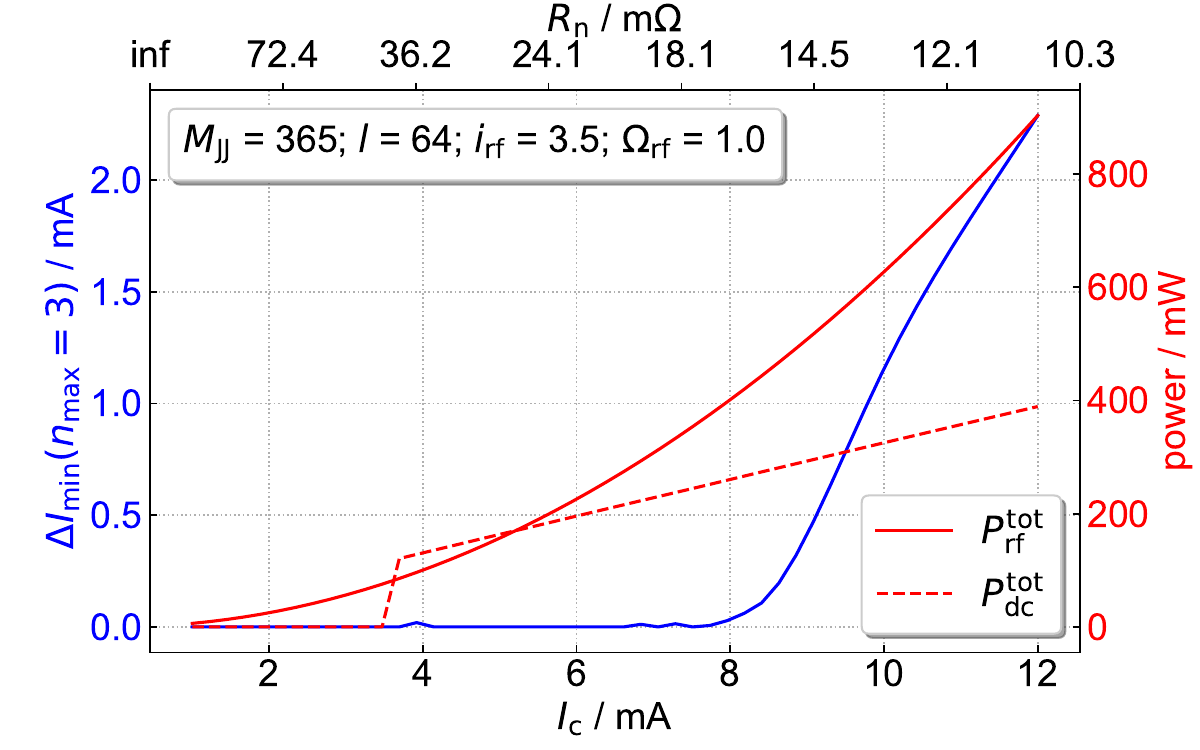}
		\caption{}
		\label{fig:EU-PJVS-nmax3-Ic_Ptot_for_1mA-Nstrips_64-irf_3.5-Orf_1.0}
	\end{subfigure}    
	\begin{subfigure}{0.48\textwidth}
		\centering
		\includegraphics[width=\linewidth]{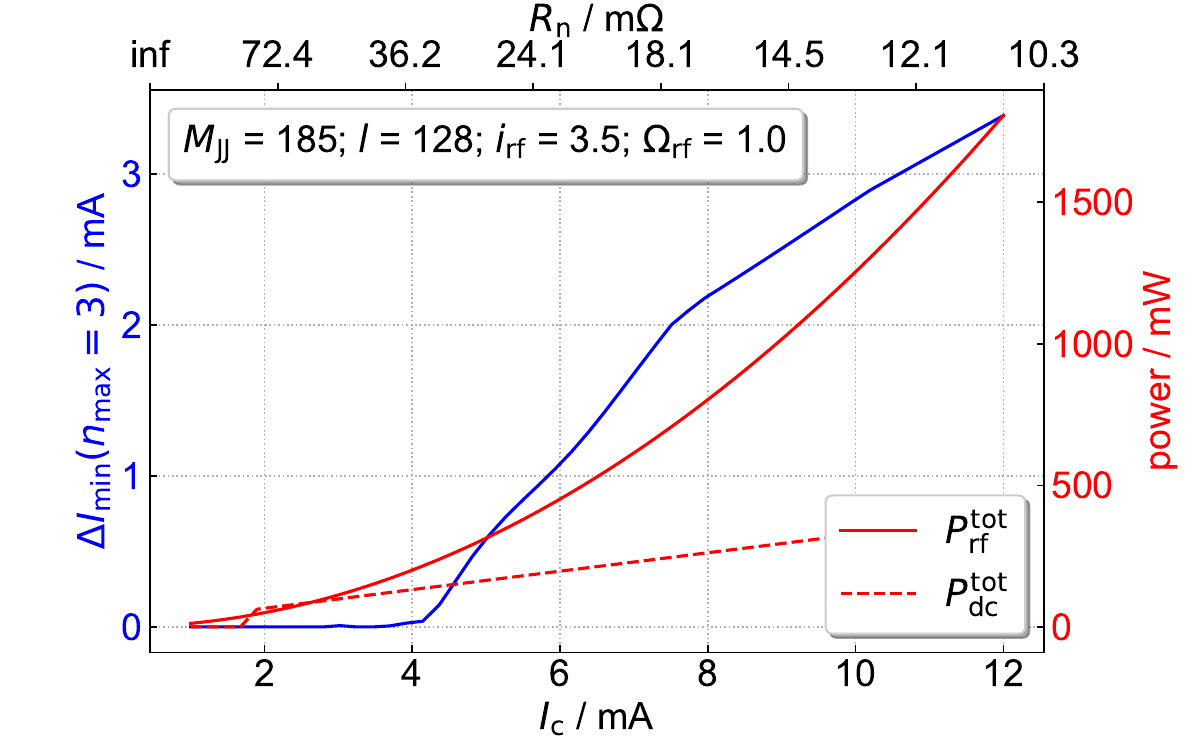}
		\caption{}
		\label{fig:EU-PJVS-nmax3-Ic_Ptot_for_1mA-Nstrips_128-irf_3.5-Orf_1.0}
	\end{subfigure}
	\caption{Analysis of the \qty{10}{\volt} EU-PJVS: minimum step width $\Delta I_\mathrm{min}(n_\mathrm{max}=3)$ (blue curve, left y-axis), total microwave power $P_\mathrm{rf}^\mathrm{tot}$ and total dc power $P_\mathrm{dc}^\mathrm{tot}$ (red curves, right y-axis) as a function of nominal critical current $I_\mathrm{c}$ (bottom x-axis) and normal resistance $R_\mathrm{n}$ (top x-axis) for $n_\mathrm{max}=3$ with $i_\mathrm{rf}=3.5$ and $\Omega_\mathrm{rf}=1$. a) $M_\mathrm{JJ}=365$ and $l=64$; b) $M_\mathrm{JJ}=185$ and $l=128$.}
	\label{fig: Ic_DeltaI_Ptot_Rn_nmax3_non_optimal}
\end{figure}

As discussed for the case of $n_{\text{max}}=2$ (Sec.~\ref{subsec:eu-pjvs-nmax2}), the possibility of using the unoptimized present junction technology has been investigated. From the single junction case (Figure~\ref{fig:min_step_colormaps_nmax3}), the $\Omega_{\text{rf}}= 1$ horizontal line  crosses the two main $\Delta i_\mathrm{min}$ crests at $i_\text{rf}\simeq 2.3$ and $i_{\text{rf}}\simeq 3.5$. However, when considering thousands of JJs (Figure~\ref{fig:EU-PJVS_5-7-9mA_nmax3}), the first crest disappears entirely for $\Omega_{\text{rf}}=1$. Therefore, the potential operating point must lie on the second crest at $i_{\text{rf}}=3.5$, which leads to a significant amount of microwave power. Figure~\ref{fig: Ic_DeltaI_Ptot_Rn_nmax3_non_optimal} depicts the correlation between step width and critical current under this non-ideal condition, emphasizing the excessive amount of dc and rf power (exceeding \qty{600}{\milli\watt}) that must be dissipated within the device.

\section{\qty{10}{\volt} US-PJVS with $n_{\mathrm{max}}=3$}
\begin{figure}[htp]
	\centering
	\begin{subfigure}{0.48\textwidth}
		\centering
		\includegraphics[width=\linewidth]{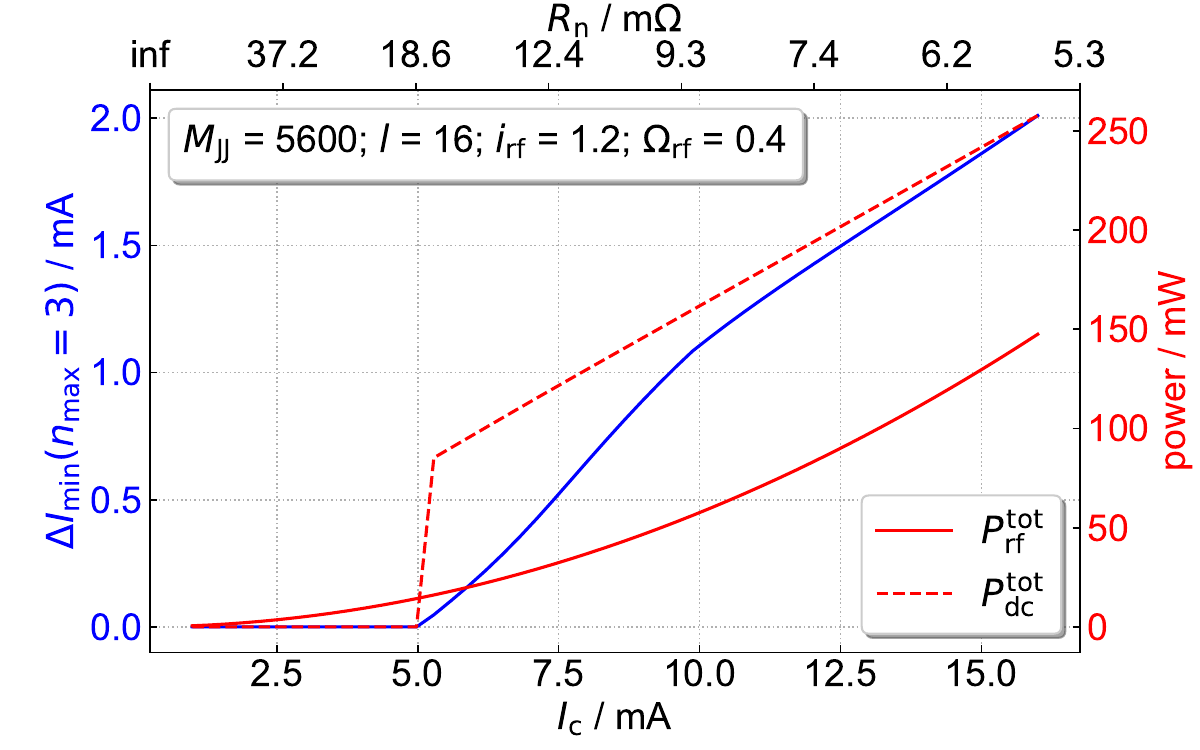}
		\caption{}
		\label{fig:US-PJVS-nmax3-Ic_Ptot_for_1mA-Nstrips_16-irf_1.2-Orf_0.4}
	\end{subfigure}    
	\begin{subfigure}{0.48\textwidth}
		\centering
		\includegraphics[width=\linewidth]{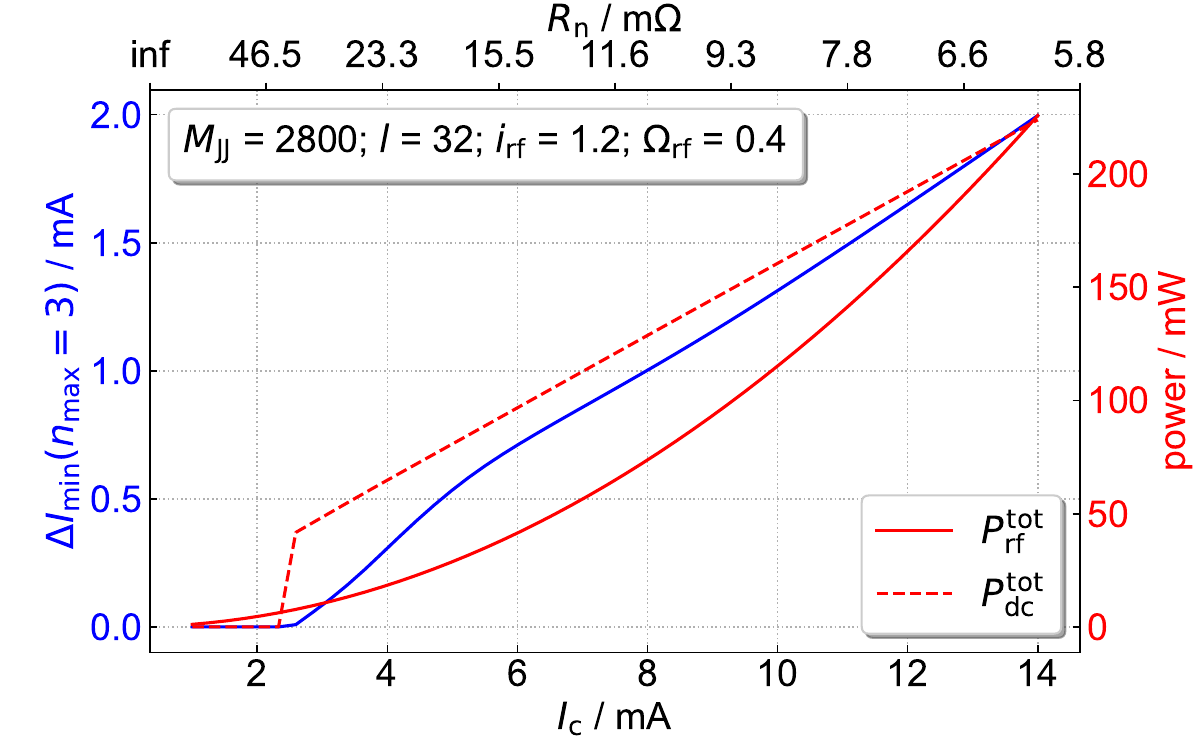}
		\caption{}
		\label{fig:US-PJVS-nmax3-Ic_Ptot_for_1mA-Nstrips_32-irf_1.2-Orf_0.4}
	\end{subfigure}
	\caption{Analysis of the \qty{10}{\volt} US-PJVS: minimum step width $\Delta I_\mathrm{min}(n_\mathrm{max}=3)$ (blue curve, left y-axis), total microwave power $P_\mathrm{rf}^\mathrm{tot}$ and total dc power $P_\mathrm{dc}^\mathrm{tot}$ (red curves, right y-axis) as a function of nominal critical current $I_\mathrm{c}$ (bottom x-axis) and normal resistance $R_\mathrm{n}$ (top x-axis) for $n_\mathrm{max}=3$ with $i_\mathrm{rf}=1.2$ and $\Omega_\mathrm{rf}=0.4$. a) $M_\mathrm{JJ}=5600$ and $l=16$; b) $M_\mathrm{JJ}=2800$ and $l=32$.}
	\label{fig:US-PJVS-Ic_DeltaI_Ptot_Rn_nmax3_optimal}
\end{figure}

\begin{figure}[tbh]
	\centering
	\begin{subfigure}{0.48\textwidth}
		\centering
		\includegraphics[width=\linewidth]{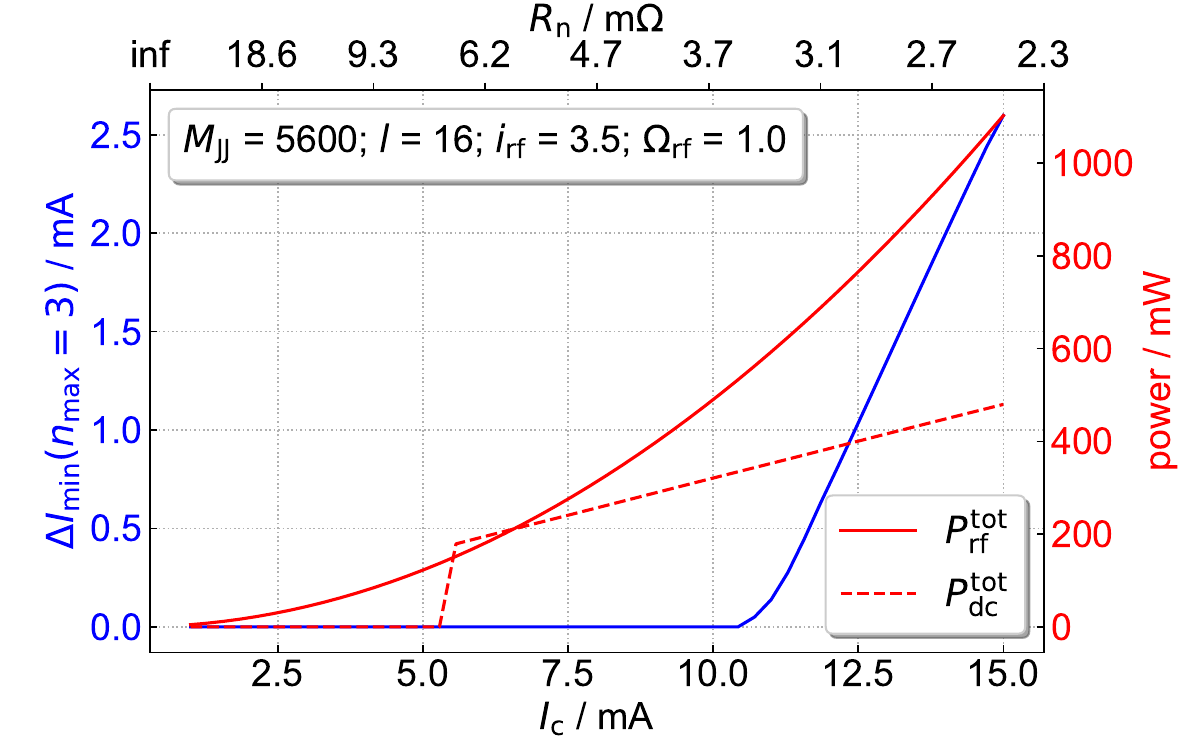}
		\caption{}
		\label{fig:US-PJVS-nmax3-Ic_Ptot_for_1mA-Nstrips_16-irf_3.5-Orf_1.0}
	\end{subfigure}    
	\begin{subfigure}{0.48\textwidth}
		\centering
		\includegraphics[width=\linewidth]{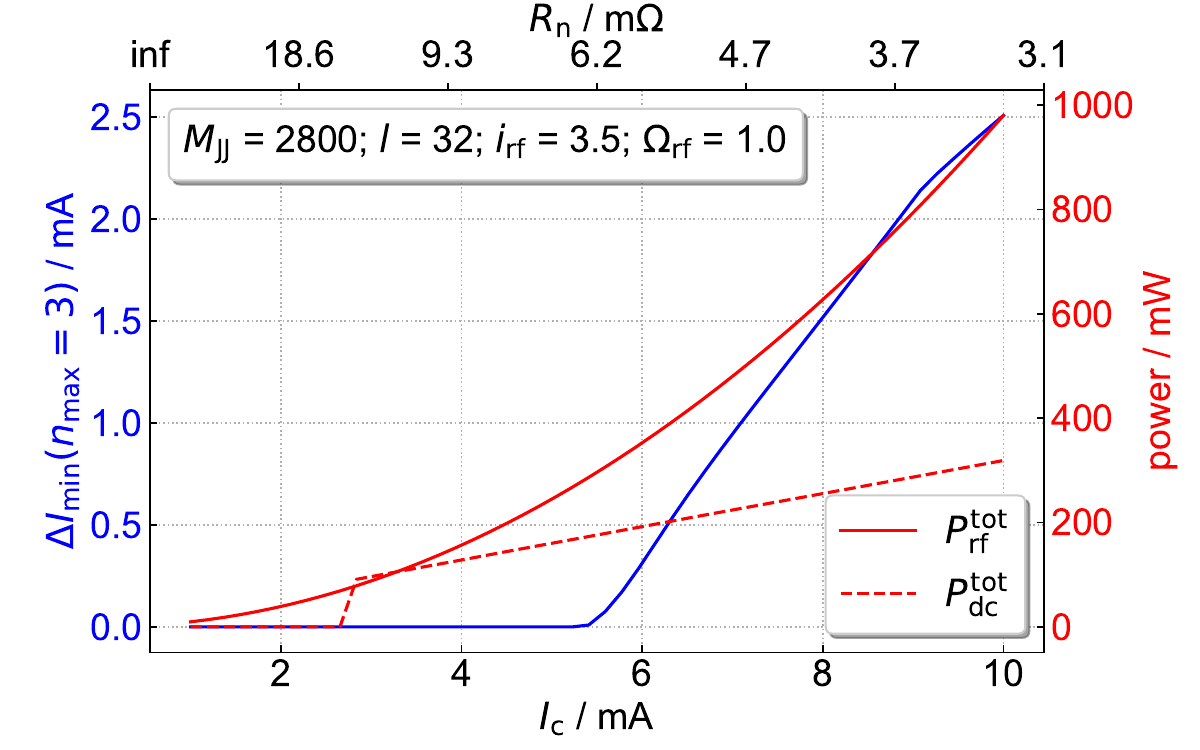}
		\caption{}
		\label{fig:US-PJVS-nmax3-Ic_Ptot_for_1mA-Nstrips_32-irf_3.5-Orf_1.0}
	\end{subfigure}
	\caption{Analysis of the \qty{10}{\volt} US-PJVS: minimum step width $\Delta I_\mathrm{min}(n_\mathrm{max}=3)$ (blue curve, left y-axis), total microwave power $P_\mathrm{rf}^\mathrm{tot}$  and total dc power $P_\mathrm{dc}^\mathrm{tot}$ (red curves, right y-axis) as a function of nominal critical current $I_\mathrm{c}$ (bottom x-axis) and normal resistance $R_\mathrm{n}$ (top x-axis) for $n_\mathrm{max}=3$ with $i_\mathrm{rf}=3.5$ and $\Omega_\mathrm{rf}=1$. a) $M_\mathrm{JJ}=5600$ and $l=16$; b) $M_\mathrm{JJ}=2800$ and $l=32$.}
	\label{fig:US-PJVS-Ic_DeltaI_Ptot_Rn_nmax3_non_optimal}
\end{figure}

By extending the usability of the step order to \( n=3 \) with US-PJVS design, the number of junctions per line for \( l=32 \) and \( l=16 \) decreases to approximately \( M_\mathrm{JJ} \simeq \num{2800} \) and \( M_\mathrm{JJ} \simeq \num{5600} \), respectively. Both scenarios are examined here, and the results are illustrated in Figure~\ref{fig:US-PJVS-Ic_DeltaI_Ptot_Rn_nmax3_optimal}. 
With optimal values of ($i_\mathrm{rf}$, $\Omega_\mathrm{rf}$) = (1.2, 0.4), usable step widths are achieved by increasing both the microwave power to approximately \( \sim \qty{60}{\milli\watt} \) and the minimum critical current to around \( \qty{10}{\milli\ampere} \) for \( l=16 \) and \( \qty{7}{\milli\ampere} \) for \( l = 32 \). 
Conversely, under the non-optimized parameters \( \Omega_\mathrm{rf}=1 \) and \( i_\mathrm{rf}=3.5 \), the required power values become unsustainable (\( > \qty{700}{\milli\watt} \)), as shown in Figure~\ref{fig:US-PJVS-Ic_DeltaI_Ptot_Rn_nmax3_non_optimal}.

The main results for both EU and US-PJVS types with \( n_{\text{max}}=3 \) are summarized in Table~\ref{tab:comparison_PJVS_nmax3}.

\begin{table}[H]
	\centering
	\caption{Comparison of simulated EU-PJVS and US-PJVS for $n_\mathrm{max}=3$ and minimum $\Delta I_\mathrm{min}=\qty{1}{\milli\ampere}$ for different normalized rf frequencies $\Omega_\mathrm{rf}$, rf currents $i_\mathrm{rf}$, number of transmission lines $l$ (as well as junctions per line $M_\mathrm{JJ}$ and number of splitting stages $s$).}
	
	\label{tab:comparison_PJVS_nmax3}
	\begin{adjustbox}{width=\textwidth}
		
		\begin{tabular}{r|c c|c c|c c|c c}
			\toprule
			& \multicolumn{4}{c|}{\textbf{EU-PJVS}} & \multicolumn{4}{c}{\textbf{US-PJVS}} \\
			\cmidrule(lr){2-5} \cmidrule(lr){6-9}
			& \multicolumn{2}{c|}{$M_\mathrm{JJ}= 365 $, $l=64$, $s=6$} & \multicolumn{2}{c|}{$M_\mathrm{JJ}=185$, $l=128$, $s=7$} 
			& \multicolumn{2}{c|}{$M_\mathrm{JJ}=5600$, $l=16$, $s=4$} & \multicolumn{2}{c}{$M_\mathrm{JJ}=2800$, $l=32$, $s=5$} \\
			\midrule
			& \makecell{$\Omega_\mathrm{rf}=0.4$\\$i_\mathrm{rf} = 1.2$} & \makecell{$\Omega_\mathrm{rf}=1$\\$i_\mathrm{rf} = 3.5$} & 
			\makecell{$\Omega_\mathrm{rf}=0.4$\\$i_\mathrm{rf} = 1.2$} & \makecell{$\Omega_\mathrm{rf}=1$\\$i_\mathrm{rf} = 3.5$} &
			\makecell{$\Omega_\mathrm{rf}=0.4$\\$i_\mathrm{rf} = 1.2$} & \makecell{$\Omega_\mathrm{rf}=1$\\$i_\mathrm{rf} = 3.5$} & 
			\makecell{$\Omega_\mathrm{rf}=0.4$\\$i_\mathrm{rf} = 1.2$} & \makecell{$\Omega_\mathrm{rf}=1$\\$i_\mathrm{rf} = 3.5$} \\
			\midrule
			$I_\mathrm{c}$ / \qty{}{\milli\ampere}  & 7.7 & 9.8 & 7.1 & 5.8 & 9.5& 12.4& 8 & 7.1\\			
			$R_\mathrm{n}$ / \qty{}{\milli\ohm}                     & 47 & 14.8 & 51 & 25 & 9.8 & 3 & 11.6& 5.2\\
			$P_\mathrm{rf}^\mathrm{tot}$ / \qty{}{\milli\watt} & 43 & 601 & 73& 425& 52& 759& 74& 492\\			
			$P_\mathrm{dc}^\mathrm{tot}$ / \qty{}{\milli\watt}                    & 125 & 319 & 116 & 191& 153 & 398 & 129 & 226 \\
			$P_\mathrm{rf}^\mathrm{tot} + P_\mathrm{dc}^\mathrm{tot}$ / \qty{}{\milli\watt}& 168 & 920 & 189 & 616& 205 & 1157 & 203 & 718\\			
			\bottomrule
		\end{tabular}
	\end{adjustbox}
\end{table}


%

\newpage

\section*{References}


\providecommand{\newblock}{}

\end{document}